\documentstyle[12pt]{article}
\newlength{\abstractwidth}
\renewcommand{\thanks}[1]{\footnote{#1}} 

\newcommand{\be}{\begin{equation}}
\newcommand{\bea}{\begin{eqnarray}}
\newcommand{\eea}{\end{eqnarray}}
\newcommand{\ee}{\end{equation}}
\newcommand{\N}{{\cal N}}
\newcommand{\<}{\langle}
\renewcommand{\>}{\rangle}
\def\ba{\begin{eqnarray}}
\def\ea{\end{eqnarray}}

\def\A{{\cal A}}

\def\D{{\cal D}}
\def\E{{\cal E}}

\def\N{{\cal N}}
\def\O{{\cal O}}

\def\X{{\cal X}}

\def\Im{{\rm Im}}
\def\tr{{\rm tr}}
\def\det{{\rm det}}
\def\sdet{{\rm sdet}}
\def\str{{\rm str}}
\def\half{ {1\over 2}}

\def\p{\partial}
\def\pz{\partial _z}
\def\pv{\partial _v}
\def\pw{\partial _w}
\def\w{{\bf w}}

\def\z{{\bf z}}
\def\tet{\vartheta}
\def\dwplus{\D _+ ^\w}

\def\dzplus{\D _+ ^\z}
\def\chiz{{\chi _{\bar z}{} ^+}}
\def\chiw{{\chi _{\bar w}{} ^+}}
\def\chiu{{\chi _{\bar u}{} ^+}}
\def\chiv{{\chi _{\bar v}{} ^+}}
\def\chix{{\chi _{\bar x}{} ^+}}

\def\hO{\hat\Omega}

\def\o{\omega}

\begin{document}
\baselineskip=16pt

\begin{flushright}
UCLA/01/TEP/25 \\
Columbia/Math/01
\end{flushright}

\bigskip

\begin{center}
{\Large \bf TWO-LOOP SUPERSTRINGS \ II}
 \\
\bigskip
{\large \bf The Chiral Measure on Moduli Space\footnote{Research supported in
part by National Science Foundation grants PHY-98-19686 and DMS-98-00783,
and by the Institute for Pure and Applied Mathematics under NSF grant
DMS-9810282.}}

\bigskip\bigskip

{\large Eric D'Hoker$^a$ and D.H. Phong$^b$} \\ 

\bigskip

$^a$ \sl Department of Physics and \\
\sl Institute for Pure and Applied Mathematics (IPAM) \\
\sl University of California, Los Angeles, CA 90095 \\
$^b$ \sl Department of Mathematics \\ 
\sl Columbia University, New York, NY 10025

\end{center}

\bigskip\bigskip

\begin{abstract}
 
A detailed derivation from first principles is given for the
unambiguous and slice-independent formula for the two-loop 
superstring chiral measure which was announced in the first
paper of this series. Supergeometries are projected onto
their super period matrices, and the integration over odd
supermoduli is performed by integrating over the fibers of this
projection. The subtleties associated with this procedure are 
identified. They require the inclusion of some new finite-dimensional
Jacobian superdeterminants, a deformation of the worldsheet 
correlation functions using the stress tensor, and perhaps 
paradoxically, another additional gauge choice, ``slice 
$\hat\mu$ choice", whose independence also has to be established.
This is done using an important correspondence between   
superholomorphic notions with respect to a supergeometry 
and holomorphic notions with respect to its super period matrix. 
Altogether, the subtleties produce precisely the corrective terms
which restore the independence of the resulting gauge-fixed
formula under infinitesimal changes of gauge-slice.
This independence is a key criterion for any gauge-fixed
formula and hence is verified in detail. 
 
\end{abstract}

\vfill\eject

\baselineskip=15pt
\setcounter{equation}{0}
\setcounter{footnote}{0}

\vfill\eject

\section{Introduction}

In string theory, Feynman rules correspond to a chiral measure on the
moduli space of Riemann surfaces. At a given loop order $h$, the
worldsheet is a surface of genus $h$, and string scattering amplitudes
are given by integrals over all geometries on  this surface. The chiral
string measure results from factoring out all symmetries from these
integrals.

\medskip

For superstrings, the problem of determining the  chiral string
measure has remained intractable to this day. The main difficulty
is the presence of {\sl supermoduli} for worldsheets of non-trivial
topology \cite{fms, superm}. The usual gauge fixing methods express 
superstring amplitudes as measures on {\sl supermoduli space}, which
incorporates odd variables over moduli space~\cite{dp88}. The odd
variables have to be integrated out in order to produce the desired
chiral string measure on moduli space. However, all such attempts so far 
have run into serious problems.

\medskip

A recurrent problem is the occurrence of apparent ambiguities,
which appear as total derivatives on local coordinate patches
of moduli space. In the early Ansatz proposed by Friedan, Martinec, and
Shenker \cite{fms} based on BRST invariance, the chiral measure includes
$2h-2$ picture-changing operators, inserted at $2h-2$ arbitrary points.
Although the Ansatz should be invariant under changes of insertion 
points, it actually changes by total derivatives.
In subsequent attempts to derive the chiral measure from
first principles by gauge fixing the integrals over all geometries,
similar total derivatives arise from changes of gauge slices in the
gauge fixing process \cite{vv}. These derivatives have been
attributed to ambiguities in the general theory of
fermionic integrals \cite{ars}.

\medskip

The total derivatives pose serious difficulties, because they
are defined only on local coordinate patches, and cannot
be reduced to boundary terms by Stokes' theorem. There have been many attempts to
overcome these difficulties.  Since the total derivative
ambiguities occur on local patches and are reminiscent of Cech cohomology,
a program was begun in \cite{v} for the construction of counterterms by
a series of descent equations similar to gauge anomalies.
Another approach has been to assume that geometric conditions
may exist under which the ambiguities are global exact
differentials on moduli space, and analyze which contributions
can arise in this way from the boundary of moduli space \cite{ams}.
Yet another approach is to put the insertion points at
certain privileged points on the worldsheet, and hope that this 
produces the correct answer. A possible choice is the unitary
gauge, where the insertion points are put at the $2h-2$
zeroes of a holomorphic Abelian differential \cite{lp}. 
The unitary gauge has the advantage of producing an explicit
cancellation between ghost and longitudinal degrees of freedom,
but introduces a new arbitrariness in the choice of Abelian
differential. A related approach is to work directly in light-cone gauge, 
and put the insertion points at the branch points of the corresponding
Mandelstam diagrams \cite{mand}. In genus $2$, several Ans\" atze have
been proposed in the hyperelliptic representation, with a possible
resolution left ultimately to factorization conditions \cite{fact}.
Operator methods \cite{opf}
as well as group theoretic constructions  \cite{neveu}
of string amplitudes have
also been developed.
In more radical departures, powerful tools from algebraic geometry have
been  brought to bear, assuming relations between string amplitudes and
deep geometric properties such as slopes of effective divisors on moduli
space \cite{mhns} or invoking formal
constructions from super algebraic geometry
\cite{asg}. Finally, there have been
suggestions to resolve the ambiguities by shifting
the superstring background and appealing to the
Fischler-Susskind mechanism \cite{ln}.
All these attempts have led to different, competing
expressions for the string chiral measure, with none
emerging as the more cogent choice. Worse still,
at the most fundamental level,
ambiguities are simply unacceptable, since they would signal
a breakdown of local gauge invariance.

\medskip

The purpose of this series of papers
is to show that, at least in genus $h=2$
and contrary to earlier worries,
superstring scattering amplitudes do not suffer from any ambiguity,
and in fact can be evaluated completely explicitly in terms
of modular forms and sections of vector bundles over the moduli
space of Riemann surfaces. 
The case of genus $h=2$
is the simplest case when supermoduli difficulties
must be addressed in all scattering amplitudes. 
Actually, our  methods are quite general, and
should apply to arbitrary genus $h$. It is the complexity of the actual
calculations which restricts presently our implementation to the
case of genus 2. In paper I of the series \cite{I},
we had provided a summary of the main formulas we obtained.
In the present paper II, we provide the detailed derivation of
the first step in our approach, namely a careful new gauge-fixing
process. This gauge-fixing process results in the expression
(\ref{finamp}) below. This expression together with the proof of its
invariance under infinitesimal changes $\delta (\chi_{\alpha})_ {\bar
z}{}^+ =-2\partial_{\bar z}\xi_{\alpha}^+$ of the gauge slice
$\chiz = \sum_{\alpha=1}^2
\zeta^{\alpha} (\chi_{\alpha})_ {\bar z}{}^+$ are the main results of II.

\medskip

The source of all the earlier difficulties turns out to be an ill-defined
projection from supermoduli space to moduli space. The local worldsheet
supersymmetry of the string scattering amplitudes requires both a zweibein
$e_m{}^a$ and a gravitino field $\chi_m{}^{\alpha}$. Together,
they correspond to a superzweibein $E_M{}^A$ in Wess-Zumino gauge. The
earlier gauge fixing procedures had been implicitly based on the
projection
\be
\label{bpr}
E_M{}^A\longrightarrow e_m{}^a
\ee
which seemed the obvious way of descending from superzweibeins to
zweibeins. However, this projection is actually ill-defined from
supermoduli space to moduli space,  because superzweibeins equivalent
under supersymmetry may project to zweibeins with distinct complex
structures. In practice, gauge fixing procedures
based on (\ref{bpr}) resulted in a
dependence on the gravitino component $\chi_m{}^{\beta}$ of the gauge
slice chosen. We shall refer to this as {\sl slice $\chi$ dependence}.
The remedy proposed in this series of papers is to
use instead the well-defined projection (once a canonical homology
basis has been chosen)
\be
\label{gpr}
E_M{}^A\longrightarrow\hat\Omega_{IJ}
\ee
Here $\hat\Omega_{IJ}$ is the super period matrix, namely the
modification of the period matrix of $e_m{}^a$ which is invariant under
worldsheet supersymmetry
\be
\label{spm1}
\hat\Omega_{IJ}
=
\Omega_{IJ}-{i\over 8\pi}\int\int d^2z\,d^2w\ 
\omega_I(z)\chiz\hat S_{\delta}(z,w) \chiw\omega_J(w)
\ee
The $\omega_I(z)$ are the dual basis of holomorphic differentials,
and $\hat S_{\delta}(z,w)$ is a modified Szeg\"o kernel,
whose precise definition is given in (\ref{sze}). For genus 2, the
modified Szeg\" o kernel $\hat S_\delta (z,w)$ coincides with the ordinary
Szeg\" o kernel
$S_\delta (z,w)$. Technically, the gauge fixing based on (\ref{gpr})
introduces a number of significant additional complications which we
explain next; but all these can and will be resolved, and the payoff will
be that the resulting formula for the chiral measure can be verified 
explicitly to be free of any ambiguity.

\medskip

A first issue requiring some care is chiral splitting.
The correct degrees of freedom of string theory require a worldsheet
formulation with Minkowski signature and $\chi_m{}^{\beta}$
two-dimensional Majorana-Weyl spinors. In the present Euclidian worldsheet
formulation of the worldsheet $\Sigma$, $\chi_m{}^{\beta}$ includes fields
$\chiz$ and $\chi_z{}^-$ of both chiralities. Thus we need chiral
splitting, that is, a process for separating and retaining in correlation
functions only the contributions of the chiral half $\chi_{\bar z}{}^+$
of the field $\chi_m{}^{\beta}$. Chiral splitting is an essential
step in the implementation of the Gliozzi-Scherk-Olive (GSO) projection
\cite{gso}, which projects out the tachyon and insures space-time
supersymmetry. For Type II superstrings \cite{gs82}, the GSO projection
must be enforced independently on left and right worldsheet chiralities,
while in the heterotic string \cite{heter}, only one of the two
worldsheet chiralities is retained. Thus, for both Type II and heterotic
strings, the key building blocks are the chiral measure and amplitudes,
and these are common to both types of superstring theories.

\medskip

The rules for chirally splitting superfields have been obtained in
\cite{dp89}. Applied to the superstring measure, they give us
the following first formula for the chiral measure on supermoduli
space (see (\ref{gf4}) below
for a detailed explanation of all the ingredients of this formula)
\bea
\label{csp}
{\bf A}^{chi} [\delta](p_I^{\mu})
&=&
\prod_A dm_A \exp(ip_I^{\mu}\hat\Omega_{IJ}p_J^{\mu}) \A [\delta]
\nonumber \\
\A [\delta] 
&=&
\bigg \< \prod_A \delta(\<H_A|B\>)
\exp \biggl ({1\over 2\pi}\int _\Sigma \! \! d^2z
\chiz S(z) \biggr ) \bigg \> _+
\eea
Here, $p_I ^\mu$, $I=1,\cdots , h$ are $h$ independent internal loop
momenta; a $(3h-3|2h-2)$ dimensional slice ${\cal S}$
for supermoduli space has been chosen,
which is parametrized by supermoduli $m^A$ 
with the label ranging over $A=1,
\cdots , (3h-3|2h-2)$; the $H_A$
are the corresponding Beltrami superdifferentials;
$S(z)$ is the total supercurrent; and $B$ is the ghost superfield. The
expectation value $\< \cdots \>_+$ is taken using effective rules for
chiral worldsheet fields,
in the background metric
$g_{mn}=e_m{}^ae_n{}^b\delta_{ab}$ on the worldsheet.
Nevertheless, we would like to stress
that it is the super period matrix $\hat
\Omega _{IJ}$ and not the period matrix $\Omega_{IJ}$
of the metric $g_{mn}$
which appears as the covariance matrix
of the internal loop momenta $p_I^{\mu}$
in the correct chiral splitting prescription. This is the first
clue that the projection (\ref{gpr}) is on the right track \cite{dp88}.

\medskip

The main problem in superstring perturbation theory is to descend from the
preceding measure (\ref{csp}) over supermoduli space to a measure 
$d\mu[\delta]$ over moduli space. It is this deceptively simple step of
integrating out the odd supermoduli $\zeta^{\alpha}$, $1\leq\alpha\leq
2h-2$, which has caused problems in the past and which has to be carried
out with particular care. In this series of papers, our main guiding
principle is to view this integration as an integration along the fibers
of the projection (\ref{gpr}). With this principle, the basic even
parameters are local parameters for the super period matrix
$\hat\Omega_{IJ}$, the odd parameters $\zeta^{\alpha}$ are independent
variables, and we can write for the chiral measure $d\mu[\delta]$ over
moduli space
\be
\label{mu}
d\mu[\delta]
=\prod_{a=1}^{3h-3}dm^a\int \prod_{\alpha=1}^{2h-2}d\zeta^{\alpha}
{\cal A}[\delta]
\ee

\medskip

We come now to the additional complications in gauge-fixing
inherent to the projection
(\ref{gpr}). There are essentially three of them:

\medskip

$\bullet$
Since the basic moduli parameters are now $\hat\Omega_{IJ}$,
the whole amplitude ${\cal A}[\delta]$ has to be re-expressed first
in terms of $\hat\Omega_{IJ}$ and $\zeta^{\alpha}$ before the
fiber integration can be carried out. Now, as can be seen in (\ref{csp}), 
superstring amplitudes are built out of correlation functions in conformal
field theory, with respect to the background complex structure
corresponding to the period matrix $\Omega_{IJ}$. 
While all anomalies cancel in the full amplitudes, whose
dependence is therefore only on the moduli $\Omega_{IJ}$, 
the contributions from the amplitudes'
individual building blocks, such as the matter, ghost and gauge fixing
parts each require specifying a metric, and not just a complex structure.
Thus, in practice, to re-express the amplitudes in terms of the super
period matrix $\hat\Omega_{IJ}$, we require a choice  
of zweibein $\hat
e_m{}^a$ (or equivalently a choice of metric 
$\hat g_{mn}=\hat e_m{}^a\hat e_n{}^b\delta_{ab}$) with
$\hat\Omega_{IJ}$ as its period matrix.  
This should be
viewed as an additional gauge choice,
and ultimately, we have to show that
it is immaterial. Another way
of describing this new gauge choice is the following. 
Let $\hat\mu_{\bar z}{}^z$ be the Beltrami differential
corresponding to the variation of complex structure
from $\Omega_{IJ}$ to $\hat\Omega_{IJ}$
\be
\label{mud}
\Omega_{IJ}
-
\hat\Omega_{IJ}
=i\int d^2z \,\hat\mu\, \omega_I(z)\omega_J(z)
\ee
The variation of complex structure is only a finite-dimensional
constraint on $\hat\mu$,
which identifies only the equivalence class $[\hat\mu]$, modulo
gauge transformations of the form $\hat\mu\to\hat\mu+\partial_{\bar z}v^z$.
But the evaluation of the individual blocks in the superstring
amplitudes involves the full $\hat\mu$. The final amplitudes have to
be shown to be independent of $\hat\mu$, as long as it satisfies
(\ref{mud}).
We shall refer to this additional gauge slice independence as {\sl slice
$\hat \mu$ independence}.

\medskip

$\bullet$
The second complication resides with the Beltrami superdifferentials
$H_A$, which are the tangent vectors to the slice within the space
of superzweibeins. In Wess-Zumino gauge, the Beltrami superdifferential
$\delta H_-{}^z\sim E_-{}^M\delta E_M^z$
corresponding to a variation $\delta E_M{}^A$ 
can be written as
\be
\label{bsd}
\delta H_-{}^z=\bar\theta(\delta\mu_{\bar z}{}^z-\theta\delta\,
\chi_{\bar z}^+)
\ee
with $\delta\mu_{\bar z}{}^z= - e_{\bar z}{}^m\delta e_m{}^z$ and
$\delta\chi_{\bar z}{}^+$ controlling respectively the variations of
moduli and of gravitino fields. 
With the naive projection (\ref{bpr}), slices
can be chosen with split Beltrami superdifferentials, that is,
superdifferentials  for either a change of moduli or a change
of gravitino field, but not both. This is no longer the case for the
Beltrami superdifferentials associated with the projection (\ref{bpr}),
since 
\be
\delta\hat\Omega_{IJ}=0
\ee
for deformations along the fiber of this projection.
In view of (\ref{spm1}), this can only hold if a variation of $\Omega_{IJ}$
is accompanied with a compensating variation of $\chiz$.

\medskip

$\bullet$ 
Finally, the correlation functions in (\ref{csp}) have to be rewritten in
terms of the background geometry $\hat e_m{}^a$, representing moduli $\hat
\Omega _{IJ}$. This is a problem of deformation of complex structures,
and has to be addressed in conformal field theory by 
adding terms involving repeated insertions of
the stress tensor $T(z)$.

\medskip

\noindent
With the above points taken into account,
we obtain our first formula for the genus $h=2$
gauge-fixed
amplitude\footnote{All two-dimensional integrations will be over the
compact orientable worldsheet $\Sigma$, whose dependence will henceforth
be suppressed: we let $ \int _\Sigma \to \int$. A collection of useful
explicit formulas for holomorphic and meromorphic differentials and
Green's functions is given in Appendix~A and their superspace
counterparts are given in Appendix~B.}, which is the following
\bea
\label{gen2}
\A [\delta] 
&=& 
{\< \prod _a b(p_a) \prod _\alpha \delta (\beta (q_\alpha)) \>
\over \det \Phi _{IJ+} (p_a) \det \< H_\alpha | \Phi ^* _\beta\>}
\biggl \{ 
1 -  {1 \over 8 \pi ^2} \int \! d^2 \! z \chiz \! \int \! d^2\! w \chiw
\< S(z) S(w)\> 
\nonumber \\
&& \hskip 1.95 in
+ {1 \over 2 \pi} \int \! d^2\! z \hat \mu (z)  \<T(z)\> \biggr \}
\eea
Here $\hat\mu(z)$ is the Beltrami differential deforming the zweibein from
$\hat e_m{}^a$ to $e_m{}^a$; $\Phi _{IJ}$ and $\Phi ^* _\beta$ are a
specific basis of odd and even superholomorphic 3/2 forms. The correlation
functions on the worldsheet are evaluated with respect to the background
geometry $\hat g_{mn}$ corresponding to $\hat\Omega_{IJ}$.
The points $p_a$, $q_{\alpha}$ are arbitrary generic points, introduced 
merely as a computational device. By construction, the measure is independent of
these points, as may be checked explicitly. The term $\<S(z)S(w)\>$ is
common to the earlier and the present approaches. The stress tensor
correlator and the finite-dimensional determinants on the right hand side
are the key new terms; each separately is slice dependent, but their
combined effect is to restore slice independence to the entire
expression.  For example, under changes of slices, the 
short distance singularities of
the supercurrent correlator
$\<S(z)S(w)\>$ as $z\to w$, whose significance had been obscure before,
are now manifestly cancelled by the stress tensor correlator term
$\<T(z)\>$. Similarly, it will turn out that the arbitrariness in $\hat\mu$
will be compensated by a related arbitrariness in the Beltrami
superdifferentials $H_{\alpha}$.

\medskip

The gauge fixed formula (\ref{gen2}) illustrates well the main features
of the method of parametrizing the even supermoduli by the super period 
matrix. However, as a formula for a measure on moduli space,
it is not yet satisfactory, since it still involves 
supergeometric notions such as $\Phi_{IJ}$ and $H_{\alpha}$,
and its independence from the choice of Beltrami differential $\hat\mu$
is not manifest. The key to overcoming this difficulty is a deep
relation between superholomorphic notions with respect to the
supergeometry $(e_m{}^a,\chi_m{}^{\alpha})$ and holomorphic notions
with respect to the super period matrix $\hat\Omega_{IJ}$.
This is an important issue which we shall revisit in detail in the
later papers of this series. For the present paper, we require only the 
simplest example of this correspondence, which says that the
superholomorphic differentials $\hat\omega_I$ dual to the canonical
$A$-homology basis are given by
\be
\hat\omega_I=\theta\omega_I(\hat\Omega,\chi=0)+{\cal D}_+\Lambda_I
\ee
where $\Lambda_I$ is a superscalar. Exploiting such relations,
we can actually express everything in terms of $\hat\Omega_{IJ}$
moduli only, and make the $\hat\mu$ independence manifest.
We arrive in this way at
the following formula, which 
is the first main result of the present
paper, and the starting point of the
subsequent ones 
\be
\label{finamp}
{\cal A}[\delta] 
=
i \ {\< \prod _a b(p_a) \prod _\alpha \delta (\beta (q_\alpha)) \>
\over \det \bigl (\omega _I \omega _J (p_a) \bigr )
 \cdot \det \< \chi _\alpha | \psi ^* _\beta\>}
\biggl \{ 1  + {\cal X}_1 + {\cal X}_2 + {\cal X}_3 + {\cal X}_4 +  {\cal
X}_5 + {\cal X}_6 \biggr \}
\ee
Here all correlation functions, $\tet$-functions, and complex variables
are  written with respect to the $\hat\Omega_{IJ}$ complex structure.  
The expressions $\psi_{\beta}^*$ are the holomorphic $3/2$
differentials normalized at the points $q_{\alpha}$ by $\psi ^* _\beta
(q_\alpha)=\delta _{\alpha \beta}$, and $\omega _I$ are holomorphic
1-forms canonically normalized on $A_J$ cycles, for a given choice of
canonical homology basis. The terms $\X _i$, $i=1,\cdots ,6$ are defined as
follows, with $p_a$, $a=1,2,3$ and $q_\alpha$, $\alpha =1,2$ arbitrary
generic points on the worldsheet $\Sigma$,
\bea
\label{Xes}
{\cal X}_1
&=&
 - {1 \over 8 \pi ^2} \int \! d^2z \chiz \int \! d^2 w \chiw \< S(z) S(w)\> 
\nonumber \\
{\cal X}_2
+
{\cal X}_3
&=&
+{1\over 16\pi^2}
\int d^2z\int d^2w\,\chiz\chiw\,T^{IJ}\omega_I(z)S_{\delta}(z,w)
\omega_J(w)
\nonumber\\
{\cal X} _4 
&=& + {1 \over 8\pi ^2} \int \! d^2w \ \p _{p_a} \pw \ln E(p_a,w) \chiw 
\int \! d^2u S_\delta (w,u) \chiu \varpi ^* _a(u)
\nonumber \\
{\cal X} _5 
&=& + {1 \over 16 \pi ^2} \int \! d^2u \int \! d^2v S_\delta (p_a,u) \chiu  
\p _{p_a} S_\delta (p_a,v) \chiv \varpi _a  (u,v) 
\nonumber \\
{\cal X} _6
&=& + {1 \over 16 \pi ^2} \int \! d^2z \chi _\alpha ^* (z) \int \!
d^2w G_{3/2} (z,w) \chiw \int \! d^2 v \chiv \Lambda _\alpha (w,v)  
\eea
Here $\chi_{\beta}^*(z)$ is the linear combination of the
$\chi_{\alpha}(z)$ characterized by $\<\chi_{\beta}^*|\psi_{\alpha}^*\> =
\delta_{\beta\alpha}$, $S_{\delta}(z,w)$ is the Szeg\"o kernel, and the
Green's functions $G_{3/2}(z,w)$ and $G_2(z,w)$ are tensors of type
$(3/2,-1/2)$ and $(2,-1)$ respectively, normalized by $G_{3/2}(q_\alpha,w)
= G_2(p_a,w)=0$. The object  $T^{IJ}$ may be defined in terms of the
holomorphic quadratic differential 
\bea
\label{teeIJ}
T^{IJ}\omega_I\omega_J(w)
&=&
\< T(w) \prod_{a=1}^3b(p_a)\prod_{\alpha=1}^2 \delta(\beta(q_{\alpha}))\>
\bigg / 
\< \prod_{a=1}^3b(p_a)\prod_{\alpha=1}^2 \delta(\beta(q_{\alpha}))\>
\nonumber \\ &&
-2\sum_{a=1}^3\p_{p_a}\p_w\ln\,E(p_a,w)\varpi^* _a (w)
\nonumber\\
&&
+\int d^2z\chi_{\alpha}^*(z) \biggl (
-{3\over 2}\p_wG_{3/2}(z,w)\psi_{\alpha}^*(w)
-{1\over 2}G_{3/2}(z,w)(\p\psi_{\alpha}^*)(w)
\nonumber\\
&&
\hskip 1.05in
+G_2(w,z)\p_z\psi_{\alpha}^*(z) +{3\over
2}\p_zG_2(w,z)\psi_{\alpha}^*(z) \biggr )
\eea
Furthermore, $\Lambda_{\alpha}$ is defined by
\be
\Lambda_{\alpha}(w,v)
=
2G_2(w,v)\p_v\psi_{\alpha}^*+3\p_vG_2(w,v)\psi_{\alpha}^*(v)
\ee
Finally, $\varpi_a (u,v)$ and $\varpi^* _a (u)$ are holomorphic
1-forms constructed as ratios of finite dimensional determinants and may
be defined by $\varpi ^* _a (u) \equiv \varpi _a (u,p_a)$ and 
\bea
\label{varpi}
\varpi  _a (u,v) &=& {\det \omega _I \omega _J(p_b \{ u,v;a \} )
\over \det \omega _I\omega _J(p_b)}
\nonumber \\
\omega _I \omega _J (p_b\{ u,v; a\} ) 
&=& \left \{ \matrix{
\omega _I \omega _J(p_b) & \qquad b\not= a\cr
\half (\omega _I(u)\omega _J(v) + \omega _I(v) \omega _J(u) )  &
\qquad b=a
\cr}
\right .
\eea
\medskip

The second main result of this paper
is to show that the gauge fixed amplitude (\ref{finamp}) satisfies the crucial requirement
of invariance under infinitesimal deformations of the gauge slice 
\be
\label{gsli}
\delta_{\xi} (\chi_{\alpha})_{ \bar z}{}^+=-2\,\p_{\bar z}\xi_{\alpha}^+
\ee
where $\xi_{\alpha} ^+$ are spinor fields, generators of local supersymmetry
transformations. In (\ref{finamp}), the term which is familiar from the early
literature in superstring perturbation theory and leads essentially to the picture
changing operator is ${\cal X}_1$. It was known to generate ambiguities under
changes of slices. In the present formula however, these ambiguities will
be cancelled out by the combination of all the remaining terms.  
Since the issue of slice dependence has led to much confusion in the
past, we present the proof in some detail.

\medskip

This paper is organized as follows. The starting point is the well-known
gauge fixed measure on supermoduli space \cite{dp88}. This measure is
recalled in Section 2, together with a brief review of the ambiguity
difficulties encountered in earlier approaches. Section 3 is devoted to
the general gauge fixing procedure based on the projection (\ref{gpr}) on
super period matrices. It is presented for genus $h=2$, but much can
be extended to general genus.
This includes the chiral splitting of the partition function and measure
giving (\ref{csp}), and a discussion of gauge slices and Beltrami 
superdifferentials giving (\ref{gen2}). The evaluation
of the genus 2 chiral string
measure is begun in Section 4; for this, the supercurrent and stress
tensor correlators are determined explicitly,
with the finite-dimensional
determinants left to the later sections.
Section 5 is devoted to slice independence.
Since the slice independence of the chiral string measure (and amplitudes)
constitutes a key test of the validity of the proposed expressions,
we present a careful and detailed proof of both the
{\sl slice $\hat \mu$
independence} and the {\sl slice $\chi$ independence} under 
infinitesimal variations. 
For this, the variations of the finite-dimensional determinants
are required, and they are derived using relations between superholomorphic
superdifferentials and holomorphic differentials with respect to
the super period matrix.
Finally, in Section 6, the formulas (\ref{finamp}) and
(\ref{Xes}) are obtained by expressing all superspace quantities in
components and arranging the result in a manifestly slice $\hat \mu$
independent way. Technical tools are collected in the Appendices for the
reader's convenience. Appendix A presents a summary of formulas on
ordinary Riemann surfaces for holomorphic differentials, Green's
functions and $\tet$-functions as well as for their variations under
changes in the worldsheet metric. Appendix B collects analogous formulas
for $\N=1$ supergeometry. Appendix C gives useful formulas for
superdeterminants. Appendix D gives a summary of forms, vector fields and
Beltrami differentials associated with a slice.

\vfill\eject

\section{The Gauge-Fixed Measure on Supermoduli Space}
\setcounter{equation}{0}

Our starting point is the gauge-fixed measure for superstring perturbation
theory, expressed as a measure on supermoduli space \cite{dp88}. We
recall it briefly in this section, with particular emphasis on the issues
which will play an important role in the sequel.

\subsection{Superstring Propagation}

The worldsheet for superstring propagation in $10$-dimensional space-time
at loop order $h$ is a closed orientable compact surface $\Sigma$ of genus
$h$.  In the Ramond-Neveu-Schwarz formulation \cite{rns},
the scattering amplitudes are constructed out of a two-dimensional
superconformal field theory on the worldsheet, consisting of $10$
``matter" scalar superfields  $X_{\mu}$, $\mu =0,1,\cdots ,9$ coupled to
two-dimensional supergravity,
\be
I_m = {1\over 4\pi} \int d^{2|2} z 
\,E\,
{\cal D}_+X_{\mu}{\cal D}_-X^{\mu}
\qquad \qquad 
E \equiv \sdet E_M {}^A
\ee
Here the two-dimensional supergravity
\footnote{
A summary of worldsheet supergeometry, two-dimensional supergravity, and
some useful formulas is given in Appendix B. We use the following
conventions for spinors. The Clifford algebra is generated by $\{\gamma
^a, \gamma ^b \} = - \delta ^{ab}$. Spinor indiced $\alpha = (+-)$ are
raised and lowered according to $\psi _+ = - \psi ^-$, $\psi _- = \psi
^+$ and all contractions where indices are not exhibited explicitly are
taken as $\phi \psi \equiv \phi ^\alpha \psi _\alpha = - \phi _\alpha \psi
^\alpha$. We have $\gamma ^z _{++}=\gamma ^{\bar z} _{--} =1$, all
other components of $\gamma ^a$ being zero, and $(\gamma _5)_+ {}^+ =
-(\gamma _5)_- {}^- = i$. Finally, we use the following convention
for the measure $d^{2|2} z = d^2 z d\theta d\bar \theta$.}  
fields are given by a superzweibein $E_M{}^A$ and a U(1) superconnection
$\Omega_M$ with Wess-Zumino constraints \cite{superg},  and the
worldsheet $\Sigma$ has been equipped with a spin structure $\delta$. The
theory is invariant under super Weyl rescalings sWeyl($\Sigma$), super
U(1) Lorentz transformations sU(1), and local super reparametrizations or
sDiff(M). Local super reparametrizations are particularly important to
our considerations. Infinitesimal super-reparametrizations are generated
by vector fields $\delta V^M$
\be
\label{susy}
E_A{}^M\delta E_M{}^B
=
{\cal D}_A\delta V^B -\delta V^CT_{CA}{}^B +\delta V^C\Omega_CE_A{}^B.
\ee
In components, the supercoordinates $\z$ and $\bar \z$ decompose into
$\z = (z,\theta)$ and $\bar \z = (\bar z , \bar \theta)$, where $\theta
^\alpha = (\theta, \bar \theta)$ are Grassmann coordinates. The
superfields $X^{\mu}$ and supergeometries $E_M{}^A$ in Wess-Zumino gauge
decompose accordingly into
\bea
X^{\mu} (\z, \bar \z)
&=& x^{\mu}(z, \bar z) + \theta \psi_+^{\mu}(z, \bar z)
+\bar \theta \psi_-^{\mu}(z, \bar z) + i \theta \bar \theta F^\mu (z,\bar
z),
\nonumber\\
E_m{}^a (\z, \bar \z)
&=&
e_m{}^a (z,\bar z)
+\theta \gamma^a \chi_m{} (z, \bar z)
-{i\over 2}\theta\bar\theta e_m{}^a A (z, \bar z),
\eea
where $e_m{}^a$ and $\chi_m{}^{\beta}$ are the zweibein and gravitino
fields on the worldsheet $\Sigma$, and $F^\mu$ and $A$ are auxiliary
scalar fields. Super vector fields $\delta V^M$ decompose accordingly as
\be
\delta V^m = \delta v^m-\theta \gamma^m \xi
-{1\over 2} \theta \bar\theta \chi_n \gamma^m \gamma^n  \xi
\ee
The vector field $\delta v^m$ generates reparametrizations,
while the spinor $\xi $ generates local supersymmetry
transformations
\bea
\label{susyc}
\begin{array}{cl}
{\rm Diff(M):} & \delta _v  e_m{}^a=\delta
v^n\p_ne_m{}^a+e_n{}^a\p_m\delta v^n
\\
{\rm SUSY:}    & \delta _\xi e_m{}^a=\xi \gamma^a \chi_m
\end{array}
\eea

The zweibein $e_m{}^a$ equips the worldsheet $\Sigma $ with a complex
structure. Let $z,\bar z$ be local holomorphic coordinates on $\Sigma $
with respect to this complex structure. Then the relevant components of
$\chi_m{}^{\beta}$ are given by $\chi_{\bar z}{}^+$, $\chi_z{}^-$,
while the others decouple by super Weyl invariance. These components can
be viewed as the Euclidian counterparts of chiral Majorana-Weyl spinors
on a two-dimensional worldsheet with Minkowski signature. Using the
component expressions of the covariant derivates $\D _\pm$ and of $E$,
given in Appendix B, and after elimination of the auxiliary field $F^\mu$
(carried out for example in \cite{dp88}), the action $I_m$ may be
expressed in components in the local coordinates $z, \bar z$ as
\bea
I_m &=& I_m ^0 + {1\over 2\pi}
\int \! d^2z \biggl (
-\chi_{\bar z}{}^+ S_m (z)
- \chi_z{}^- \overline {S_m (z)}
-{1\over 4} \chi_{\bar z}{}^+\chi_z{}^-\psi_+^{\mu}
\psi_-^{\mu} \biggr )
\nonumber \\
I_m ^0 &=& {1\over 4\pi}
\int \! d^2z \biggl (
\partial_zx^{\mu} \partial_{\bar z}x^{\mu}
- \psi_+^{\mu}\partial_{\bar z}\psi_+^{\mu}
- \psi_-^{\mu}\partial_z\psi_-^{\mu} \biggr )
\eea
Here, $I_m ^0$ is the ``free matter action" and $S_m$ is the matter 
worldsheet supercurrent,
\be
S_m \equiv - \half \psi ^\mu _+ \p _z x_\mu \, .
\ee

\subsection{The Non-Chiral Supermoduli Measure}

The partition function of the worldsheet superconformal field theory for
the scalar superfields $X^\mu$ at given spin structure $\delta$ is given
by
\be
\label{partition}
{\bf A} [\delta] = \int DE_M{}^A D\Omega _M
\delta (T) \int DX^{\mu} e^{-I_m}
\ee
To obtain the correct degrees of freedom of the superstring, we have to
extract out of this expression the contributions of the chiral half
$\chiz$ of the gravitino field $\chi_m{}^{\beta}$ and ultimately carry
out the GSO projection, which involves a summation over spin structures
$\delta$. In this paper, we discuss only the partition function
(\ref{partition}) in order to construct the chiral measure, leaving the
case of scattering amplitudes and vertex operators to a later paper.

\medskip
  
Let $s{\cal M}_h$ be the supermoduli space of supergeometries on
surfaces of genus $h$, that is, the quotient space
\be
s{\cal M}_h
=\{E_M{}^A,\Omega_M \ {\rm obeying \ torsion \ constraints}\}
\bigg /
s{\rm Weyl}\times s{\rm U(1)}
\times s{\rm Diff}.
\ee
In view of its invariance under super Weyl rescalings, super U(1)
gauge transformations, and super-reparametrizations, the partition
function ${\bf A} [\delta]$ reduces to an integral over $s{\cal M}_h$.

\medskip

For genus $h\geq 2$, $s{\cal M}_h$ is a superspace of dimension
$(3h-3|2h-2)$. Since it is a quotient space, there are no canonical
coordinates which we may use. However, locally, we can always parametrize
it by a $(3h-3|2h-2)$-dimensional slice ${\cal S}$ of supergeometries
which is transversal to the orbit of all the symmetry transformations.
Let $m^A =(m^a|\zeta^{\alpha})$, with $a=1, \cdots , 3h-3$ and
$\alpha = 1, \cdots, 2h-2$ be parameters for such a slice ${\cal S}$.
Let $H_A$ be the Beltrami superdifferentials
tangent to the gauge slice. They are given by\footnote{The sign factor in
front is + unless $A$ is odd and $M$ is even, when it is $-$. This factor
arises from using (\ref{bsd}) for the variation and pulling out the
differential $\delta m^A$ to obtain the superderivative. It was omitted
in \cite{dp88} but correctly included in \cite{dp90}.} 
\be
\label{superbeltrami}
(H_A)_-{}^z = (-)^{A(M+1)} E_-{}^M {\partial E_M{}^z\over\partial m^A}
\ee
Then the symmetry groups can be factored out of the non-chiral partition
function (\ref{partition}), giving the following gauge-fixed expression
(see \cite{dp88}, p. 967)
\be
\label{gf}
{\bf A} [\delta]
=
\int_{\cal S}\prod_A |dm^A|^2
\int D(XB\bar BC\bar C)
\prod_A |\<\delta (H_A|B) \>|^2 e^{-I_m - I_{gh}}
\ee
Here, $B$ and $C$ are the ghost superfields of sU(1) weight $3/2$ and $-1$
respectively, and given in components by
\bea
B(\z, \bar \z) & = &
 \beta(z, \bar z) + \theta b(z,\bar z) + \bar \theta B_2 (z,\bar z)
+ i \theta \bar \theta B_3 (z,\bar z)
\nonumber \\ 
C (\z, \bar \z) & = & 
c(z, \bar z) + \theta \gamma (z,\bar z) + \bar \theta C_2 (z,\bar z) + i
\theta \bar \theta C_3 (z,\bar z)
\eea
and the ghost action is given by
\bea
I_{gh}
=
{1\over 2\pi} \int d^{2|2}z\, E
\,\biggl ( B\D_- C + \bar B \D _+ \bar C \biggr )
\eea
Here, $B_2$, $B_3$, $C_2$ and $C_3$ are auxiliary fields, which may be
eliminated following \cite{dp88}. The ghost action in components then
becomes
\bea
I_{gh} & = & 
I_{gh} ^0 + {1\over 2\pi} \int d^2z \bigl (- \chiz S_{gh} - \chi _z {}^-
\overline{S_{gh}} \bigr )
\nonumber \\
I_{gh}^0 &=&
{1\over 2\pi} \int d^2z \bigl ( b\p_{\bar z} c + \beta \p_{\bar z} \gamma
+\bar b \p_z \bar c + \bar \beta \p_z \bar \gamma \bigr )
\eea
Here, $I_{gh}^0$ is the ``free superghost action" and $S_{gh}$ is the
ghost worldsheet supercurrent,
\bea
S_{gh} ={1\over 2}b\gamma-{3\over 2}\beta\p_zc -(\p_z\beta)c \, .
\eea
The gauge-fixed expression (\ref{gf}) is independent
of the choice of gauge slice ${\cal S}$. A change of gauge
slice ${\cal S}$ results in compensating changes in
the measure $|\prod_A dm^A|^2$ and in the insertion
$|\prod_A\<\delta(H_A|B)\>|^2$.

\subsection{The Main Difficulties}

The main difficulties in superstring perturbation theory
are how to chirally split the gauge-fixed amplitudes,
and how to descend from a measure on supermoduli space to 
a measure on moduli space. 
These two difficulties turn out to be related.
However, chiral splitting
has been considerably clarified in \cite{dp89}, and we
shall discuss it in detail for the partition function
in the next section, and for general scattering
amplitudes in a later paper of this series.
We focus now on the specific problems in descending from
supermoduli space to moduli space.

\subsubsection{The Picture-Changing Ansatz}

Early on, an Ansatz for chiral superstring amplitudes 
as integrals over moduli space had been
proposed by Friedan, Martinec, and Shenker \cite{fms},
based on BRST invariance
\be
\label{brst}
{\bf A}^{chi} _{BRST} [\delta]
=
{1\over (2\pi)^{2h-2}}
\int \prod_{a=1}^{3h-3} dm^a
\int D(xbc\beta\gamma)e^{-I_m^0 - I_{gh}^0}
\prod_{a=1}^{3h-3}
\<\mu_a|b\>\prod_{\alpha=1}^{2h-2} Y(z_{\alpha})
\ee
Here, the integration is over
a $3h-3$ dimensional slice of metrics $g_{mn}$
on $\Sigma$, parametrized by $3h-3$ complex parameters $m^a$,
and $2 \mu_a=g_{z\bar z} \partial g^{ z z}/\partial m^a$
are the Beltrami differentials tangent to the slice. 
The operator $Y(z)$ is the picture-changing operator, which is formally
BRST invariant. It can be expressed as \cite{vv}
\be
Y(z)=\oint_{C_z} dw\, j_{BRST}(w)H(\beta(z))
\ee
$C_z$ is a small contour surrounding the point $z$,
$H$ is the Heaviside function, $T(w)$ is the stress tensor, and
$j_{BRST}(w)=c(w)(T(w) + b \p c)+{1\over 2}\gamma\p x^{\mu}\psi_+^{\mu}
+{1\over 4}\gamma^2b$ is the BRST current.

\medskip

The main problem with the above Ansatz is that ${\bf A}^{chi} _{BRST} [\delta]$
should be independent of the insertion points $z_{\alpha}$.
However, this property does not hold for the right hand
side of (\ref{brst}), although it is tantalizingly close.
Indeed, under a change of insertion points,
$Y(z)$ changes by a BRST transform. Deforming the contour $C_z$
away, we pick up the BRST transform of the $b$ field insertions, which is
the stress tensor. This produces derivatives with respect to the moduli
parameters $m^a$. Such derivatives could be harmless, if they were
defined {\sl globally} over moduli space, so that the invariance
under changes of $z_{\alpha}$ can be restored when
the boundary of moduli space does not contribute.
But, as noted already in \cite{vv},
this is not the case: the derivatives are defined only
over local coordinate patches in moduli space, where it makes sense
to pick $2h-2$ points, i.e., $2h-2$ sections of the universal curve.

\subsubsection{Integrating Odd Supermoduli}

The natural way of reducing an integral over supermoduli space to an
integral over moduli space is to integrate out the odd supermoduli $\zeta
^{\alpha}$. This approach was tried by many authors \cite{vv}, and ran
as follows.

Choose again a $3h-3$ slice of metrics $g_{mn}$ on $\Sigma$,
parametrized by local coordinates $m^a$. Let $\zeta ^{\alpha}$ be $2h-2$
anticommuting parameters, choose for each metric $2h-2$ generic gravitino
fields $\chi_{\alpha} =(\chi_{\alpha})_m{}^{\beta}$, $\alpha = 1,\cdots,
2h-2$, and set $\chi=\sum_{\alpha=1}^{2h-2} \zeta ^{\alpha}
\chi_{\alpha}$. Then 
\be
\label{section}
m^A\to (g_{mn},\chi)
\ee
can be viewed as a $(3h-3|2h-2)$ space ${\cal S}$ of 
supergeometries in Wess-Zumino gauge, which is a slice 
for supermoduli space. The Beltrami superdifferentials
$H_A$ which are tangent to ${\cal S}$ are readily evaluated
\be
\label{bel1}
H_a=\bar\theta \mu_a,
\quad\quad
H_{\alpha}
=-\bar\theta\theta \chi_{\alpha},
\ee
where $\mu_a$ is as before the Beltrami differential tangent
to the slice of metrics $g_{mn}$. The gauge-fixed formula (\ref{gf}) becomes
\bea
\label{gfs}
{\bf A}_{BRST} [\delta]
&=&
\int_{S} |\prod_{a=1}^{3h-3} dm^a|^2
\int |\prod_{\alpha=1}^{2h-2}d \zeta ^ {\alpha}|^2
\int D(x\psi _\pm b \bar b c \bar c \beta \bar \beta \gamma \bar\gamma)
\bigg | \prod_{a=1}^{3h-3}\<\mu_a|b\>
\prod_{\alpha=1}^{2h-2}\delta(\<\chi_{\alpha}|\beta\>) \bigg | ^2
\nonumber\\
&&
\quad\quad
\times
\exp \biggl ( -I_m^0 - I_{gh}^0 + {1\over 2\pi}
\int(\chi S+\bar\chi\bar S)-{1\over 8\pi}\int \chi\bar\chi\psi_+\psi_-
\biggr )
\eea
where $S(z) =S_m(z)+S_{gh}(z)$ is the total supercurrent.
The naive Ansatz to chirally split this expression is just to drop the
term $\chi\bar\chi\psi_+\psi_-$ in the action, and to keep as the chiral
contribution of $\chiz$ at each point $m^a$ on moduli space
\bea
\label{gf2}
{\bf A}^{chi} _{BRST} [\delta](m^a)
&=&
\prod_{a=1}^{3h-3}dm^a
\int \prod_{\alpha=1}^{2h-2} d \zeta ^{\alpha}
\int D(x_+ \psi_+ b c\beta\gamma)
\prod_{a=1}^{3h-3}\<\mu_a|b\>
\prod_{\alpha=1}^{2h-2}\delta(\<\chi_{\alpha}|\beta\>)
\nonumber\\
&&
\quad\quad\quad\quad
\times
\exp \biggl ( -I_m ^0 - I_{gh} ^0 + {1\over 2\pi}
\sum_{\alpha=1}^{2h-2} \zeta ^{\alpha} \int\chi_{\alpha} S \biggr )
\eea
The integral over the anticommuting supermoduli $\zeta ^{\alpha}$
can now be carried out
\be
\label{gf3}
{\bf A}^{chi} _{BRST} [\delta]
=   
\prod_{a=1}^{3h-3} dm^a
\int D(x_+ \psi_+ b c\beta\gamma)
\prod_{a=1}^{3h-3} \<\mu_a|b\>
\prod_{\alpha=1}^{2h-2}{1\over 2\pi} \delta (\<\chi_{\alpha}|\beta\>)\<
\chi_{\alpha} |S\>
e^{-I_m ^0 - I_{gh} ^0}
\ee
Choosing the gravitino fields $\chi_{\alpha}(z)$ to be Dirac
$\delta$-functions, $\chi_{\alpha}(z)=\delta(z,z_{\alpha})$, the
$\chi _\alpha$-dependent parts of the preceding formula become
\be
\delta (\<\chi_{\alpha}|\beta\>)\< \chi_{\alpha} |S\> 
\ \ \to \ \
\delta \big( \beta (z_{\alpha})\big) S(z_{\alpha}) \equiv Y(z_\alpha)
\ee
and we recapture in this way the formula (\ref{brst}) proposed in
\cite{fms}.

\medskip

This derivation is attractive, since it seems to derive the BRST
invariant formula (\ref{brst}) from first principles. However,
the outcome suffers then from the same ambiguities as (\ref{brst}).
This is surprising, since we started from a manifestly slice-invariant
gauge-fixed formula (\ref{gf}) on supermoduli space. The
problem has been traced back to subtleties in integration
over supermanifolds \cite{ars}. But as we shall see later,
it lies in the present case with both the naive chiral splitting used
in the derivation, as well as with the slice parametrization
(\ref{section}), which is not compatible with local supersymmetry.

\subsubsection{Unitary Gauge}

If one admits the inherent ambiguity in the preceding gauge-fixed formulas
(\ref{gf3}), a possible strategy would be to try and guess a choice
of insertion points which would lead to physically acceptable string
amplitudes. There are some natural choices: for example, the points
$z_{\alpha}$ can be taken to be the $2h-2$ zeroes of a holomorphic
abelian differential $\omega(z)$. This is called the unitary gauge, and
has many practical advantages, including a concrete cancellation between
longitudinal and ghost degrees of freedom. However, an arbitrariness in
the choice of the differential $\omega(z)$ has been introduced, together
with a host of other delicate issues such as modular invariance and
factorization properties. Although there can be consistency checks such as
a vanishing cosmological constant, there is no answer to the basic
question of why the gauge-fixed amplitudes (\ref{gf3}) should be
ambiguous, and if they do turn out to be so,  of why a unitary gauge
should be the correct one.

\vfill\eject

\section{The Super Period Matrix Gauge Fixing Method}
\setcounter{equation}{0}

We describe now a new procedure for descending from supermoduli 
to moduli space, which
will not produce the ambiguities encountered in
the previous section.

\subsection{Projection on the Super Period Matrix}

Our starting point is the gauge fixed formula (\ref{gf}) which gives an
integral over supermoduli space, parametrized locally by a slice ${\cal
S}$ with arbitrary local coordinates $m^A=(m^a|\zeta ^{\alpha})$.
We consider again the procedure outlined in Section 2.3.2, where $m^a$ are
coordinates for the moduli of the geometry $e_m{}^a$, and $\zeta
^{\alpha}$ are the fermionic coordinates of the gravitino field $\chi(z) =
\sum_{\alpha=1}^{2h-2} \zeta ^{\alpha}\chi_{\alpha}(z)$. Ignoring for the
moment the subtleties of chiral splitting, the main problem with this
procedure is the fact that the integral over the odd supermoduli
corresponds then to the integral along the fibers of the
projection (\ref{bpr}). However, this projection does not descend to a
projection from supermoduli space, since a supersymmetry transformation
(\ref{susyc}) on $E_M{}^A$ will change the complex structure of $e_m{}^a$
\bea
\label{bpr1}
\begin{array}{ccc}
E_M{}^A & \sim  & E_M{}^A + \delta E_M{}^A \\
\downarrow & {}   &\downarrow\\
e_m{}^a& \not\sim & e_m{}^a+\delta e_m{}^a
\end{array}
\eea
Thus the fiber of the projection (\ref{bpr}) is not well-defined,
and the complex moduli of $e_m{}^a$ itself is not an intrinsic
notion in supersymmetry.

\medskip

Although there is no known modification of the zweibein $e_m{}^a$ 
invariant under supersymmetry, there is a natural supersymmetric
invariant generalization of the period matrix $\Omega_{IJ}$. This is the
super period matrix $\hat\Omega_{IJ}$ \cite{dp88,dp89}, which can be
constructed as follows.
The first construction is in terms of supergeometry, and hence
manifestly supersymmetric. Fix a canonical homology basis $(A_I,B_I)$,
$I=1,\cdots ,h$ for $\Sigma$, $\#(A_I\cap B_J)=\delta_{IJ}$. Then there
exists a unique basis of superholomorphic odd differentials $\hat
\omega_I ({\bf z})$ of U(1) weight $1/2$ which satisfy
\be
{\cal D}_-\hat\omega_I=0,
\quad\quad
\oint_{A_J}dzd\theta\,\hat\omega_I=\delta_{IJ}
\ee
Here $\hat\omega_I({\bf z})=\hat\omega_{I0}+\theta\hat\omega_{I+}$, and
${\cal D}_-$ is the covariant derivative on forms of weight $1/2$
(cf. Appendix B for explicit formulas)
The super period matrix $\hat\Omega_{IJ}$ is defined then by
\be
\label{spm}
\hat\Omega_{IJ}=\oint_{B_J}dzd\theta\,\hat\omega_I
\ee
The second construction of $\hat\Omega_{IJ}$ is in components, and relates
it to the period matrix $\Omega_{IJ}$ for the metric $g_{mn} =
e_m{}^ae_n{}^b\delta_{ab}$. Recall that $z,\bar z$ are conformal
coordinates for $g_{mn}$, and let $\omega_I$ be the basis of holomorphic
differentials with respect to $g_{mn}$ which is dual to the homology
cycles $A_I$. Then
\be
\oint_{A_I}\omega_J(z)dz=\delta_{IJ}
\quad\quad
\oint_{B_I}\omega_J(z)dz=\Omega_{IJ}
\ee
The super period matrix can be equivalently defined by
\be
\hat\Omega_{IJ}
=
\Omega_{IJ}-{i\over 8\pi}\int\int d^2z\,d^2w\ 
\omega_I\chiz\hat S_{\delta}(z,w) \chiw\omega_J
\nonumber
\ee
The modified Szeg\"o kernel $\hat S_{\delta}(z,w)$ is a $-1/2$ form in
both $z$ and $w$, and is defined as the unique solution to the
integral-differential equation
\be
\label{sze}
\p_{\bar z}\hat S_{\delta}(z,w)
+{1\over 8\pi}\chiz\int d^2x\chix \p_z\p_x\ln E(z,x)\hat S_{\delta}(x,w)
=2\pi\delta(z,w)
\ee
and may clearly be generated explicitly from the standard Szeg\" o kernel
$S_{\delta}(z,w)$ in a perturbative series in $\chiz$, which terminates
since $\chiz$ is anticommuting and contains only a finite number of
odd Grassmann variables. (For genus 2, $\hat S_\delta (z,w) = S_\delta
(z,w)$.) Thus $\hat\Omega_{IJ}$ can be also be generated explicitly from
$\Omega_{IJ}$ in a finite series in
$\chiz$. With the help of the component representation of the super line
integral,
\be
\label{sli}
\int_{\bf w}^{\bf z}
dzd\theta\,\hat\omega_I
=\int_w^z \biggl ( dz\,\hat\omega_{I+}-{1\over 2}d\bar
z\,\chiz\hat\omega_{I0} \biggr )
+\theta_z\hat\omega_{I0}(z)-\theta_w\hat\omega_{I0}(w),
\ee
the component result (\ref{spm1}) may be recovered from (\ref{spm}).

\medskip

Returning to supermoduli space, we can now define a projection
which is invariant under supersymmetry and more generally,
infinitesimal super reparametrizations
\be
\label{gpj}
E_M{}^A \longrightarrow \hat\Omega_{IJ}
\ee
This projection is well-defined for any genus $h$. The prescription
for descending from supermoduli to moduli is then to integrate along
the fibers of (\ref{gpj}). In the Introduction,
we had described in detail the subtleties associated with
using the projection (\ref{gpj}). If we compare with the
earlier faulty way of descending in Section 2.3.2, they
can be summarized as follows.

\begin{itemize}

\item
The even coordinates $m^a$ should be taken as moduli for 
$\hat\Omega_{IJ}$ instead of $\Omega_{IJ}$ as in~(\ref{section});

\item
Now individual correlation functions in conformal field theory cannot
be written just with respect to a complex structure: they require
a metric. Thus, we must choose a metric $\hat g_{mn}$ whose period matrix
is $\hat\Omega_{IJ}$. This is similar to a choice of slice, and
introduces an arbitrariness which must be shown to be immaterial in the
final expressions for the measure and amplitudes; 

\item
The Beltrami superdifferentials $H_A$ of (\ref{superbeltrami})
are changed accordingly. In particular, $H_A$ will no longer
be split as in (\ref{bel1}). Instead, in Wess-Zumino gauge, we have
\be
\label{H}
H_A = \bar\theta (\mu_A - \theta \chi _A)
\ee
where all components of $\mu_A$ and $\chi _A$ will be non-vanishing;

\item
All correlation functions in the worldsheet supergeometry have to be
expressed in terms of the $\hat\Omega_{IJ}$ moduli instead
of $\Omega_{IJ}$. This is a deformation of the background geometry, and
will require an appropriate insertion of the stress tensor.

\end{itemize}

We shall see how this procedure generates compensating terms which
eliminate the ambiguities of the original picture-changing
formula.

\subsection{Chiral Splitting} 

Our first step in carrying out the super period matrix gauge fixing
procedure outlined above is a careful chiral splitting
of the contribution of each chiral half $\chiz$ or $\chi_z{}^-$
in the correlation functions (\ref{gf}).

\medskip

The contributions of the superghosts $B,C$ are manifestly chiral,
so the main difficulty in chiral splitting resides with the
scalar superfields $X^{\mu}$. The scalar fields $x^{\mu}$ are not split 
because of zero modes, and fields of both chiralities couple in the
action $I_m$  and the super covariant derivatives ${\cal D}_\pm$.
Nevertheless, as shown in \cite{dp89}, the chiral contributions
of $\chiz$ in the $X^{\mu}$ scalar superfield theory can be identified by
a simple effective prescription. We provide it for even spin structure
$\delta$, which is all we need in the present paper. The modifications
required when $\delta$ is odd can also be found in \cite{dp89}.

\medskip

For general vertex operators of the form\footnote{It is understood that
the actual vertex operators for the NS-NS part of the supergraviton
multiplet are recovered by expanding each vertex operator to linear
order in $\epsilon _\mu$ and linear order in $\bar \epsilon _{\bar \mu}$.
The present form is especially useful since the derivatives occur in the
exponential as linear sources, see \cite{dp88}.} 
\be
V({\bf z};k,\epsilon)
=
\exp\bigg (ik_{\mu}X^{\mu}
+\epsilon_{\mu}{\cal D}_+X^{\mu}({\bf z})
+\bar \epsilon_{\bar\mu}{\cal D}_-X^{\bar\mu}({\bf z})\bigg )
\ee
the non-chiral amplitudes can de decomposed as follows into chiral
amplitudes
\be
\label{chispl}
\<\prod_{i=1}^N\,V({\bf z}_i;k_i,\epsilon_i\>_{X^{\mu}}
=
\int dp_I^{\mu}\
\bigg|\<\prod_{i=1}^N
V^{chi}(z_i,\theta_i;k_i,\epsilon_i, p_I^{\mu})\>_+ \bigg |^2
\ee
where the effective chiral vertex operators
$V^{chi}(z_i, \theta_i;k_i,\epsilon_i,p_I^{\mu})$ are contracted with the
help of the effective rules of Table 1.

\begin{table}[h]
\begin{center}
\begin{tabular}{|c||c|c|} \hline 
 $X^{\mu}$ Superfields & {\rm Non-Chiral} & {\rm Effective Chiral} 
                \\ \hline \hline
              {\rm Bosons}  
            & $x^{\mu}(z)$ 
            & $x_+^{\mu}(z)$        
             
 \\ \hline
              Fermions  
            & $\psi_+^{\mu}(z)$ 
            & $\psi_+^{\mu}(z)$
\\ \hline
Action
& $I_m$
& $ - {1\over 2\pi}\int d^2z \chiz S_m$
\\ \hline
 Internal Loop momenta
& None
& ${\rm exp}(p_I^{\mu}\oint_{B_I}dz\partial_zx_+^{\mu})$       
            
 \\ \hline
              $x$-propagator  
            & $\<x^{\mu}(z)x^{\nu}(w)\>$ 
            & $-\delta^{\mu\nu}{\rm ln}\,E(z,w)$        
 \\ \hline
              $\psi_\pm$-propagators  
            & $\<\psi_\pm^{\mu}(z)\psi_\pm^{\nu}(w)\>$ 
            & $- \delta^{\mu\nu}S_{\delta}(z,w)$        
 \\ \hline
              Covariant Derivatives  
            & $\D _+, \ \D_-$ 
            & $\partial_{\theta}+\theta\partial_z, \
               \partial_{\bar \theta} + \bar \theta
               \partial_{\bar z}$        
 \\ \hline
\end{tabular}
\end{center}
\caption{Effective Rules for Chiral Splitting}
\label{table:1}
\end{table}
In Table 1, all correlators in the effective chiral formulation are
computed in terms of the effective chiral fields $x_+^{\mu}$ and
$\psi_+^{\mu}$. There, the $S_m$-dependent effective action is to be
inserted, as is the internal loop momentum dependent exponential, and the
corresponding expectation values will be indicated by $\< \cdots \>_+$. 
As suggested in \cite{vv} for bosonic scalar fields, the parameters
$p_I^{\mu}$ can be interpreted as internal loop momenta.

\medskip
 
In this paper, we shall focus on the partition function, in order to
construct the chiral measure. For scalar superfields, the preceding chiral
splitting prescription yields
\be
\label{1}
\<1\>_{X^{\mu}}
=
\int dp_I^{\mu}
\bigg | \bigg \< \exp \biggl ( 
{1\over 2\pi}\int d^2z\chiz S_m(z) + p_I^{\mu}\oint_{B_I}dz\p_z x_+^{\mu} 
\biggr ) \bigg \> _+ \bigg|^2
\ee
The chiral blocks on the right hand side are easily evaluated
using the identities (\ref{prf}) for the prime form.
We obtain in this way the following basic formula
for the chiral block of the partition function 
\be
\label{block}
\biggl \< \exp \biggl ( {{1\over 2\pi}\int d^2z\chiz S_m
+p_I^{\mu}\oint_{B_I}dz\p_z x_+^{\mu}} \biggr ) \biggr \>_+
=
e^{i \pi p_I ^\mu \hat \Omega_{IJ} p_J^{\mu}}
\,
\biggl \< \exp \biggl (
{{1\over 2\pi}\int d^2z\chiz S_m} 
\biggr ) \biggr \>_+
\ee
where $\hat\Omega_{IJ}$ is the super period matrix introduced in the
previous subsection. Integrating in $p_I^{\mu}$, this implies immediately
the following formula for the scalar partition function
\be
\label{blockk}
\<1\>_{X^{\mu}}
=
(\det\,\Im\hat\Omega)^{-5}\ \bigg | \biggl \< \exp \biggl ( 
{{1\over 2\pi}\int d^2z \chiz  S_m(z)} \biggr ) \biggr \>_+ \bigg|^2
\ee
With the basic formula (\ref{block}) for chiral blocks, we 
now return to the construction of the chiral superstring
amplitude. Assembling the chiral blocks of matter and ghost
fields, we  define the chirally split partition function ${\bf A}^{chi}
[\delta ](p^\mu _I)$ in terms of a basic correlator $\A [\delta]$,
\bea
\label{gf4}
{\bf A}^{chi}[\delta](p_I^{\mu})
&=&
\prod_A dm^A \
\exp ( i \pi p_I^{\mu}\hat \Omega_{IJ} p_J^{\mu}) \
\A [\delta ]
\nonumber \\
\A [\delta ] & \equiv &
\biggl \< \prod_A \delta(\<H_A|B\>)
\exp \biggl ( {1\over 2\pi}\int d^2z \chiz S(z) \biggr ) \biggl \> _+
\eea
Earlier prescriptions for chiral splitting had missed the appearance of
the super period matrix $\hat\Omega_{IJ}$. But perhaps more important,
its  appearance in chiral splitting is a confirmation that it is the
variable which should be used in gauge fixing the superstring.

\subsection{Parametrizations of Supermoduli as Fiber Space}

The next step in our gauge fixing procedure is to provide
suitable coordinates $m^A$ in which the fiber of the supersymmetric
projection (\ref{gpj}) is conveniently parametrized. A detailed
discussion of the choices of slices and associated Beltrami differentials
is given in Appendix~D.

Henceforth, we shall restrict our discussion to the case of genus $h=2$.
This case is simpler, because all matrices in the Siegel upper half-space
are then period matrices of a metric, and because the construction below
of gravitino fields $\chi_{\alpha}$ does not require any iteration.
In genus $h=2$, it is convenient (although by no means necessary
for our arguments) to take $\{\hat\Omega_{IJ}\}_{I\leq J}$
as local holomorphic coordinates for the space of matrices $\hat
\Omega_{IJ}$. Choose a corresponding $3$-dimensional slice $\hat S$ of
zweibeins  $\hat e_m{}^a$ whose period matrices are the matrices
$\hat\Omega_{IJ}$. For each point on the slice $\hat S$, choose $2$
generic gravitino sections $\hat \chi_{\alpha}$, $\alpha =1,2$, and set
$\hat\chi=\sum_{\alpha=1}^{2h-2}\zeta^{\alpha}\chi_{\alpha}$, for $2h-2$
anticommuting parameters $\zeta ^{\alpha}$. We may choose then as follows
a $(3|2)$-dimensional slice of supergeometries $(e_m{}^a,\chi)$ whose
corresponding period matrix $\Omega_{IJ}$ satisfies the relation
(\ref{spm}). First, we observe that the relation (\ref{spm1}) implies
that $\Omega_{IJ}$ differs only from $\hat\Omega_{IJ}$ by terms of
order $O(\zeta\zeta)$. We may choose then zweibeins $e_m{}^a$ with period
matrix $\Omega_{IJ}$ which differ from $\hat\Omega_{IJ}$ by terms
of order $O(\zeta\zeta)$. But the gravitino sections $\hat\chi_{\alpha}$
can then be considered as gravitino sections $\chi_{\alpha}$
with respect to $e_m{}^a$, since any corrections would be of order 
strictly greater than $2$ and hence vanish.  

\medskip

Altogether, we have obtained a $(3|2)$-dimensional slice of 
supergeometries
\be
(\hat\Omega_{IJ},\zeta^{\alpha})
=m^A\longrightarrow (e_m{}^a,\chi=\sum_{\alpha=1}^2\zeta^{\alpha}
\chi_{\alpha})
\ee
with the fiber of the projection (\ref{gpj}) given precisely
by $m^a={\sl constant}$. It is convenient to introduce the Beltrami
differential $\hat\mu_{\bar w}{}^w$ which deforms the
metric $\hat g_{mn}$ to the metric $g_{mn}$
\be
\label{bd}
\hat\mu_{\bar w}{}^w
={1\over 2}\hat g_{w\bar w} \biggl ( g^{ w w} - \hat g^{ w w} \biggr ) 
\ee
If $w,\bar w$ are conformal coordinates for the metric $\hat g_{mn}$,
then we may set $\hat g_{ww}=\hat g_{\bar w\bar w}=0$.

\medskip
 
The main consequence of a choice of a slice is the corresponding Beltrami
superdifferentials $H_A$. We have already stressed that, unlike the slice
used in Section 2.3.3 for the derivation of the picture-changing formula
(\ref{brst}), a generic slice ${\cal S}$ based on $\hat\Omega_{IJ}$ will
lead in general to Beltrami superdifferentials which in Wess-Zumino gauge
assume the simplified form (\ref{H}), $ H_A=\bar\theta (\mu_A - \theta
\chi _A)$, which have both non-vanishing $\mu _A$ and $\chi _A$ 
components. For the slice we have just constructed
using $\hat\Omega_{IJ}$, the Beltrami superdifferentials $H_A$ have the
following properties:

\medskip

$\bullet$ 
Let $\Phi_{IJ}$ be the following basis of odd superholomorphic 3/2 
superdifferentials (see Appendix B)
\be
\label{Phis} 
\Phi _{IJ} = -{i \over 2}\biggl ( \hat \omega _J \D _+ \hat \omega
_I + \hat \omega _I \D _+ \hat \omega _J \biggr )
\ee
Then the matrix $\<H_a|\Phi_{IJ}\>$ has maximal rank. In fact, since
we have chosen the even supermoduli to be $m^a=\hat\Omega_{IJ}$,
$I\leq J$, we have
$\<H_a|\Phi_{IJ}\>=\delta_{a,IJ}$, as shown in (\ref{varsuperperiod}).
Geometrically, the tangent vectors to supermoduli space corresponding to
$H_a$ modulo super reparametrizations are dual to the cotangent vectors
$\Phi_{IJ}$.

$\bullet$ 
The even component of $H_{\alpha}=\bar \theta (\mu _\alpha - \theta
\chi _\alpha)$ is given by
\be
\chi _{\alpha} (z) ={\p\chiz (z) \over \p \zeta ^{\alpha}},
\qquad
\alpha = 1, 2
\ee
The odd Beltrami differential $\mu _\alpha$ is associated with the
dependence of the metric on $\zeta^\alpha$ and is related to the Beltrami
differential $\hat \mu $ corresponding to the deformation of the metric
$\hat g_{mn}$ to the metric $g_{mn}$ by the following formula
\be
\label{mualpha}
\mu _\alpha = {\p \hat \mu  \over \p \zeta ^\alpha}  
\ee
where $\hat \mu$ is the Beltrami differential deforming $\hat g_{mn}$ to
$g_{mn}$. 

$\bullet$ 
The $H_\alpha$ obey an orthogonality condition with $\Phi _{IJ}$
which guarantees that $H_\alpha$ produce no variations in $\hat \Omega
_{IJ}$,
\be
\< H_\alpha |\Phi _{IJ}\>=0\, .
\ee
This relation determines the conformal class $[\mu _\alpha]$ of $\mu
_\alpha$ but leaves the precise form of $\mu_\alpha$ subject to the same
choice of slice that exists for $\hat \mu $ itself. Similarly,
the exact values of $\mu_a$, $\chi _a$ depend on the many choices
which entered the construction of the slice ${\cal S}$.

\subsection{Auxiliary Dirac $\delta$ Beltrami Superdifferentials}

The insertion of the superghost $\delta(\<H_A|B\>)$ functions in
Wess-Zumino gauge for generic Beltrami differentials $H_A$
produces an enormous complication of the superstring measure,
\bea
\label{badinsertion}
\prod _A \delta (\< H _A | B \>) 
= 
\prod _{a=1}^{3} 
\biggl ( \< \mu_a | b \> - \<\chi _a |\beta\> \biggr )
\prod _{\alpha=1}^{2} 
\delta \biggl ( \< \mu _\alpha | b \> + \< \chi _\alpha | \beta \>
\bigg )\, .
\eea
The product over the bosonic index $a$ will produce already $8$ terms
(and $2^{3h-3}$ terms in genus $h$). It is then much more convenient basis to
work with a basis $ H ^* _A$ where this proliferation does not take
place. First, we derive the formulas for changing bases.

\medskip

It is simplest to return to the chirally symmetric expression
for the superghost contributions. For any given set of Beltrami
superdifferentials $H_A$, we have (for genus $h\geq 2$)
\be
\int D(B\bar BC\bar C) |\prod_A \delta(\<H_A|B\>)|^2 e^{-I_{gh}}
=
|\sdet\,{\cal D}_+{\cal D}_-^{(3/2)}|^2
{|\sdet\,\<H_A| \Phi _C \>|^2
\over
\sdet\,\<\Phi_A|\Phi_C \>}
\ee
Here the upper index $3/2$ indicates that the super Laplacian ${\cal
D}_+{\cal D}_-^{(3/2)}$ acts on fields of U(1) weight $3/2$, and $\Phi_C$
is any $(3h-3|2h-2)$ dimensional basis of superholomorphic $3/2$
superdifferentials. As an immediate consequence, we see that the
inner product of $H_A$ effectively is always taken with a holomorphic
form. Using the behavior of the $\delta$ function under a change of
basis, we readily obtain an expression involving the $\delta$ function
for any other set of Beltrami superdifferentials $H ^* _A$, 
\be
\label{chb}
\prod_A \delta(\<H_A|B\>)
=
{\sdet\,\<H_A|\Phi_B\>
\over
\sdet\,\< H ^* _A |\Phi_B\>}
\prod_A \delta(\< H ^* _A|B\>)
\ee
Thus we can exchange the correlation functions of
$\prod_A\delta(\<H_A|B\>)$ for the potentially much simpler correlation
functions of $\prod_A\delta(\< H ^* _A|\Phi_B\>)$, at the cost of
introducing a ratio of {\sl finite dimensional} superdeterminants.

\medskip

A convenient choice for the new basis $H ^* _A$ of Beltrami
superdifferentials is generic $\delta$-functions. Let $p_a$ and $q_\alpha$
be generic distinct points on the surface. By setting
\bea
\label{superbeltrami1}
H^* _a (z,\theta) = \bar \theta \delta (z,p_a) 
\qquad && \qquad a=1,\cdots ,3
\nonumber \\
H^* _\alpha (z,\theta) = \bar \theta \theta \delta (z,q_\alpha)
\qquad && \qquad \alpha = 1,2\, ,
\eea
we obtain the following simpler effective insertion formula
\bea
\label{si}
\prod _A \delta (\< H _A | B \>) 
=
{\sdet \< H_A | \Phi _B \> \over \sdet \< H ^* _A | \Phi _B\>} 
\prod _a b(p_a) \
\prod _\alpha \delta \big ( \beta (q_\alpha) \big )\, .
\eea
It is important to realize that with above insertions of $b(p_a)$ and
$\delta (\beta (q_\alpha))$, the ghost and superghost Green's functions
$G_2$ and $G_{3/2}$ will automatically be normalized at the points $p_a$
and $q_\alpha$ by $G_2(p_a,w)= G_{3/2}(q_\alpha,w)=0$, a fact that
represents a very considerable simplification as compared with the
general insertions of (\ref{badinsertion}).

\subsection{Evaluation of the Finite Dimensional
Superdeterminant}

With the previous formula (\ref{si}), all the dependence on the
choice of slice ${\cal S}$ for supermoduli space is concentrated
in the background geometry of the effective correlation
functions and in the finite dimensional superdeterminant
\be
\label{sdeterminant}
{\sdet \< H_A | \Phi _B \> \over \sdet \< H^*  _A | \Phi _B\>}
\ee
where $\Phi_B$ is an arbitrary basis of superholomorphic
$3/2$ superdifferentials. We shall make use of two such bases,
$\Phi_A$ and $\Phi_A^*$, which are dual respectively
to the Beltrami superdifferentials $H_A$ and $H^* _A$.

\medskip

$\bullet$ 
The first basis $\Phi_A$ of superholomorphic $3/2$
superdifferentials is defined by duality 
\be
\<H_A|\Phi _B\> = \delta _{AB} \, ,
\ee
Henceforth, we always choose, locally, the bosonic coordinates $m^a$ of
the slice ${\cal S}$ as a subset of the variables $\hat\Omega_{IJ}$. In
this case, the odd superholomorphic superdifferentials $\Phi_a$ are given
by the 3/2 superdifferentials $\Phi_{IJ}$ of (\ref{Phis}),
$\Phi_a=\Phi_{IJ}$, $a=1, \cdots , 3h-3$. There is no such simple
expression for the even superdifferentials $\Phi_{\alpha}$, since they
depend on the gauge slice~${\cal S}$.

\medskip

$\bullet$ 
The second basis $\Phi_A^*$ of superholomorphic 3/2 differentials is
defined instead by normalization conditions at the points
$p_a$, $q_{\alpha}$. If we write
$\Phi_A^*=\Phi_{A0} ^* + \theta \Phi_{A+} ^* $, these normalization
conditions are\footnote{Explicit formulas for $\Phi_A^*$ in terms of 
$\tet$-functions and Green's functions can be found in Appendix B.}
\bea
\label{pqn}
\Phi _{\alpha 0} ^* (q_\beta) &=& \delta _{\alpha \beta} 
\qquad \qquad 
\Phi _{\alpha +} ^* (p_b) \ = \ 0
\nonumber \\
\Phi _{a +} ^* (p_b) &=& \delta _{ab}
\qquad \qquad \
\Phi _{a 0} ^* (q_\beta)\ =\ 0 \, .
\eea
In particular, we have
\be
\<H^* _A|\Phi_B^*\>=\delta_{AB}
\ee

\medskip

Returning to the superdeterminant (\ref{sdeterminant}), it suffices to
evaluate $\sdet\,\<H^* _A|\Phi_B\>$. Since both $\Phi_A$ and
$\Phi_A^*$ are bases of superholomorphic 3/2 superdifferentials,
$\Phi _\beta$ can be expressed as a linear combination of the basis of
even holomorphic differentials $\Phi ^* _\beta$ as well as the odd
holomorphic differentials $\Phi _{IJ}$
\be
\label{lincomb}
\Phi _\beta (\z) 
= \Phi ^* _\gamma (\z) \ C^\gamma {}_\beta 
+ \Phi _{IJ} (\z) \ D^{IJ} {}_\beta
\ee  
where the coefficients $C$ and $D$ are independent of $\z$, $C$ even and
$D$ odd. The superdeterminant $\sdet \< H^* _A |\Phi _B\>$ may now be
evaluated as follows
\bea
\label{superdet}
\sdet \< H^* _A | \Phi _B \> 
&=&
\sdet \left (\matrix{
\< H^* _a | \Phi _{IJ} \>     & \< H^* _a | \Phi _\beta \> \cr 
&\cr
\< H^* _\alpha |\Phi _{IJ} \>& \< H^* _\alpha |\Phi _\beta \>
}\right )
\nonumber \\ 
&& \nonumber \\
&=&
\sdet \left (\matrix{
\< H^* _a | \Phi _{IJ} \> & \< H^* _a | \Phi ^* _\gamma \>
C^\gamma {}_\beta + \< H^* _a | \Phi _{IJ} \> D^{IJ} {}_\beta
\cr  &\cr
 \< H^* _\alpha |\Phi _{IJ} \> & \< H^* _\alpha |\Phi _\gamma ^*
\> C^\gamma {}_\beta +  \< H^* _\alpha |\Phi _{IJ} \> D^{IJ} {}_\beta}
\right )
\, .
\eea
Now we make use of the fact that the addition of multiples of columns in
the superdeterminant is immaterial (shown in Appendix C).
As a result, the shift by $D$ in (\ref{superdet}) is immaterial, just as
for ordinary determinants. We can simplify
the resulting formula further by using the above duality
relations between $H^* _A$ and $\Phi_B^*$
\be
\sdet \< H^* _A | \Phi _B \> 
=
\sdet \left (\matrix{
\< H^* _a | \Phi _{IJ} \>     & 0 \cr 
&\cr
\< H^* _\alpha |\Phi _{IJ} \>& C^\alpha {}_\beta
}\right ) ={\det\, \<H^* _\alpha |\Phi _{IJ} \> \over \det\, C}
\, .
\ee
Finally, taking the inner product of (\ref{lincomb}) against $H_\alpha$,
we obtain 
\be
\<H_\alpha |\Phi _\beta \> = \delta ^\alpha{} _\beta = \<H_\alpha |\Phi ^*
_\gamma \> C^\gamma {}_\beta
\ee
and conclude with the following final formula in terms of finite 
dimensional determinants
\be
{\sdet \< H_A | \Phi _B \> \over \sdet \< H^*  _A | \Phi _B\>}
= \biggl ( 
 \det \, \Phi _{IJ+}(p_a) \cdot \det \< H_\alpha |\Phi ^*_\beta\>
\biggr ) ^{-1}\, .
\ee

\subsection{First Summary}

It is convenient to summarize here our formula for the chiral superstring
measure, for fixed spin structure $\delta$
\bea
\label{gf5}
{\bf A}^{chi}[\delta](p_I^{\mu})
&=&
\prod _a dm^a \prod_\alpha d\zeta ^ {\alpha}
\exp(i \pi p_I^{\mu}\hat\Omega_{IJ}p_J^{\mu}) \A [\delta]
\nonumber \\
\A [\delta ] &=&
{ \bigg \< \prod_a b(p_a) \prod_ \alpha \delta(\beta(q_{\alpha}) 
\exp \biggl ( {1\over 2\pi}\int d^2z \chiz S(z) \biggr ) \bigg \>_+ (g)
\over 
\det \, \Phi _{IJ+}(p_a) \cdot \det\, \< H_\alpha |\Phi ^*_\beta\>}
\eea
In this formula, we should stress that $p_a$, $q_{\alpha}$ are arbitrary
generic points, unrelated to the slice ${\cal S}$. As we saw in the above
derivation, they are a computational device, and the amplitude
(\ref{gf5}) manifestly does not depend on them. The dependence on $g=g_{mn}$ is
made explicit as a reminder that the correlation function is with respect
to the metric $g_{mn}$. 
In our approach, $\hat\Omega_{IJ}$ is the only intrinsic notion,
and thus the metric $g_{mn}$ is slice dependent. So is $\chiz$.
As pointed out before, we can change metric
backgrounds from $g_{mn}$ to $\hat g_{mn}$ by using the stress tensor.
After this is properly done,
the slice dependence of the correlation functions
will have to cancel out with the slice
dependence of the finite dimensional determinants.

\vfill\eject

\section{The Genus 2 Chiral Superstring Measure}
\setcounter{equation}{0}

In this section, we show how to evaluate the gauge-fixed formula
(\ref{gf5}) explicitly. Our method is quite general, but the calculations
are much simpler in genus $h=2$, since there are then only
$2$ supermoduli. We shall obtain in this case a formula which can be
proven independently to be invariant under infinitesimal changes of the
gauge slice ${\cal S}$. Thus there are no ambiguities as had occurred
earlier in the picture-changing formula (\ref{brst}).

\subsection{Formulation in terms of $\hat g_{mn}$}

We begin by making more explicit the deformation of background
metric from $g_{mn}$ to $\hat g_{mn}$. First, in genus $h=2$, the
gravitino field $\hat\chi$ is identical with $\chi$, since
$\hat\chi - \chi$ is of order $\zeta \zeta \zeta$, and must thus vanish.
Next, the Beltrami differential of (\ref{bd}), given by $\hat\mu (z)
={1\over 2}\hat g_{z \bar z} g^{zz}$ is of order $\zeta \zeta$,
so that deformations can be obtained exactly by a single insertion of the
stress tensor
\bea
&&
\biggl \< \prod_a b(p_a) \prod_\alpha
\delta \big (\beta(q_{\alpha}) \big )
\exp \biggl ( {1\over 2\pi}\int d^2z \chiz S(z) \biggr ) \biggr \>_+ (g)
\nonumber\\
&&
\hskip .5 in
=
\biggl \< \prod _a b(p_a) \prod _\alpha
\delta \big ( \beta(q_{\alpha}) \big )
\exp \biggl ( {1\over 2\pi}\int d^2z \chiz S(z) \biggr ) \biggr \>_+
(\hat g)
\nonumber\\
&&
 \hskip .7in
+\int _\Sigma d^2z \hat \mu (z) \biggl \<  T(z)
\prod _a b(p_a) \prod _ \alpha \delta \big ( \beta(q_{\alpha}) \big)
\biggr \>_+ (\hat g)
\eea
In the second term on the right hand side, the supercurrent
contribution has been dropped, since the remaining factors are already
of order $\zeta \zeta$. Henceforth, we shall consider only correlation
functions with respect to the background metric $\hat g_{mn}$,
and denote them by $\<\cdots \>$, dropping the subscript $+$ and the
dependence $\hat g$.
Similarly, $z$ will denote henceforth a holomorphic coordinate
for $\hat g_{mn}$ (and no longer for $g_{mn}$, as had been the
case up to this point). The chiral superstring measure can
then be expressed as
\bea
\label{compamp}
{\cal A} [\delta] 
& = & 
{\< \prod _a b(p_a) \prod _\alpha \delta (\beta (q_\alpha)) \>
\over \det \Phi _{IJ+} (p_a) \cdot \det \< H_\alpha | \Phi ^* _\beta\>}
\biggl \{ 1 - \half {1 \over (2 \pi )^2} \int \! d^2z \chiz \int \! d^2 w
\chiw
\< S(z) S(w)\> 
\nonumber \\
&& \hskip 2in+ {1 \over 2 \pi} \int d^2 z \hat \mu (z) \< T(z)\> \biggr \}
\eea
where the supercurrent and stress tensor correlators are defined as usual
\bea
\< S(z) S(w) \> & = & {\< S(z) S(w) \prod _a b(p_a) \prod _\alpha \delta
(\beta (q_\alpha)) \> \over \< \prod _a b(p_a) \prod _\alpha \delta
(\beta (q_\alpha)) \>}
\nonumber \\
&& \nonumber \\
\< T(z) \> &=& {\< T(z) \prod _a b(p_a) \prod _\alpha \delta (\beta
(q_\alpha)) \> \over \< \prod _a b(p_a) \prod _\alpha \delta (\beta
(q_\alpha)) \>}
\eea
and all moduli are $\hat \Omega _{IJ}$. However, the finite-dimensional
determinant prefactors $\det\,\Phi_{IJ}(p_a)$ and $\sdet\<H_{\alpha}|
\Phi_{\beta}^*\>$ are supergeometric notions, and as such, are still
formulated in the supergeometry $(g_{mn},\chiz)$.

\subsection{Evaluation of the Correlators}

The chiral partition function may be expressed as \cite{vvv}
\bea
{\cal Z} 
\equiv 
 \< \prod _a b (p_a) \prod _\alpha \delta (\beta (q_\alpha)) \> 
&=& 
{\tet [\delta ](0)^5 \tet (D_b) \prod _{a<b} E(p_a, p_b) \prod _a \sigma
(p_a)^3 \over Z^{15}  \tet [\delta ](D_\beta) \prod _{\alpha <
\beta } E(q_\alpha , q_\beta) \prod _\alpha \sigma (q_\alpha )^2}
\nonumber
\eea
where the ghost and superghost divisors are defined by
\be
D_b = \sum _a p_a -3\Delta 
\qquad \qquad 
D_\beta = \sum _\alpha q_\alpha -2\Delta\, ,
\ee
and the scalar partition function $Z$ is given by
\be
Z^3 = {\tet (\sum _I z_I -w -\Delta) \over \sigma (w) \prod _I E(z_I,w)}
{ \prod _{I<J} E(z_I,z_J) \prod _I \sigma (z_I) \over \det \omega _I(z_J)}
\ee
a formula in which neither side depends upon $z_I$ or $w$. For
further useful formulas on $\tet$-functions, differentials and Green's
functions, see Appendix A.

\subsubsection{The Supercurrent Correlators}

Everything we need can be deduced from the correlator of two supercurrents
\be
\< S(z) S(w) \prod _a b(p_a) \prod _\alpha \delta (\beta (q_\alpha)) \>
= C(z,w) \< \prod _a b(p_a) \prod _\alpha \delta (\beta
(q_\alpha)) \>\, ,
\ee
where the expectation values $\< \cdots \>$ are taken over all the chiral
fields $x_+, \psi_+, b, c, \beta $ and $\gamma$, and the dependence on
$p_a$ and $q_\alpha$ in $C(z,w)$ is suppressed.  The total supercurrent is
defined by $S=S_m +S_{gh}$ with 
\be
S_m = - \half \psi _+ ^\mu \p _z x^\mu 
\qquad \qquad
S_{gh} = \half b \gamma - {3 \over 2} \beta \p _z c - (\p _z \beta) c
\ee
and the relevant correlators are given by
\bea
\< \psi _+ (z) \psi _+ (w) \> & = & - G_{1/2}[\delta] (z,w) = - S_\delta
(z,w)
\nonumber \\
\< \p_z x _+ (z) \p_w x _+ (w) \> & = & -\p_z \p_w \ln E(z,w)
\nonumber \\
\< b(z) c(w) \> &=& +G_2 (z,w) \nonumber \\
\< \beta (z) \gamma (w) \> &=& -G_{3/2}[\delta] (z,w) 
\eea
The Green's functions are given by (see \cite{fay} and \cite{vvv})
\be
G_n[\delta](z,w)  
=  {\tet [\delta ] (z-w+D_n) \over \tet [\delta ](0) E(z,w) }
\prod _{i=1} ^n {E(z,z_i) \over E(w,z_i)} {\sigma (z)^{2n-1} \over \sigma
(w)^{2n-1}} 
\ee
where the divisor $D_n = \sum _i z_i -(2n-1)\Delta$ and $\Delta$ is the
Riemann vector. When no confusion is possible, the dependence on
$[\delta]$ will not be exhibited.

\medskip

Since $\< S_m (z) S_{gh} (w)\>=0$, we may split the calculation of
$C(z,w)$ into matter and ghost parts $C(z,w) = C_m(z,w)+C_{gh}(z,w)$.
With the help of the above propagators, the matter contribution is found
to be
\be
4 C_m(z,w) = 10 S_\delta (z,w) \ \p_z \p_w \ln E(z,w)
\ee
while the ghost contribution is obtained from the following manipulations
\bea
\label{ghostcorrelator}
4C_{gh}(z,w) 
&=&
 \< b\gamma (z) \bigl ( -3 \beta \p_w c(w) -2 (\p_w \beta) c(w) \bigr ) \>
- (z
\leftrightarrow w)
 \\
&=& 
-3 \< b(z) \p_w c(w)\> \<\gamma (z) \beta (w)\> 
-2 \< b (z) c(w) \> \< \gamma (z) \p_w \beta (w)\> -(z\leftrightarrow
w)
\nonumber \\
&=&
3 \p_w G_2(z,w) G_{3/2}(w,z) +2 G_2 (z,w) \p_w G_{3/2}(w,z) -(z
\leftrightarrow w)
\nonumber
\eea

\subsubsection{The Stress Tensor Correlators}

In view of the $\N=1$ superconformal structure of both the matter and
ghost parts of the superstring, we have the following operator product
relation involving the supercurrent $S(z)$ and the stress tensor $T(z)$,
\be
S(z) S(w) =  {2c/3 \over (z-w)^3} + \half {  T(w) \over z-w} + {\rm
regular}\, .
\ee
For the superstring, the total central charge $c$ vanishes (as may be
checked explicitly by adding the cubic poles of $C_m$ and $C_{gh}$) and
the stress tensor term produces the leading and only singularity as $z\to
w$. It is convenient to calculate the stress tensor correlator $\<T(z)\>$
from $C(z,w)$ by picking up the limit as $z\to w$. To extract
$T(w)$, we need the expansion of the Green's functions $G_n(z,w)$ up to
order $\O (z-w)$ included. Denoting the coefficients as follows
\be
G_n (z,w) = {1 \over z-w} + f_n(w) +(z-w) \{g_n(w) -T_1(w) \} +\O(z-w)^2
\ee
where the chiral scalar boson stress tensor $-T_1$ is defined by
\bea
T_1(z) &=& \lim _{w\to z} \half 
\biggl (\pz x(z) \pw x(w) + {1 \over (z-w)^2} \biggr )
\nonumber \\
E(z,w) &=& (z-w) + (z-w) ^3 T_1(w) +\O(z-w)^4
\eea
Using the explicit formulas for $G_n(z,w)$, we find
\bea
\label{fandg}
f_n(w) & = & \omega _I (w) \p _I \ln \tet [\delta ] (D_n) + \p _w \ln
\biggl ( \sigma (w) ^Q \prod _i E(w,z_i) \biggr )
\nonumber \\
g_n(w) & = & \half \omega _I (w) \omega _J (w) \p _I \p _J \ln \tet
[\delta ] (D_n) + \half f_n(w)^2 + \half \p_w f_n (w)
\eea
and from this the full stress tensor
\be
T_n(w) = g_n(w) - n \pw f_n(w) -T_1(w)
\ee
Note that $f_n(w)$ is the same as $\< j_w \>$ in Verlinde and Verlinde
\cite{vvv}, formula (7.12).

\medskip

The full stress tensor may be extracted as the residue of the $z=w$
pole and we find
\be
T= 5T_{1/2} - 10 T_1 - T_{3/2} + T_2
\ee
where each of these tensors is given by
\bea
T_{1/2} (w) & = & 
 g_{1/2}(w) -T_1(w)  \nonumber \\
&=& \half \omega _I\omega _J(w) \p _I \p _J \ln \tet
[\delta ](0) - T_1 (w)
\nonumber \\
T_{3/2} (w) & = & 
 g_{3/2} (w) -{3\over 2} \p _w f_{3/2}(w) -T_1(w)
\nonumber \\
& = & \half \omega _I\omega _J(w) \p _I \p _J \ln \tet [\delta
](D_\beta ) + \half f_{3/2}(w)^2 - \p_w f_{3/2}(w) - T_1 (w)
\nonumber \\
T_2(w) &=&
 g_2(w) -2  \p_w f_2 (w) -T_1 
\nonumber \\
&=& \half \omega _I\omega _J(w) \p _I \p _J \ln \tet [\delta
](D_b ) + \half f_2 (w)^2 - {3 \over 2}\p_w f_2 (w) - T_1 (w)\, .
\eea
Combining all of the above, we find for the full stress tensor
\bea
\label{stresstotal}
T(w) 
&= &
-15 T_1(w) + \half f_2 (w)^2 - {3 \over 2}\p_w f_2 (w) 
- \half f_{3/2}(w)^2 + \p_w f_{3/2}(w)
 \\
&& +\half \omega _I\omega _J(w) \biggl ( 
5\p _I \p _J \ln \tet [\delta ](0) - \p _I \p _J \ln \tet
[\delta ](D_\beta ) +   \p _I \p _J \ln \tet [\delta ](D_b ) \biggr ) 
\nonumber
\eea

\vfill\eject

\section{Slice Independence}
\setcounter{equation}{0}

One of the most fundamental criteria for our gauge fixed formulas is their
independence of the choices of gauge slices. Infinitesimally, this
independence is equivalent to the invariance of the formulas under local 
diffeomorphisms (which would vary the choice of metrics $g_{mn}$, that
is, the choice of the Beltrami differential $\hat\mu_{\bar z}{}^z$) and
under local supersymmetry transformations (which would vary the choice of
$\chi_{\alpha}$). As the issue of slice independence has caused much
confusion in previous approaches, we shall provide here careful and
detailed accounts of both proofs.

\medskip

One key ingredient is a deep relation between superholomorphic notions
with respect to the supergeometry $(e_m{}^a,\chi_m{}^{\alpha})$ and
holomorphic notions with respect to the super period matrix
$\hat\Omega_{IJ}$. This is an important issue which we shall revisit in
detail in the later papers of this series. For the present paper, we
require only the  simplest example of this correspondence, which we
presently discuss.

\subsection{Superholomorphicity and Holomorphicity}

First kind Abelian superdifferentials $\hat \omega _I$ and ordinary first 
kind  Abelian differentials on a surface with period matrix $\hat
\Omega_{IJ}$ have  the same homology integrals and thus they must differ
by an exact form. Here,  we work out this result in detail and compute
the exact form for the simplest  case of genus 2 and even spin structure. 

\medskip

The differential equations defining $\hat \omega _I$ are (see Appendix B,
(\ref{superdiffeq}) for $n=\half$),
\bea
\nabla _{\bar z} \hat \omega _{I+} + \half \nabla _z (\chiz \hat    
\omega _{I0})  &=& 0 \nonumber \\
\nabla _{\bar z} \hat \omega _{I0} + \half \chiz \hat \omega _{I+} &=& 0
\eea 
Now, let $\hat \mu (z) \equiv \hat \mu_{\bar z}{}^z = \half g_{z\bar z}
g^{zz}$ be a Beltrami differential that  accounts for the
variation of the metric $\hat g$ to the metric $ g$, so that
\be
\label{diffomega}
\Omega _{IJ} - \hat \Omega _{IJ} = i \int \! d^2 z\,  \hat \mu (z) \omega
_I \omega _J (z)
\ee
then the covariant derivatives $\nabla$ with respect to $g$ may be 
expressed in  terms of the covariant derivatives $\hat \nabla$ with
respect to $\hat g$ as  follows
\be
\nabla _{\bar z} ^{(n)} = \hat \nabla _{\bar z} ^{(n)} +  \hat \mu \nabla
_z  ^{(n)} + n (\nabla _z \hat \mu)\, .
\ee
The equation written with respect to the metric $\hat g$ is now
\be
\hat \nabla _{\bar z} ^{(1)} \hat \omega _{I+} + \nabla _z ^{(-1)}
(\mu 
\hat \omega _{I+} + \half \chiz \hat \omega _{I0})=0\, .
\ee
The form $\hat \mu \hat \omega _{I+} + \half \chiz \hat \omega _{I0}$ has
vanishing  inner  product with every holomorphic 1-form $\omega _J$, as
can be seen by
\be
\int \omega _J (\hat \mu \hat \omega _{I+} + \half \chiz \hat \omega
_{I0}) = \int \hat \mu \omega _I \omega _J + \half \int \omega _J \chiz
\hat
\omega _{I0} =0
\ee
and thus there exists a (well-defined, single valued) scalar function 
$\lambda  _I(z)$ such that
\be
\hat \mu \hat \omega _{I+} + \half \chiz \hat \omega _{I0} = - \p _{\bar
z}
\lambda _I
\ee
The function $\lambda _I(z) $ itself may be recovered up to an additive 
constant
\be
\lambda _I (z) = \lambda _I (z_0) + {1 \over 2 \pi}
\int d^2w \pw \ln {E(w,z) \over E(w,z_0)} \biggl ( \hat \mu _{\bar w}
{}^w  
\omega  _I (w) + \half \chiw \hat \omega _{I0} (w) \biggr )\, .
\ee
The full relation may now be written as follows
\bea
\hat \omega _I  &=& \theta \omega _I(\hat \Omega, \chi=0)  
+ \D _+ \Lambda _I
\nonumber \\
\Lambda _I (z,\theta) &=& \lambda _I (z) + \theta  \hat \omega _{I0}(z) \, .
\eea
Here, $\omega _I(\hat \Omega , \chi =0)$ stands for the ordinary first kind 
Abelian differential on a surface with period matric $\hat \Omega$.
Of the Beltrami differential $\hat \mu$, only the class is known. The
effect of  a  change of Beltrami differential within the class produces a
simple  transformation on $\lambda _I$ by
\bea
\hat \mu            & \to & \hat \mu + \p _{\bar z} v^z \nonumber \\
\lambda _I (z) & \to & \lambda _I(z) - v^z \omega _I(z)\, ,
\eea
a formula that will be very useful later on.

\medskip

The effect of this reformulation on the holomorphic (odd) 3/2 
superdifferentials is also easily worked out and we have
\be
\Phi _{IJ}  = 
-{i \over 2} \biggl ( \hat \omega _I \D_+ \hat \omega _J + 
\hat \omega _J \D_+ \hat \omega _I \biggr )
\ee
and the components are given by 
\bea
i \Phi _{IJ0} &=&  \half (\hat \omega _{I0} \omega _J + \hat \omega _{J0}
\omega _I) 
\nonumber \\
i\Phi _{IJ+} &=& \omega _I \omega _J + \omega _I \pz \lambda _J +
\omega _J \pz  \lambda _I - \half \hat \omega _{I0}\pz \hat \omega _{J0} -
\half \hat \omega _{J0} \pz \hat \omega _{I0}
\eea
where $\omega _I = \omega _I (\hat \Omega , \chi =0)$.

\subsection{Slice $\hat\mu$ Independence: Diffeomorphism Invariance}

The change of metric $\hat g \to g$ associated with the change of moduli
$\hat \Omega _{IJ} \to \Omega _{IJ}$ (at fixed choice of $\chi _\alpha$)
is governed by the Beltrami differential $\hat \mu $ defined by
(\ref{bd}), namely
\be
\hat \mu (z) = \hat \mu _{\bar z} {}^z = \half \hat g_{z \bar z} \bigl (
g^{zz} - \hat g ^{zz} \bigr )\, .
\ee
The Beltrami differential $\mu _\alpha$, which is part of the super
Beltrami differential  $H_\alpha = \bar \theta (\mu _\alpha - \theta \chi
_\alpha)$, is related to $\hat \mu$ by $\mu _\alpha = \p \hat \mu /\p
\zeta ^\alpha$, as explained in (\ref{mualpha}). While the conformal
classes of $\hat \mu$ and
$\mu _\alpha$ are determined by the associated moduli deformation $\Omega
_{IJ} - \hat \Omega _{IJ}$, the representative in the class is not
determined and depends upon the choices of the metrics $g$ and $\hat g$.
A change of representative is given by a diffeomorphism vector field
$v^z$, 
\bea
\delta _v (\hat \mu ) _{\bar z}{}^z  & = & \p _{\bar z} v^z
\nonumber \\
\delta _v (\mu _\alpha) _{\bar z}{}^z  & = &  \p _{\bar z} v_\alpha ^z 
\qquad \qquad 
v_\alpha ^z = {\p v^z \over \p \zeta
^\alpha}\, .
\eea
Since $\hat \mu$ is of order $\zeta \zeta$, consistency requires that
$v^z$ itself be also of order $\zeta \zeta$. 

\medskip

In this subsection, we shall show that the superstring measure $\A
[\delta]$, expressed in terms of the moduli $\hat \Omega _{IJ}$, as given
in (\ref{compamp}) is independent of the slice $\hat \mu$ provided $\hat
\mu$ and $\mu _\alpha$ transform as above and $\chi _\alpha $ is kept
fixed. Since $v$ is already of order $\zeta \zeta$, the supercurrent
correlator term is invariant by itself. Thus, the variation of ${\cal
A}[\delta]$ under a change in slice by $\delta _v$ reduces to
\be
\delta _v \ln {\cal A} [\delta]
=
 {1\over 2 \pi} \int \! d^2 z \p _{\bar z} v^z \< T(z)\>
-\delta _v \ln \det \Phi _{IJ} (p_a) - \delta _v \ln \det \<H_\alpha
|\Phi ^* _\beta\> 
\ee
and we now compute each of these terms in turn.

\subsubsection{Rank 2 Differential Contribution}

The variation under $\delta _v$ of $\Phi _{IJ+}$ is deduced from the
variation of $\hat \omega _I$.  We express $\hat \omega _I$ in terms of
the form $\theta \omega _I (\hat \Omega)$ at the period matrix $\hat
\Omega _{IJ}$ plus an exact form as follows
\bea
\hat \omega _I &=& \theta \omega _I(\hat \Omega) + \D_+ \Lambda _I
\nonumber \\
\Lambda _I & = & \lambda _I + \theta \hat \omega _{I0}\, .
\eea
Since $v^z$ is of order $\zeta \zeta$, the transformation properties are
simple
\be
\delta _v \omega _I (\hat \Omega ) = \delta _v \hat \omega _{I0} =0
\qquad \qquad 
\delta _v \lambda _I= - v^z \omega _I
\ee
Using the expression $2i \Phi _{IJ} = \hat \omega _I \D_+ \hat \omega _J + 
\hat 
\omega _J \D_+
\hat \omega _I$, 
the +  component is readily evaluated
\be
i \Phi _{IJ+} = \omega _I(\hat \Omega ) \omega _J(\hat \Omega)
+\omega _I (\hat \Omega ) \pz \lambda _J + \omega _J (\hat \Omega ) \pz
\lambda _I - \half (\hat \omega _{I0} \pz \hat \omega _{J0} + \hat \omega
_{J0} \pz \hat \omega _{I0} )
\ee
and its variation is given by
\be
\label{varPhi}
\delta _v \Phi _{IJ+} = -v^z \pz \Phi _{IJ+} - 2 (\pz v^z) \Phi _{IJ+}
\ee
as would be expected for a two-form under a diffeomorphism. 
The variation of the logarithm of the determinant is given by
\be
\delta _v \ln \det \Phi _{IJ+} (p_a)
=
\sum _b \biggl (
- v^{p_b} \p _{p_b} \ln \det \Phi _{IJ+} (p_a) 
-2 (\p _{p_b} v^{p_b})  \biggr )\, .
\ee
We clarify the derivation of this expression. Since $v$ is already of
order $\zeta \zeta$, $\Phi _{IJ+}$ on the rhs of (\ref{varPhi}) reduces
to  $\Phi _{IJ+}(z) = -i \omega _I \omega _J (z)$, and is thus effectively
a holomorphic 2-form. Consider the following ratio of $3 \times 3$
determinants,
\be
{\det \Phi _{IJ+} (p_a[w;b]) \over \det \Phi _{IJ+}(p_a)}
=
\phi ^{(2)*} _b(w)
\ee
where $p_a[w;b]= p_a$ when $a\not=b$ and $p_a[w;a]=w$. The form $\phi
^{(2)*}_b(w)$ is clearly a holomorphic 2-form in $w$, and satisfies
$\phi ^{(2)*} _b(p_a) =\delta _{ab}$, so it is the normalized holomorphic
2-form introduced in (\ref{diffexp}). We thus obtain our final expression
\be
\label{fin1}
-\delta _v \ln \det \Phi _{IJ+} (p_a)
=
\sum _b \biggl (
 v^{p_b} \p  \phi ^{(2)*}_b( p_b) 
+ 2 (\p  v^{p_b}) (p_b)  \biggr )  \, .
\ee

\subsubsection{Rank 3/2 Differential Contribution}

Expressing the inner product $\<H_\alpha |\Phi ^* _\beta \>$ in
components, we have 
\be
-\<H_\alpha |\Phi ^* _\beta \> = \< \mu _\alpha | \Phi ^* _{\beta +}\>
+\< \chi _\alpha | \Phi _{\beta 0}^*\>\, .
\ee
Its variation under $v$ is given by
\be
-\delta _v \<H_\alpha |\Phi ^* _\beta \> = \< \p _{\bar z} v^z _\alpha |
\Phi ^* _{\beta +}\> +\< \chi _\alpha | \delta _v \Phi _{\beta 0}^*\>\, .
\ee
The first term is easily computed using the differential equation
(\ref{superdiffeq}) with $n=3/2$ for $\Phi ^*_{\beta +}$ and the fact that
$v^z$ is of order $\zeta \zeta$,
\be
\< \p _{\bar z} v^z _\alpha | \Phi ^* _{\beta +}\>
=
\int \! d^2 z v^z _\alpha  \biggl ( \half \chiz \pz \psi _\beta ^*
+{3 \over 2} (\pz \chiz) \psi ^*_\beta \biggr )
\ee
The second term must be computed with some care and we use the variational
formulas (\ref{vardiff}) for holomorphic forms\footnote{The $\chi$
- dependent corrections in $\Phi ^*_{\beta 0}$ are immaterial since  $v$
is already of order $\zeta \zeta$.}
\be
\delta _v \Phi ^* _{\beta 0}(z) = \delta _v \psi ^* _\beta (z)
=
{1 \over 2 \pi} \int \! d^2w \ \p _{\bar w} v^w \ \delta _{ww} \psi
^*_\beta (z)\, ,
\ee
with the variations given by
\be
\delta _{ww} \psi ^* _\beta (z) = {3 \over 2} \pw G_{3/2} (z,w) \psi
^*_\beta (w) + \half G_{3/2} (z,w) \pw \psi ^*_\beta (w)\, .
\ee
Taking the $\p _{\bar w}$ derivative 
\bea
\p _{\bar w} \delta _{ww} \psi ^* _\beta (z) 
&=&
{3 \over 2} \p _w \biggl ( -2 \pi \delta (z,w) + 2\pi \sum _\alpha 
\psi ^*_\alpha (z) \delta (w,q_\alpha) \biggr ) \psi _\beta ^* (w)
\nonumber \\
&& + \half \biggl ( -2 \pi \delta (z,w) + 2\pi \sum _\alpha 
\psi ^*_\alpha (z) \delta (w,q_\alpha) \biggr ) \p _w \psi ^* _\beta
(w)\, ,
\eea
integrating by part and regrouping terms, we obtain
\bea
\delta _v \Phi ^* _{\beta 0}(z) = \delta _v \psi ^* _\beta (z)
&=& -v^z \pz \psi ^*_\beta (z) - {3 \over 2} (\pz v^z) \psi ^*_\beta (z)
\nonumber \\
&& + \sum _\alpha \psi ^*_\alpha (z) \biggl (
v^{q_\alpha} \p _{q_\alpha} \psi ^*_\beta (q_\alpha) + {3 \over 2} (\p
_{q_\alpha } v^{q_\alpha} ) \psi ^* _\beta (q_\alpha) \biggr )\, .
\eea
This expression should have been expected : it states that
$\psi ^*_\beta(z)$ transforms as a form of rank 3/2 in $z$, $-3/2$ in
$q_\beta$ and of rank 0 in the remaining points $q_\alpha $, $\alpha
\not=\beta$.

\medskip

Assembling all contributions, we have 
\bea
\delta _v \<H_\alpha |\Phi ^* _\beta \> 
&=&
-\int \! d^2z v^z _\alpha \biggl ( \half \chiz \pz \psi _\beta ^*
+{3 \over 2} (\pz \chiz) \psi ^*_\beta \biggr )
+ \int \! d^2z \chi _\alpha \biggl (v^z \pz \psi ^*_\beta  + {3
\over 2} (\pz v^z) \psi ^*_\beta \biggr )
\nonumber \\
&&- \sum _\gamma \< \chi _\alpha | \psi ^* _\gamma\>
\biggl (
v^{q_\gamma } \p _{q_\gamma } \psi ^*_\beta (q_\gamma) + {3 \over 2} (\p
_{q_\gamma} v^{q_\gamma} ) \psi ^* _\beta (q_\gamma) \biggr )\, .
\eea
The first two terms on the rhs will cancel upon using the fact that $v^z$
is of order $\zeta \zeta$, a fact that allows us to introduce a unique
$\zeta$-independent quantity $\bar v^z$, such that 
\be
v^z \equiv \zeta ^1 \zeta ^2 \ \bar v ^z
\qquad \Rightarrow  \qquad 
v^z_\alpha = {\p v^z \over \p \zeta ^\alpha} = \epsilon _{\alpha \beta } \zeta 
^\beta
\bar v ^z\, , \qquad \epsilon _{12} =1\, .
\ee
Indeed, pulling out their $\zeta $-dependence, the first two terms
become
\bea
- \epsilon _{\alpha \gamma} \zeta ^\gamma \zeta ^\delta 
\int \! d^2z \ \bar v^z 
\biggl ( \half \chi _\delta  \pz \psi _\beta ^* 
+{3 \over 2} (\pz \chi _\delta) \psi ^*_\beta \biggr ) 
+ \zeta ^1 \zeta ^2  \int \! d^2z \ \chi _\alpha
\biggl (\bar v^z \pz \psi ^*_\beta  + {3 \over 2} (\pz \bar v^z) \psi
^*_\beta \biggr )
\nonumber
\eea
Using the elementary relation $-\epsilon _{\alpha \gamma} \zeta ^\gamma
\zeta ^\delta = \delta _\alpha {} ^\delta \zeta ^1 \zeta ^2$, and
integrating by parts in $z$ so as to leave $\chi _\alpha$ without
derivatives acting, the above two terms cancel as announced,
\bea
 \zeta ^1 \zeta ^2
\int \! d^2z  \biggl ( \half \bar v^z \chi _\alpha  \pz \psi _\beta ^* 
-{3 \over 2} \chi _\alpha \pz( \bar v^z \psi ^*_\beta ) 
+ \chi _\alpha \bar v^z \pz \psi ^*_\beta  
+ {3 \over 2} \chi _\alpha (\pz \bar v^z) \psi ^*_\beta \biggr ) 
=0
\nonumber
\eea
The remaining change is given by
\bea
\label{variation}
\delta _v \<H_\alpha |\Phi ^* _\beta \> 
&=& - \sum _\gamma \< \chi _\alpha | \psi ^* _\gamma\> \biggl (
v^{q_\gamma } \p _{q_\gamma } \psi ^*_\beta (q_\gamma) + {3 \over 2} (\p
_{q_\gamma} v^{q_\gamma} ) \psi ^* _\beta (q_\gamma) \biggr )
\eea
The change in the determinant is computed with the standard formula
\be
\delta _v \ln \det \< H_\alpha | \Phi ^*_\beta\>
= \tr \biggl ( \< H_\alpha | \Phi ^*_\beta\>)^{-1} \delta _v \< H_\alpha |
\Phi ^*_\beta\> \biggr )\, .
\ee
Using the fact that the variation $\delta _v \< H_\alpha |
\Phi ^*_\beta\>$ is already of order $\zeta \zeta$, we see that the
inverse matrix $\< H_\alpha | \Phi ^*_\beta\>)^{-1}$ may be taken at
$\zeta =0$ and thus reduces $(-\<\chi _\alpha | \psi ^*_\beta\> )^{-1}$.
But this factor now cancels the matrix $\<\chi _\alpha | \psi ^*_\beta\>$
in (\ref{variation}) and we are left with
\be
\label{fin2}
- \delta _v \ln \det \< H_\alpha | \Phi ^*_\beta\>
= - \sum _\beta \biggl ( v^{q_\beta  } \p \psi ^*_\beta (q_\beta) + {3
\over 2} (\p v^{q_\beta} ) (q_\beta) \biggr )
\ee

\subsubsection{Stress Tensor Contribution}

The stress tensor correlator was computed in (\ref{stresstotal}) and is
given by
\bea
T(w) 
&= &
-15 T_1(w) + \half f_2 (w)^2 - {3 \over 2}\p_w f_2 (w) 
- \half f_{3/2}(w)^2 + \p_w f_{3/2}(w)
 \\
&& +\half \omega _I\omega _J(w) \biggl ( 
5\p _I \p _J \ln \tet [\delta ](0) - \p _I \p _J \ln \tet
[\delta ](D_\beta ) +   \p _I \p _J \ln \tet [\delta ](D_b ) \biggr ) 
\nonumber
\eea
where all ingredients in the above formula were discussed in Section
4.2.3. The singularities of $T(w)$ are derived from the knowledge of
the singularities of $f_n(w)$, and the part of the expansion needed
here is given by
\bea
f_n(w) & = & {1 \over w-z_i}
+  \p \phi _i ^{(n)*} (z_i) + \O (w-z_i) 
\nonumber \\
\p \phi _i ^{(n)*} (z_i)  &=&
 \omega _I(z_i) \p _I
\ln \tet [\delta ](D_n) +\p_{z_i} \ln \bigl ( \sigma (z_i)^Q \prod
_{j\not= i} E(z_i,z_j)
\bigr ) 
\eea
plus terms holomorphic at $w\sim z_i$. Here, $\phi _i ^{(n)*} (w)$ are the
holomorphic $n$-forms normalized on the points $z_i$ as usual
$\phi _i ^{(n)*} (z_j) = \delta _i ^j$.
The singular terms in $T(w)$ are now readily identified
\be
T(w) = 
\sum _a  \biggl ( {2 \over (w-p_a)^2} + {\p \phi ^{(2)*}_a (p_a) \over
w-p_a} \biggr )
-
\sum _\alpha   \biggl ( {3/2 \over (w-q_\alpha)^2} + {\p \phi
^{(3/2)*}_\alpha (q_\alpha ) \over w-q_\alpha} \biggr ) + {\rm reg.}
\ee
As a result, and using the previous notation $\psi _\alpha ^*(w) = \phi
^{(3/2)*}_\alpha (w)$, we have 
\bea
\label{fin3}
{1\over 2 \pi} \int \! d^2 z \p _{\bar z} v^z \< T(z)\>
&=& 
\sum _\beta \biggl ( v^{q_\beta  } \p  \psi ^*_\beta (q_\beta) 
+ {3 \over 2} (\p v^{q_\beta})  (q_\beta) \biggr )
\nonumber \\
&&
- \sum _b \biggl ( v^{p_b} \p \phi ^{(2)*}_b(p_b) 
+ 2 (\p  v^{p_b}) (p_b) \biggr ) \, .
\eea
Assembling the partial results (\ref{fin1}), (\ref{fin2})
and (\ref{fin3}), we see that $\delta _v {\cal A} [\delta] =0$, and thus
the chiral measure is invariant under infinitesimal changes of $\hat \mu$
- slice.

\subsection{Slice $\chi _\alpha$ Independence  : Worldsheet supersymmetry}

Local supersymmetries act as follows
\be
\label{susytfon}
\delta _\xi \chiz = -2 \p_{\bar z} \xi ^+
\qquad \qquad 
\delta _\xi \hat \mu _{\bar z} {}^z  = \xi ^+ \chiz
\ee
where $\hat \mu$ is the shift in metric accompanying the shift in complex
structure from $\Omega $ to $\hat \Omega $. As this supersymmetry should
correspond to a change in $\chi _\alpha$ - slice, the supersymmetry
parameter $\xi ^+$ should be viewed as being of order $\zeta$. It is
convenient to separate the $\xi$-variations of $\ln \A [\delta]$ into
those arising from the correlators $\ln \A _{{\rm corr}}[\delta]$ and
those from the rank 2 and rank 3/2 differentials.

\subsubsection{Correlator contributions}

The supersymmetry transformation of the correlator terms is given by 
\bea
\delta _\xi \ln {\cal A}_{{\rm corr}} [\delta]
=
 {1 \over 2 \pi ^2} \int \! d^2z \p_{\bar z} \xi ^+
\int \! d^2w \chiw
\< S(z) S(w)  \>
+{1 \over 2 \pi} \int \! d^2 z  \xi ^+ \chiz
\< T(z)  \>\, .
\eea
The pole at $z=w$ in the $S(z)S(w)$ correlator is precisely cancelled by
the corresponding contribution from the stress tensor term, using the fact
that
\be
\p_{\bar z} S(z) S(w) = 2 \pi \delta (z,w) \half T(z) + {\rm \
other \ than \ }z=w
\ee
and an integration by parts in $\p _{\bar z}$. The remaining poles now
arise only from the ghost contributions, given through
(\ref{ghostcorrelator}),
\be
\delta _\xi \ln {\cal A}_{{\rm corr}} [\delta]
= -{1 \over 2 \pi ^2} \int \! d^2z \ \xi ^+ (z)
\int \! d^2w \ \chiw \ \p _{\bar z} C_{gh}(z,w) \bigg |_{z\not= w}
\ee
The calculation of $\p_{\bar z} C_{gh}(z,w)$ is simplified by the fact
that we are instructed to ignore the $z=w$ pole, so that effectively
$\p_{\bar z} G_2(z,w)=\p_{\bar z} G_{3/2}(z,w)=0$ and 
\bea
\p_{\bar z} G_{3/2} (w,z) 
&=& 
2 \pi \sum _\alpha \delta (z,q_\alpha) \phi ^{(3/2)*}_\alpha (w)
\nonumber \\
\p_{\bar z} G_2 (w,z) 
&=& 
2 \pi \sum _a \delta (z,p_a) \phi ^{(2)*}_a (w)
\eea
Thus, we find
\bea
4\p_{\bar z} C_{gh}(z,w) \big |_{z\not=w}
& = &
+3 \p_w G_2(z,w) \p_{\bar z}G_{3/2}(w,z) +2 G_2 (z,w) \p_{\bar z} \p_w
G_{3/2}(w,z) 
\nonumber \\
&& -3 \p_{\bar z} \p_z G_2(w,z) G_{3/2}(z,w) -2 \p _{\bar z} G_2
(w,z) \p_z G_{3/2}(z,w) \qquad
\eea
and using the $\pz G$ formulas above, we have
\bea
4\p_{\bar z} C_{gh}(z,w) \big |_{z\not=w}
&=&
- 2\pi \sum _a \phi _a ^{(2)*} (w) \biggl (3\pz \delta (z,p_a)
G_{3/2}(z,w) + 2 \delta (z,p_a) \pz G_{3/2}(z,w)  \biggr )
\nonumber \\
&&  + 2\pi \sum _\alpha \delta (z,q_\alpha) \biggl ( 3\phi _\alpha
^{(3/2)*} (w) \pw G_2 (z,w) +2 \pw \phi _\alpha ^{(3/2)*} (w) G_2(z,w)
\biggr )
\nonumber
\eea
Substituting this result into $\delta _\xi \ln {\cal A}_{{\rm corr}}
[\delta]$, we find first for the $\alpha $-sum that
\bea
&&-{1 \over 2 \pi}
\sum _\alpha \xi ^+(q_\alpha) 
\int d^2w \chiw \biggl (G_2(q_\alpha,w) \pw \phi _\alpha ^{(3/2)*} (w)
+{3\over 2} \pw G_2 (q_\alpha , w) \phi _\alpha ^{(3/2)*}(w) \biggr )
\nonumber \\
&&= 
{1 \over 2 \pi}
\sum _\alpha \xi ^+(q_\alpha) 
\int d^2w G_2(q_\alpha ,w)  \biggl (\half  \pw \phi _\alpha
^{(3/2)*} (w) \chiw +{3\over 2} \pw \chiw \phi _\alpha ^{(3/2)*}(w)
\biggr )
\nonumber
\eea
We now recall the definitions of the components of the (even)
superholomorphic 3/2 forms (\ref{evenodd}) and recognize that these are
precisely the combinations that occur here. Putting all together, we have 
\be
\label{correlatorvar}
\delta _\xi \ln {\cal A}_{{\rm corr}} = 
-  \sum _\alpha \xi ^+(q_\alpha) \Phi _{\alpha +} ^* (q_\alpha) 
+  \sum _a \biggl (\xi ^+ (p_a) \p \Phi _{a0} ^* (p_a) 
+ 3 (\p \xi ^+)(p_a) \Phi _{a0} ^* (p_a) \biggr )
\ee
a form that is suggestive of the effect of supersymmetry Ward identities.

\subsubsection{Rank 2 Differentials Contribution}

To evaluate $\delta _\xi \ln \det \Phi _{IJ+}(p_a)$, we compute the
transformation properties of $\Phi_{IJ+}$, 
\be
i\Phi _{IJ+} = \omega _I \omega _J + \omega _I \pz \lambda _J + \omega
_J \pz \lambda _I - \half \hat \omega _{I0} \pz \hat \omega _{J0} -
\half \hat \omega _{J0} \pz \hat \omega _{I0}\, ,
\ee 
where the transformation laws of each of the ingredients is given by
\bea
\delta _\xi \omega _I (\hat \Omega) &= & 0
\nonumber \\
\delta _\xi \hat \omega _{I0} (z)  &= & \xi ^+(z) \omega _I(z)
\nonumber \\
\delta _\xi \lambda _I(z)  & = & \xi ^+(z) \hat \omega _{I0} (z)\, .
\eea
Collecting all terms and using the fact that $\xi$ is of order $\zeta$,
the total transformation law may be recast in the following form,
\bea
\delta _\xi \Phi _{IJ+}(z)
=
\xi ^+ (z) \pz \Phi _{IJ0}(z) + 3 (\pz \xi^+) \Phi _{IJ0}(z)\, .
\eea 
The variation of the determinant is computed as follows
\be
\delta _\xi \ln \det \Phi _{IJ+}(p_a)
=
\sum _b \biggl (\xi ^+ \pw F_b(w) + 3 (\pw \xi ^+) F_b(w) \biggr )
_{w=p_b}
\ee
where the quantity $F_b(w)$ is defined as follows
\be
F_b(w) = {\det \Phi _{IJ} (p_a[w,b]) \over \det \Phi _{IJ+}(p_a)}
\qquad \qquad
\Phi _{IJ}(p_a[w,b]) = 
\left \{ \matrix{\Phi _{IJ+}(p_a) & a\not= b\cr
       \Phi _{IJ0}(w)   & a=b \cr} \right .
\ee
It remains to compute $F_b(w)$. In view of the differential equation
(\ref{superdiffeq}), needed here for $n=3/2$, 
\be
\p_{\bar w} \Phi _{IJ0}(w) = {i \over 2} \chiw \omega _I \omega _J (w)\, ,
\ee
the 3/2 form $F_b(w)$ obeys a simple differential equation 
\bea
\p _{\bar w} F_b(w) &=& -\half \chiw  \phi ^{(2)*} _b (w)
\nonumber \\
\phi ^{(2)*} _b (w) &=& {\det \omega _I \omega _J(p_a[w,b])
\over \det \omega _I\omega _J(p_a)}
\qquad \qquad
p_a[w,b] = \left \{ \matrix{p_a & a\not= b\cr
       w  & a=b \cr} \right .
\eea
The form $\phi ^{(2)*}_b(w)$ is readily recognized to be holomorphic
of rank 2 in $w$ and to be normalized so that $\phi ^{(2)*}_b(p_a) =
\delta _{ab}$. Hence, $F_b(w)$ obeys the same differential equation as
the quantity $\Phi  ^* _{b0}(w)$, and must thus differ from it only by a
holomorphic 3/2 form~:
\be
F_p (w) =  \Phi ^* _{b0}(w) +\sum _\alpha \psi ^*_\alpha (w) R^\alpha {}_b 
\,  .
\ee
The coefficients $R ^\alpha {}_b$ are independent of $w$, first order in
$\zeta$ and may be determined by evaluating the equation at $w=q_\alpha$,
using the fact that $\Phi _{b0} ^* (q_\alpha)=0$, so that 
\be
R ^\alpha {}_b
= F_b(q_\alpha) 
= {\det \Phi _{IJ}(p_a[q_\alpha ,b]) \over \det \Phi _{IJ+}(p_a)}\, .
\ee
In terms of the matrices 
\be
\label{mandn}
M ^a {}_{IJ} \equiv \Phi _{IJ+}(p_a)
\qquad \qquad 
N^\alpha {}_{IJ} \equiv \Phi _{IJ0}(q_\alpha)\, ,
\ee
we simply have $R ^\alpha {}_b = (NM^{-1})^\alpha {}_b$.
Assembling the entire contribution, we have 
\bea
\label{twodeterminant}
- \delta _\xi \ln \det \Phi _{IJ+}(p_a)
&=&
- \sum _b \biggl (\xi ^+(p_b) \p \Phi ^*_{b0} (p_b) + 3 (\p \xi ^+) (p_b)
\Phi ^*_{b0}(p_b) \biggr )
 \\
&& -\sum _{b , \alpha} \biggl ( 
\xi ^+(p_b) \p \psi ^*_\alpha (p_b) + 3 (\p \xi ^+)(p_b)
\psi ^* _\alpha  (p_b) \biggr ) (NM^{-1})^\alpha {}_b
\nonumber
\eea
We shall not need to make the form of the matrices $M$ and $N$ explicit
because their contribution will be cancelled later on by expressions
manifestly of the same form.

\subsubsection{Supersymmetry variations of $H_\alpha$}

The supersymmetry variation $\delta _\xi H_\alpha$ will be needed for
the calculations of the $\xi$-variation of the finite-dimensional
determinant involving $H_\alpha$. The calculation of this variation is
subtle and interesting. Furthermore, the variation evaluated with
respect to the metric $\hat g$ will present further delicate new features,
the thorough understanding of which will be crucial later on. Therefore,
we devote this subsection to these issues before working out the
$\xi$-variation of the rank 3/2 differentials.

\medskip

The starting point is the supersymmetry transformations of
(\ref{susytfon}) which, by construction, are written with respect to the
metric $g$. It is assumed that the metric $\hat g$ is independent of
$\zeta ^\alpha$ and that it transforms under $\xi$ at most by a
diffeormorphism (whose effects have been shown to be immaterial in the
preceding section and may thus be safely dropped) and that derivation with
respect to coordinates like $\zeta^\alpha$ commutes with $\delta _\xi$,
\be
\label{commute}
{\p \over \p \zeta ^\alpha} \delta _\xi = \delta _\xi {\p \over \p \zeta
^\alpha} \, .
\ee
>From these facts, we shall now prove the following transformation
formulas,
\bea
\delta _\xi \chi _\alpha
& = &
- 2 \p _{\bar z} \xi ^+ _\alpha - 2 \mu _\alpha \p _z \xi ^+
+(\p_z \mu _\alpha) \xi ^+
\nonumber \\
\delta _\xi \mu _\alpha 
& = &
\xi ^+_\alpha \ \chiz  - \xi ^+ \ \chi _\alpha   \, .
\eea
for the variation under a local susy $\xi $ of the slice
function $\chi _\alpha$, defined by
\bea
\label{zetaderivatives}
\chi _\alpha = {\p \chi ^+ _{\bar z} \over \p \zeta ^\alpha}
\hskip 1in
\mu _\alpha = {\p \hat \mu \over \p \zeta ^\alpha}
\hskip 1in
\xi ^+_\alpha = {\p \xi ^+ \over \p \zeta ^\alpha}
\eea
To do this, start from (\ref{susytfon}) and differentiate with respect to
$\zeta ^\alpha$, using (\ref{commute}),
\bea
\label{Belvariations}
\delta _\xi \chi _\alpha 
& = &
{\p \over \p \zeta ^\alpha} \delta _\xi \chiz 
=
 {\p \over \p \zeta ^\alpha} \biggl (-2 \nabla _{\bar z} ^{(-1/2)} \xi ^+
\biggr )
=
-2 \p_{\bar z} \xi ^+ _\alpha 
- 2 \biggl ( {\p \over \p \zeta ^\alpha} \nabla ^{(-1/2)} _{\bar z}
\biggr ) \xi ^+
\nonumber \\
\delta _\xi \mu _\alpha 
& = &
{\p \over \p \zeta ^\alpha} \delta _\xi \hat \mu \hskip .175in
=
 {\p \over \p \zeta ^\alpha} \biggl (\xi ^+ \chiz \biggr ) \hskip .44in
=
+ \xi ^+_\alpha \ \chiz  - \xi ^+ \ \chi _\alpha 
\eea
The second line thus establishes the desired formula for the variation of
$\mu _\alpha$. The formula for $\delta _\xi \chi_\alpha$ needs more work
and involves interesting subtleties. First, the operator $\nabla ^{(-1/2)}
_{\bar z}$ depends on $\zeta^\alpha$ through its dependence on the metric
$g$ which itself depends on $\zeta^\alpha$. (This is in contrast with the
metric $\hat g$ which is $\zeta^\alpha$-independent.) Second, the
$g$-dependence of $\nabla ^{(-1/2)} _{\bar z}$ is known and may be
expressed in terms of the differential $\hat \mu$ as follows,
\bea
\label{derdeformation}
\nabla ^{(-1/2)} _{\bar z} = \hat \nabla ^{(-1/2)} _{\bar z} +\hat \mu
\p_z - \half \p_z \hat \mu
\eea
where $\hat \nabla$ is now evaluated with respect to the
metric $\hat g$. Differentiating with respect to $\zeta ^\alpha$ and
using the second equation in (\ref{zetaderivatives}), we find
\bea
{\p \over \p \zeta ^\alpha} \nabla ^{(-1/2)} _{\bar z} =   \mu _\alpha
\p_z - \half \p_z  \mu _\alpha
\eea
Combining this result with (\ref{Belvariations}), we recover the proposed
formula for $\delta _\xi \mu _\alpha$.

\medskip

We now seek to re-express the transformation laws (\ref{Belvariations})
with respect to the metric $\hat g$ instead of $g$, as written in
(\ref{Belvariations}). The entire transformation law of $\mu _\alpha$ is
of order $\zeta$ already and thus a change from $g$ to $\hat g$ is
immaterial. Next we deal with the case of $\delta _\xi \chi _\alpha$.
>From the definitions of $\mu _\alpha$ and the fact that $\hat \mu$ is of
order $\zeta \zeta$, we have 
\bea
\mu _\alpha = \epsilon _{\alpha \beta } \zeta ^\beta \bar \mu\, ,
\qquad
\hat \mu = \zeta ^1 \zeta ^2 \bar \mu\, ,
\hskip 1in
\xi ^+ = \zeta ^\alpha \xi ^+ _\alpha\, .
\eea
We now perform a Fierz type rearrangement on the products $\mu _\alpha
\xi ^+$ occurring in (\ref{Belvariations}). Omitting irrelevant
derivatives, and using $\zeta ^\beta \zeta ^\gamma = \epsilon ^{\beta 
\gamma} \zeta ^1 \zeta ^2$, we have 
\bea
\mu _\alpha \xi ^+ = \epsilon _{\alpha \beta} \zeta ^\beta \bar \mu \zeta
^\gamma \xi ^+ _\gamma = - \hat \mu \xi ^+ _\alpha
\eea
The expression for the variation becomes (restoring now
derivatives either on $\mu _\alpha$ or on $\xi^+$)
\bea
\delta _\xi \chi _\alpha
=
-2 \p_{\bar z} \xi ^+_\alpha + 2 \hat \mu \p_z \xi ^+ _\alpha
- (\p_z \hat \mu) \xi ^+ _\alpha
\eea
Using the fact that $ g = \hat g + \hat \mu$, equation
(\ref{derdeformation}), and expressing all quantities with respect to
$\hat g$, we recognize that a remarkable simplification occurs in the
transformation rule, 
\bea
\delta _\xi \chi _\alpha = - 2 \hat \p _{\bar z} \xi ^+ _\alpha
\eea

\subsubsection{Rank 3/2 Differential Contribution}

To evaluate $\delta _\xi \ln \det \< H_\alpha | \Phi ^* _\beta \>$,
we start again from the component expression 
\be
-\< H_\alpha | \Phi ^* _\beta\>
=\< \mu _\alpha |\Phi _{\beta +}^* \> 
+ \< \chi _\alpha |\Phi _{\beta 0}^*\>
\ee
and deduce its variation
\be
-\delta _\xi \< H_\alpha | \Phi ^* _\beta\>
=\< \delta _\xi \mu _\alpha |\Phi _{\beta +}^* \> 
+ \< \delta _\xi \chi _\alpha |\Phi _{\beta 0}^*\>
+\< \mu _\alpha |\delta _\xi \Phi _{\beta +}^* \> 
+ \< \chi _\alpha |\delta _\xi \Phi _{\beta 0}^*\>\, .
\ee
We shall now evaluate each of these inner products in turn. We begin by
computing the variations of the components of $\Phi ^* _\alpha$.

\medskip

The variation of the components $\Phi ^*_{\beta+}$ of the holomorphic
superdifferentials is
\bea
\delta _\xi \Phi ^*_{\beta +} (z)
&=&
\xi ^+(z) \p \Phi ^*_{\beta 0}(z) + 3 (\p \xi ^+)(z) \Phi ^*_{\beta 0}(z)
\nonumber \\
&& -
\sum _a \biggl ( \xi ^+(p_a) \p \Phi ^*_{\beta 0} (p_a)
+3 \p \xi ^+(p_a) \Phi ^*_{\beta 0}(p_a) \biggr ) \Phi ^*_{a+}(z)\, ,
\eea
where no variation of the metric is required since the differential is
already of order $\zeta$. 

\medskip

The variation of the component $\Phi ^*_{\beta 0}$ contains a contribution
due to the variation of the metric for its leading term which is of order
zero in $\zeta$. We have
\bea
\label{deltaPhi}
\delta _\xi \Phi ^*_{\beta 0} (z) 
&=&
\delta _\xi \psi ^* _\beta (z)
-{1 \over 2 \pi} \int \! d^2w G_{3/2}(z,w) \half (\delta _\xi \chiw)
\Phi ^*_{\beta +}(w)
\nonumber \\
&&\qquad \qquad \!
-{1 \over 2 \pi} \int \! d^2w G_{3/2}(z,w) \half  \chiw
(\delta _\xi \Phi ^*_{\beta +}(w))\, ,
\eea
where the first term on the rhs of (\ref{deltaPhi}), due to the variation
of the metric, is given by
\be
\delta _\xi \psi ^* _\beta (z)
=
{1 \over 2 \pi} 
\int \! d^2w \xi ^+(w) \chiw \biggl (
{3 \over 2} \pw G_{3/2}(z,w) \psi ^*_\beta (w) +\half G_{3/2}(z,w) \pw
\psi ^*_\beta (w) \biggr )\, .
\ee
The second term on the rhs of (\ref{deltaPhi}) may be worked out more
explicitly and is given by
\bea
&& \xi ^+(z) \Phi ^*_{\beta +}(z) 
-\sum _\alpha \xi ^+(q_\alpha) 
\Phi ^*_{\beta +}(q_\alpha) \psi _\alpha ^*(z) 
\nonumber \\
&&\qquad 
 +{1 \over 2 \pi} \int \! d^2w G_{3/2}(z,w) \xi ^+(w) \biggl (
\half \chiw \pw \Phi ^*_{\beta 0}(w) +{3 \over 2} (\pw \chiw) \Phi
^*_{\beta 0}(w) \biggr )\, ,
\nonumber
\eea
while the third term on the rhs of (\ref{deltaPhi}) becomes
\bea
&& -{1 \over 2 \pi} \int \! d^2w G_{3/2}(z,w) \half \chiw \biggl (
\xi ^+(w) \pw \psi ^*_\beta (w) + 3 \pw \xi ^+(w) \psi ^*_\beta (w)
\biggr )
\nonumber \\
&& \qquad 
+ \sum _a \biggl (
\xi ^+(p_a) \p \psi ^*_\beta (p_a) + 3 \p \xi ^+(p_a) \psi ^* _\beta
(p_a) \biggr ) \Phi ^*_{a0}(z)\, .
\nonumber 
\eea
Combining all contributions, we have
\bea
\delta _\xi \Phi ^* _{\beta 0} (z)
&=&
\xi ^+(z) \Phi ^* _{\beta +} (z)
- \sum _\alpha \xi ^+ (q_\alpha) \Phi ^* _{\beta +} (q_\alpha) \Phi ^*
_{\alpha 0} (z) 
\nonumber \\
&& + \sum _a \biggl ( \xi ^+ (p_a) \p \Phi ^*_{\beta 0} (p_a) + 3 \p \xi ^+
(p_a)
\Phi ^* _{\beta 0} (p_a) \biggr ) \Phi ^* _{a0} (z)
\eea

\medskip

Assembling $ \< \delta _\xi \chi _\alpha |\Phi _{\beta 0}^*\>
+ \< \chi _\alpha |\delta _\xi \Phi _{\beta 0}^*\>$, all terms involving
the Green's function $G_{3/2}(z,w)$ cancel one another, while upon further
addition of the term $\< \delta _\xi \mu _\alpha |\Phi _{\beta +}^* \> $,
all terms involving the Green's function $G_2(z,w)$ also cancel one
another. One is thus left with (care is needed, while arranging these
formulas, to  the precise {\sl order} of the anti-commuting entries)
\bea
\< \delta _\xi \mu _\alpha |\Phi _{\beta +}^* \> 
&+& \< \delta _\xi \chi _\alpha |\Phi _{\beta 0}^*\>
+ \< \chi _\alpha |\delta _\xi \Phi _{\beta 0}^*\>
\nonumber \\
&=&
-\sum _\gamma \xi ^+(q_\gamma) \Phi ^* _{\beta+}(q_\gamma) \<\chi _\alpha
|\psi ^*_\gamma\>
-\int \! d^2 z \mu _\alpha  \bigl ( \xi ^+ \pz \psi ^*_\beta 
+ 3 \p \xi ^+ \psi ^* _\beta  \bigr )
\nonumber \\
&& + \sum _a  \biggl ( \xi ^+(p_a) \p \psi
_\beta ^* (p_a) + 3 \p \xi ^+(p_a)
\psi ^* _\beta (p_a) \biggr ) \< \chi _\alpha |\Phi ^* _{a0}\> \, .
\eea
On the other hand, one has
\bea
\< \mu _\alpha |\delta _\xi \Phi _{\beta +}^* \> 
&=&
\int \! d^2 z \mu _\alpha \bigl ( \xi ^+ \pz \psi ^*_\beta 
+ 3 \p \xi ^+ \psi ^* _\beta  \bigr )
\nonumber \\
&& + \sum _a  \biggl ( \xi ^+(p_a) \p \psi
_{\beta } ^*(p_a) +3 \p \xi^+(p_a) \psi ^*_{\beta } (p_a) \biggr ) \< \mu
_\alpha |\Phi ^* _{a+}\> 
\, .
\eea
Combining this with preceding contributions as well, and reassembling the
terms belonging to  $-\< H_\alpha |  \Phi ^* _\beta\>
=\< \mu _\alpha |\Phi _{\beta +}^* \>  + \< \chi _\alpha |\Phi _{\beta
0}^*\>$, we obtain a fairly simple formula
\bea
\delta _\xi \< H_\alpha | \Phi ^*_\beta \>
&=&
 - \sum _\gamma \xi ^+(q_\gamma) \Phi ^*_{\beta +}(q_\gamma) \<H_\alpha
|\Phi ^*_\gamma\> 
\nonumber \\
&& + \sum _a  \biggl ( \xi ^+(p_a) \p \psi
_{\beta } ^*(p_a) +3 \p \xi^+(p_a) \psi ^*_{\beta } (p_a) \biggr ) 
\< H _\alpha |\Phi ^* _a\>\, .
\eea
The variation of the determinant follows directly from this result and is 
given 
by
\bea
- \delta _\xi \ln \det \< H_\alpha | \Phi ^* _\beta \>
&=&
 +  \sum _\gamma \xi ^+(q_\gamma) \Phi ^*_{\gamma +}(q_\gamma) 
\\
&& - \sum _{a\alpha \beta}  \biggl ( \xi ^+(p_a) \p \psi
_{\beta } ^*(p_a) +3 \p \xi^+(p_a) \psi ^*_{\beta } (p_a) \biggr ) 
(\< H_\alpha | \Phi ^* _\beta \>)^{-1}
\< H _\alpha |\Phi ^* _a\>\, .
\nonumber 
\eea
To compare this result with the one for the other finite dimensional 
determinant, it is necessary to reformulate the last term as follows.
We use the decomposition of $\Phi _{IJ}$ onto $\Phi ^* _a$ and $\Phi ^* 
_\alpha$
\be
\Phi _{IJ} (\z) = \sum _a \Phi ^* _a (\z) M^a {}_{IJ} + \sum _\alpha \Phi
^* _\alpha  (\z) N^\alpha {}_{IJ}\, ,
\ee
and from projections onto $H^*  _a$ and $H^*  _\alpha$, we
recognize that the matrices $M$ and $N$ coincide with the ones introduced
in (\ref{mandn}). Taking now the inner product with $H_\gamma$, we find
\be
\< H_\gamma | \Phi ^* _a \>
= - \sum _\alpha \< H_\gamma | \Phi ^* _\alpha  \> (NM^{-1} )^\alpha {}_a
\ee
Using this result, we have
\bea
\label{threehalfdeterminants}
- \delta _\xi \ln \det \< H_\alpha | \Phi ^* _\beta \>
&=&
 +  \sum _\gamma \xi ^+(q_\gamma) \Phi ^*_{\gamma +}(q_\gamma) 
 \\
&& + \sum _{a \beta}  \biggl ( \xi ^+(p_a) \p \psi
_{\beta } ^*(p_a) +3 \p \xi^+(p_a) \psi ^*_{\beta } (p_a) \biggr ) 
(NM^{-1})^\beta {}_a \, .
\nonumber 
\eea

\subsubsection{Summary of slice $\chi$ independence}

Assembling all contributions (\ref{correlatorvar}),
(\ref{twodeterminant}), (\ref{threehalfdeterminants}) to
$\delta _\xi \ln \A _{{\rm corr}} [\delta]$, we easily see that they all
cancel one another, and thus the chiral string measure is completely
independent of all choices of slice.

\vfill\eject

\section{Manifestly Reparametrization Invariant Formulas}
\setcounter{equation}{0}

We have established in the previous sections the invariance of our gauge
fixed formulas under infinitesimal changes of gauge slice ${\cal S}$. In
this section, we shall obtain a last formula in which we make the
invariance under infinitesimal diffeomorphisms manifest, by eliminating
completely the dependence on the Beltrami differential $\hat \mu$.  The
key fact is that, unlike $\hat \mu$ itself, the conformal class
of $\hat \mu$ is known. Thus it suffices to show that the entire
dependence on $\hat \mu$ of the gauge fixed formulas resides in
its pairing with a holomorphic quadratic differential. We begin with a
more detailed discussion of the conformal class of $ \mu _\alpha$ in
subsection 6.1, and then work out the various contributions to the final
formula.

\subsection{The conformal class of  $\mu _\alpha$ for genus 2}

Having fixed the gravitino slice $\chi _\alpha$, the orthogonality
relation $\< H_\alpha | \Phi _{IJ}\>=0$ fixes the conformal class of
$\mu_\alpha$, though not the differential itself. To see this, we work
in WZ gauge, where $H_\alpha = \bar  \theta  (\mu _\alpha - \theta \chi
_\alpha)$, recall the expression for $\Phi _{IJ}$ of (\ref{Phis}),
\be
\Phi _{IJ} = -{i \over 2}\biggl ( \hat \omega _J \D _+ \hat \omega
_I + \hat \omega _I \D _+ \hat \omega _J \biggr )
\nonumber 
\ee
and use the expression for $\hat \omega _I = \hat \omega _{I0} + \theta
\hat \omega _{I+}$, neglecting the contribution from the auxiliray
field $A$, which cancels out, 
\be
\int \! d^2 z\,  \mu _\alpha \biggl (\hat \omega _{I+} \hat \omega _{J+} -
\hat \omega _{I0} \p _z \hat \omega _{J0} + I\leftrightarrow J \biggr )
= - \int \! d^2 z\,  \chi _\alpha  \biggl ( \hat \omega _{I0} \hat \omega
_{J+} + \hat \omega _{J0} \hat \omega _{I+} \biggr )\, .
\ee
For genus 2, both sides are odd and of order $\zeta $, so that $\mu
_\alpha$ is odd and first order in $\zeta ^\alpha$, while the term
$\hat \omega _{I0} \p _z \hat \omega _{J0}$ is of order $\zeta \zeta $
and may be dropped, leading to the following simplified form involving
the ordinary holomorphic differentials $\omega _I$,
\be
\label{mualphaeq}
2\int \! d^2 z \,  \mu _\alpha  \omega _I \omega _J 
= - \int \! d^2z \, \chi _\alpha  \biggl ( 
\hat \omega _{I0} \omega _J + \hat \omega _{J0} \omega _I \biggr ) \, .
\ee
This equation uniquely determines the conformal class of $\mu _\alpha$.

\medskip

Next, we show that this conformal class coincides with the one predicted
by the relation $\mu _\alpha = \p \hat \mu /\p \zeta ^\alpha$,
thus further reinforcing the general validity of our approach. To this
end, we compute in two different ways the derivative of $\p \Omega _{IJ}/
\p \zeta ^\alpha$, keeping $\hat \Omega _{IJ}$ fixed. The first way uses
the fact that $\hat \mu$ is the Beltrami differential that takes us from
metric $\hat g$ to $g$, 
\be
\Omega _{IJ} - \hat \Omega _{IJ} = i \int \! d^2 z\,  \hat \mu \omega _I
\omega _J
\quad \Longrightarrow \quad
{\p \Omega _{IJ} \over \p \zeta ^\alpha} 
= i \int \! d^2w\,  \mu _\alpha 
\omega _I \omega _J
\, .
\ee
The second way makes use of the explicit relation between $\Omega _{IJ}$
and $\hat \Omega _{IJ}$, given by
\be
\Omega _{IJ} = \hat  \Omega _{IJ} 
  + {i \over 8 \pi } \int d^2u \int d^2v \omega _I (u)  
  \chi _{\bar u} {}^+ S_\delta (u,v) \chi _{\bar v}{}^+ \omega _J(v) 
\ee
and the constancy of $\hat \Omega _{IJ}$, so that 
\bea
{\p \Omega _{IJ} \over \p \zeta ^\alpha} 
&=& {i \over 8 \pi} \int d^2 w \int d^2 v \omega _I(w)
\chi _\alpha (w) S_\delta (w,v) \chi _{\bar v} ^+ \omega _J(v) +
(I\leftrightarrow J)
\nonumber \\
&=& -{i \over 2} \int d^2 w  \chi _\alpha (w) \biggl ( 
\hat \omega _{I0} (w) \omega _J (w) + \hat \omega _{J0} (w) \omega _I (w)
\biggr )
\eea 
where we have used the following relation in passing from the first to
the second line above
\be
\hat \omega _{I0}(z) = -{1 \over 4 \pi} \int d^2w \ S_\delta (z,w) \chi
_{\bar w} {}^+ \omega _{I}(w)\, .
\ee
The agreement of both calculations of (\ref{mualphaeq}) confirms
our determination of the class of $\mu _\alpha$.

\subsection{Contributions from $\hat \mu$ : spin 2 part }

The contribution from the first finite dimensional determinant involves
$\Phi _{IJ}$ of (\ref{Phis}), 
\be
i\Phi _{IJ+} = \omega _I \omega _J + \omega _I \pz \lambda _J + \omega
_J \pz \lambda _I - \half \hat \omega _{I0} \pz \hat \omega _{J0} -
\half \hat \omega _{J0} \pz \hat \omega _{I0}\, ,
\ee 
whose entire $\hat \mu$ dependence is through $\lambda _I$, and is given
by
\be
\pz \lambda _I (z) \bigg | _{\hat \mu} 
=
{1 \over 2 \pi} \int \! d^2w \pz \pw \ln E(z,w) \hat \mu (w) \omega _I
(w) \, .
\ee
Recalling the definition of the holomorphic 1-form $\varpi ^* _a (w)$,
introduced in (\ref{varpi}) via $\varpi ^* _a (w) = \varpi  _a
(p_a,w)$ and noting that this object satisfies $\varpi ^* _a (p_a) =1$,
we have the following expansion in $\hat \mu$, which terminates to first
order, 
\be
-i {\det \omega _I \omega _J(p_a) \over \det \Phi _{IJ+}(p_a)} \bigg
|_{\hat \mu} =
 - {1 \over 2 \pi} \sum _a \int \! d^2w \p _{p_a} \pw \ln E(p_a,w) 
\hat \mu (w) 2 \omega _a ^* (w) \, .
\ee
Clearly, the 2-form integrated versus $\hat \mu$ has poles at $p_a$, but
is holomorphic everywhere else. This contribution goes naturally together
with the one from the stress tensor insertion $T_2$ involving  $f_2$. We
add here the suitable multiple of $T_1(w)$ to make this contribution
into a well-defined and single valued 2-form. Assembling all parts, we
find
\bea
&& 
-i { \det \omega _I \omega _J(p_a) \over \det \Phi _{IJ+}(p_a)} 
\bigg |_{\hat \mu}
+ { 1 \over 2 \pi } \int \! d^2w \hat \mu (w) \biggl \{-27 T_1(w) + \half
f_2 (w)^2 -{3 \over 2} \pw f_2(w) \biggr \}
\nonumber \\ 
&& \qquad 
=
 { 1 \over 2 \pi } \int \! d^2w \hat \mu (w) B_2(w) \, ,
\eea
where the meromorphic 2-form is given by
\bea
\label{beetwo}
B_2(w) = -27 T_1(w) +
 \half f_2 (w)^2 -{3 \over 2} \pw f_2(w)-2 \sum _a \p_{p_a} \pw \ln
E(p_a,w) \varpi ^* _a (w) \, .
\eea
Actually, even though this is only a partial result from the spin
2 sector only, $B_2(w)$ by itself is a holomorphic and single-valued
2-form. To see this, apply $\p_{\bar w}$, and use the asymptotic
expansion for $w \sim p_a$,
\be
f_2 (w) = {1 \over w-p_a} + \p \phi ^{(2)*} _a (p_a) + \O(w-p_a)\, .
\ee
One finds
\bea
\p _{\bar w} B_2(w) 
&=&
\sum _a  \p _{\bar w}  \biggl (
- 2 \p_{p_a} \pw \ln E(w,p_a) \varpi ^*_a(w) - 2 \pw {1 \over w-p_a} + {1
\over w-p_a} \p \phi ^{(2)*} _a (p_a) \biggr )
\nonumber \\
&=&
2 \pi \sum _a    \biggl ( 2\pw \delta (w,p_a) \varpi ^*_a(w)
-2 \pw \delta (w,p_a) + \delta (w,p_a) \p \phi ^{(2)*}_a (p_a) \biggr )\,
.
\eea
This expression vanishes using  $\varpi ^*_a(p_a) =1$ as
well as $ \p \phi ^{(2)*} _a(p_a) = 2 \p \varpi ^* _a (p_a)$.

\subsection{Contributions from $\hat \mu$ : spin 3/2 part }

This finite dimensional determinant receives contributions from $\<
\chi_\alpha | \Phi ^* _{\beta 0} \>$ and $\< \mu _\alpha | \Phi ^*
_{\beta +}\>$. The latter is given by
\be
\< \mu _\alpha | \Phi ^* _{\beta +} \> 
=
- {1 \over 4 \pi} \int \! d^2 w \mu _\alpha (w) 
\int \! d^2z G_2(w,z) \biggl (\chiz \pz \psi ^* _\beta (z) + 3 (\pz
\chiz) \psi ^* _\beta (z) \biggr )\, .
\ee
The expression may be simplified by using the $\zeta$-dependence $\hat
\mu = \zeta ^1 \zeta ^2 \bar \mu$ and
\be
\mu _\alpha (w) \chiz (z)
=
\epsilon _{\alpha \beta} \zeta ^\beta \bar \mu \zeta ^\gamma \chi _\gamma
(z)
=
- \delta _\alpha {}^\gamma \zeta ^1 \zeta ^2 \bar \mu (w) \chi _\gamma (z)
= - \hat \mu (w) \chi _\alpha (z)\, .
\ee
Using this simplification, we may assemble the $\hat \mu$ contribution to
the finite dimensional determinant as follows
\bea
-\< H_\alpha | \Phi ^*_\beta\>
&=&
\<\chi_\alpha | \Phi ^* _{\beta 0} \>
+ \< \mu _\alpha | \Phi ^* _{\beta  +}\>
\\
&=&
\<\chi_\alpha | \psi ^* _\beta  \>
+ {1 \over 2 \pi} \int \! d^2w \hat \mu (w) \int \! d^2z \chi _\alpha (z)
\biggl (
{3 \over 2} \pw G_{3/2}(z,w) \psi ^* _\beta (w)
\nonumber \\
&&  + \half G_{3/2} (z,w) \p
\psi ^* _\beta (w) - G_2 (w,z) \pz \psi ^* _\beta (z) - {3 \over 2} \pz
G_2 (w,z) \psi ^* _\beta (z) \biggr )
\nonumber
\eea
Note that in the first term, $\<\chi _\alpha | \psi ^* _\beta\>$, the
differential $\psi ^* _\beta$ is evaluated with respect to the metric 
$\hat g$, while $\chi _\alpha$ is still considered with respect to the
metric $g$.

\medskip

To compute the determinant, it is useful to change basis for the $\chi
_\alpha$, as follows
\be
\chi _\alpha (z) =  \sum _\beta \< \chi _\alpha | \psi ^* _\beta\> \chi ^*
_\beta (z)
\qquad \qquad
\< \chi ^* _\alpha | \psi ^* _\beta \> = \delta _\alpha {} ^ \beta \, .
\ee
As a result, the determinant takes on a simple form in this basis, as the
$\hat \mu$-dependent correction amounts to taking the trace.
\bea
\det \< H_\alpha | \Phi ^*_\beta\>
&=& \det  \<\chi_\alpha | \psi ^* _\beta  \>
\biggl ( 1 + {1 \over 2 \pi} \int \! d^2w \hat \mu (w) \int \! d^2z \chi
^* _\alpha (z)
\biggl [ {3 \over 2} \pw G_{3/2}(z,w) \psi ^* _\alpha (w)
\nonumber \\
&&  
+ \half G_{3/2} (z,w) \pw \psi ^* _\alpha (w) 
- G_2 (w,z) \pz \psi ^* _\alpha (z) 
- {3 \over 2} \pz G_2 (w,z) \psi ^* _\alpha (z)
\biggr ]
\biggr )\, .
\nonumber
\eea
It is natural to combine this contribution with the part of the stress
tensor insertion $T_{3/2}$ that only depends upon the form $ f_{3/2}$,
suitably augmented by a multiple of $T_1(w)$ to make this contribution 
into a well-defined form. One obtains 
\bea
&& 
{ \det  \<\chi_\alpha | \psi ^* _\beta  \> \over 
\det \< H_\alpha | \Phi ^*_\beta\>} \bigg |_{\hat \mu}
 + {1 \over 2 \pi} \int \! d^2w \hat \mu (w) \biggl \{ +12 T_1(w) - 
\half f_{3/2}(w)^2 +\pw f_{3/2}(w) \biggr \}
\nonumber \\
&& \qquad =
 {1 \over 2 \pi} \int \! d^2w \hat \mu (w) B_{3/2}(w) 
\eea
where
\bea
\label{bthreehalfs}
B_{3/2}(w) &=& 12 T_1(w) 
-\half f_{3/2}(w)^2 + \pw f_{3/2}(w) 
 \\
&& +\int \! d^2z \chi ^* _\alpha (z) \biggl (
-{3 \over 2} \pw G_{3/2} (z,w) \psi ^* _\alpha (w)
  -\half  G_{3/2} (z,w) \pw \psi ^* _\alpha (w)
\nonumber \\
&& \qquad \qquad \qquad \qquad 
+ G_2 (w,z) \pz \psi ^* _\alpha (z) + {3 \over 2} \pz G_2 (w,z) \psi
^* _\alpha (z) \biggr )
\nonumber
\eea
This 2-form is also holomorphic, as can be seen from applying $p_{\bar
w}$,
\bea
\p _{\bar w} B_{3/2}(w)
&=&
3\pi \sum _\gamma \pw \delta (w,q_\gamma) - 2 \pi \sum _\gamma \delta
(w,q_\gamma) \p \psi ^* _\gamma (q_\gamma) 
\nonumber \\
&& +  \pi \sum _\gamma \biggl ( -3  \psi ^* _\gamma (w) 
\pw \delta (w,q_\gamma) - \delta (w,q_\gamma) \pw \psi ^* _\gamma
(q_\gamma) \biggr )
\eea
which vanishes in view of the fact that $\psi ^*_\gamma (q_\gamma)=1$.

\subsection{Remaining contributions from spin 2 part}

The remaining contributions from the finite-dimensional determinant $\det
\Phi _{IJ+} (p_a)$ that do not involve $\hat \mu$ arise from the
$\lambda_I$ terms and from the terms in $\hat \omega _{I0}$ through their
$\chi$-dependence. The first terms arise from
\bea
\pz \lambda _I \bigg | _{\chi} & = &
{1 \over 4 \pi} \int \! d^2 w \pz \pw \ln E(z,w) \chiw \hat \omega_{I0}(w)
\nonumber \\
& = &
-{1 \over 16 \pi ^2} \int \! d^2w \pz \pw \ln E(z,w) \chiw \int \! d^2u
S_\delta (w,u) \chiu \omega _I(u)
\eea
Using the same object $\varpi ^*_a$ as defined above, we find that the
$\chi$-dependence in $\lambda$ produces the contribution,
\bea
-i { \det \Phi _{IJ+}(p_a) \over \det \omega _I \omega _J(p_a) } \bigg
|_{\chi \ \lambda } 
=  - {1 \over 8\pi ^2} \int \! d^2w \ \p _{p_a} \pw \ln E(p_a,w) \chiw 
\int \! d^2u S_\delta (w,u) \chiu \varpi ^* _a(u)  \qquad
\eea
This contribution gives rise to the term $\X_4$ below.

\medskip

The second term arises from the $\chi$-dependence of
\be
\hat \omega _{I0}(z) = - {1 \over 4 \pi} \int \! d^2 u S_\delta (z,u) 
\chiu \omega _I(u)
\ee
and requires the use of $\varpi _a (u,v)$ of (\ref{varpi}). Recall
that this object is a holomorphic 1-form in $u$ and in $v$ separately, it
is a scalar in $p_b$, $b\not=a$ and a $-2$ form in $p_a$. It also
satisfies
$\varpi  _a (v,u) = \varpi  _a (u,v)$ as well as $\varpi _a
(p_a,v) = \varpi ^* _a (v)$. Using this definition, the contribution of 
the second term is
\bea
-i { \det \Phi _{IJ+}(p_a) \over \det \omega _I \omega _J(p_a) }
\bigg |_{\chi \ \hat \omega _{I0}}
=  - {1 \over 16 \pi ^2} \int \! d^2u \int \! d^2v S_\delta (p_a,u)
\chiu  
\p _{p_a} S_\delta (p_a,v) \chiv \varpi _a  (u,v) \, .
\eea
This contribution gives rise to the term $\X_5$ below.

\subsection{Remaining contributions from spin 3/2 part}

The remaining contribution from the spin 3/2 part is due solely to the 
terms quadratic in $\chi$ appearing in the $\Phi ^* _\beta$ differential,
as integrated versus $\chi _\alpha$. It is given by
\bea
{\det \< H_\alpha | \Phi ^* _\beta \>
\over \det \< \chi _\alpha | \psi ^* _\beta \>} \bigg |_\chi
=
1 - {1 \over 16 \pi ^2} \int \! d^2z \chi _\alpha ^* (z) \int \! d^2w 
G_{3/2} (z,w) \chiw \int \! d^2 v \chiv \Lambda _\alpha (w,v)
\eea
where $\Lambda _\alpha$ is defined by
\bea
\Lambda _\alpha (w,v) \equiv  2 G_2(w,v) \pv \psi ^* _\alpha + 3
\pv G_2 (w,v) \psi ^*_\alpha (v) \, .
\eea
This term contribution gives rise to the term $\X_6$ below.

\subsection{Summary}

Using the results of the previous calculations of the finite-dimensional
determinants and stress tensor insertions, we may write a more explicit
result as follows.
\bea
\label{finamp2}
{\cal A} [\delta]
&=&
i \ {\< \prod _a b(p_a) \prod _\alpha \delta (\beta (q_\alpha)) \>
\over \det \bigl (\omega _I \omega _J (p_a) \bigr )
 \cdot \det \< \chi _\alpha | \psi ^* _\beta\>}
\biggl \{ 1  + {\cal X}_1 + {\cal X}_2 + {\cal X}_3 + {\cal X}_4 +  {\cal
X}_5 + {\cal X}_6 \biggr \}
\nonumber \\
{\cal X}_1
&=&
 - {1 \over 8 \pi ^2} \int \! d^2z \chiz \int \! d^2 w \chiw \< S(z) S(w)\> 
\nonumber \\
{\cal X} _2 
& = & + {i \over 4\pi} (\hat \Omega _{IJ} - \Omega _{IJ} )
\biggl (  5\p _I \p _J \ln \tet [\delta ](0) - \p _I \p _J \ln \tet
[\delta ](D_\beta ) +   \p _I \p _J \ln \tet (D_b ) \biggr )
\nonumber \\
{\cal X} _3 
&=& + {1 \over 2 \pi} \int d^2 w \hat \mu (w)  \biggl ( B_2(w) +
B_{3/2}(w)
\biggr )
\nonumber \\
{\cal X} _4 
&=& + {1 \over 8\pi ^2} \int \! d^2w \ \p _{p_a} \pw \ln E(p_a,w) \chiw 
\int \! d^2u S_\delta (w,u) \chiu \varpi ^* _a(u)
\nonumber \\
{\cal X} _5 
&=& + {1 \over 16 \pi ^2} \int \! d^2u \int \! d^2v S_\delta (p_a,u) \chiu  
\p _{p_a} S_\delta (p_a,v) \chiv \varpi _a  (u,v) 
\nonumber \\
{\cal X} _6
&=& + {1 \over 16 \pi ^2} \int \! d^2z \chi _\alpha ^* (z) \int \!
d^2w G_{3/2} (z,w) \chiw \int \! d^2 v \chiv \Lambda _\alpha (w,v)  
\eea
The various ingredients in the formula have been defined throughout the
text. 

\medskip
 
It remains to re-express the above formula (\ref{finamp2}) in terms of
the final result (\ref{finamp}) and (\ref{Xes}). This may be done by
exhibiting a detailed correspondence between the $\X _i$, $i=1,\cdots ,6$
of (\ref{Xes}) and those of (\ref{finamp2}). The quantities $\X_1$,
$\X_4$, $\X_5$ and $\X_6$ are identical in both cases already. Thus, it
simply remains to regroup $\X_2$ and $\X_3$ and to combine them into the
sum given in (\ref{finamp}).  Using (\ref{diffomega}), we represent
$\X_2$ of (\ref{finamp2}) in terms of an integral versus $\hat \mu$, of a
form similar to the integral in $\X_3$. Next, using the expressions for
$B_2$ and $B_{3/2}$ in (\ref{beetwo}) and (\ref{bthreehalfs}), and
combining those with the calculation of the full stress tensor given in
(\ref{stresstotal}), we obtain a holomorphic two-form, given by $T^{IJ}
\omega _I (w) \omega _J(w)$ integrated versus the Beltrami differential
$\hat \mu$.
The final step consists in expressing the inner product $\< \hat \mu |
\omega _I \omega _J\>$ in terms of $\Omega _{IJ} - \hat \Omega _{IJ}$ with
the help of (\ref{diffomega}) and using (\ref{defsuperperiod}) to
re-express the result solely in terms of $\chi$,
\bea
\X_2 + \X _3 & = &
{1 \over 2\pi} \int \! d^2w \hat \mu (w) T^{IJ} \omega _I (w) \omega _J(w)
\nonumber \\
&=&
- {i \over 2 \pi} T^{IJ} \biggl (\Omega _{IJ} - \hat \Omega _{IJ} \biggr )
\nonumber \\
&=&
{1 \over 16 \pi ^2} \int \! d^2 u \int \! d^2v \ T^{IJ} \omega _I(u)
\chiu S_\delta (u,v) \chiv \omega _J(v)
\eea
This result now precisely coincides with the final result given for $\X_2
+ \X_3$ in (\ref{Xes}), thereby completing the proof of this formula.

\vfill\eject
\appendix

\section{Appendix: Bosonic Riemann Surface Formulas}
\setcounter{equation}{0}

In this section, we review basic formulas, holomorphic and meromorphic
differentials and Green's functions on an ordinary Riemann surface
$\Sigma$ of genus $h$, as well as associated variational formulas.
Standard references are \cite{fay}, \cite{vvv} and \cite{dp88}.

\subsection{Basic Objects}

The basic objects on a Riemann surface
$\Sigma$, from which all others may be reconstructed, are the holomorphic
Abelian differentials, the Jacobi $\tet$-function, and the prime form. We
choose a canonical homology basis $A_I$, $B_I$, $I=1,\cdots ,h$, with
canonical intersection matrix $\# (A_I,B_J) = \delta _{IJ}$. Modular
transformations are defined to leave the intersection form invariant and
form the group $Sp(2h,{\bf Z})$.

\medskip

The {\sl holomorphic Abelian differentials} $\omega _I$ are holomorphic
1-forms which may be normalized on $A_I$ cycles, and whose integrals
on $B_I$ cycles produce the period matrix,
\bea
\oint _{A_I} \omega _J = \delta _{IJ}
\hskip 1in 
\oint _{B_I} \omega _J = \Omega _{IJ}
\eea
The Jacobian is then defined as $J(\Sigma) \equiv {\bf C}^h
/ \{ {\bf Z}^h + \Omega {\bf Z}^h\}$.

\medskip

Given a base point $z_0$, the {\sl Abel map} sends $d$ points $z_i$, with
multiplicities $q_i \in {\bf Z}$, $i=1,\cdots,d$ and divisor $D=q_1
z_1 + \cdots q_d z_d$ of degree $q_1 + \cdots + q_d$ into ${\bf C}^h$ by
\bea
\label{abelmap}
q_1 z_1 + \cdots + q_d z_d 
\equiv 
\sum _{i=1} ^d q_i \int _{z_0} ^{z_i} (\omega _1, \cdots , \omega _h)
\eea
The Abel map onto ${\bf C}^h$ is multiple valued, but it is single
valued onto $J(\Sigma)$. 

\medskip

The {\sl Jacobi $\tet$-functions} are defined on $\zeta = (\zeta _1,
\cdots, \zeta _h)^t \in {\bf C}^h$ by 
\bea
\tet [\delta] (\zeta, \Omega) 
\equiv  
\sum _{n \in {\bf Z}^h } 
\exp \biggl (i \pi (n + \delta ') ^t \Omega (n+ \delta ') + 2\pi i
(n+\delta ') ^t  (\zeta + \delta '') \biggl ) \, .
\eea
Here, $\delta  = \left ( \delta ' | \, \delta '' \right )$ is a general
characteristic, where $\delta ', \ \delta '' \in {\bf C}^h$ are
both written as a column vector. Henceforth, we shall assume that
$\delta$ corresponds to a spin structure, and thus be valued in $\delta
', \delta '' \in ({\bf Z}/2{\bf Z}) ^h$. The parity of the
$\tet$-functions depends on $\delta$ and is defined by (for $\zeta$ and
$\Omega$ such that $\tet [\delta ](\zeta, \Omega)\not=0$),
\bea
\tet [\delta ] (- \zeta , \Omega ) = (-1) ^{4 \delta ' \cdot \delta ''}
\tet [\delta ](\zeta ,\Omega)
\eea
According to whether $4\delta ' \cdot \delta ''$ is even or odd, 
$\delta$ is referred to as an {\sl even or odd spin structure}. One often
denotes $\tet (\zeta, \Omega ) \equiv \tet [0] (\zeta , \Omega)$. Upon
shifting by full periods, $M, \ N \in {\bf Z}^h$,
\bea
\tet [\delta] (\zeta + M + \Omega N, \Omega ) =
\exp \biggl (-i \pi N^t \Omega N - 2 \pi i N^t (\zeta + \delta ') + 2
\pi i M^t \delta '' \biggr ) \tet [\delta] (\zeta , \Omega)
\eea
Under a modular transformation $U \in Sp(2h,{\bf Z})$, the characteristic
$\delta = (\delta ' | \, \delta ")$ transforms as (see
for example \cite{igusa,fay})
\bea
\left ( \matrix{ \tilde \delta ' \cr \tilde \delta ''} \right )
=
\left ( \matrix{D & -C \cr -B & A} \right )
\left ( \matrix{ \delta ' \cr \delta ''} \right )
+ \half {\rm diag} \left ( \matrix{CD^t \cr AB^t} \right ) 
\hskip .7 in
U= \left ( \matrix{A & B \cr C & D} \right )
\eea
The period matrix transforms as 
\bea
 \tilde \Omega = (A\Omega + B) (C \Omega + D)^{-1}
\eea 
while the $\tet$-function transforms as (see \cite{fay, igusa}), with
$\epsilon ^8=1$,
\bea
\tet [\tilde \delta ] (\{ (C\Omega +D)^{-1} \}^t  \zeta , \tilde
\Omega) =
\epsilon (\delta, U) \det (C\Omega + D)^{\half} \tet [\delta ](\zeta ,
\Omega)
\eea

\medskip

The {\sl Riemann vector} $\Delta \in {\bf C}^h$, which depends on the base
point $z_0$ of the Abel map, enters the {\sl Riemann vanishing Theorem},
which states that $\tet (\zeta , \Omega )=0$ if and only if there
exist $h-1$ points $p_1, \cdots , p_{h-1}$ on $\Sigma$, so that $\zeta =
\Delta - p_1 \cdots -  p_{h-1}$. The explicit form of $\Delta$ may be
found in \cite{dp88}, formula (6.37) and will not be needed here.

\medskip

The {\sl prime form} is constructed as follows \cite{fay}. For any odd
spin structure
$\nu$, all the $2h-2$ zeros of the holomorphic 1-form $\sum _I \p_I
\tet [\nu ](0,\Omega) \omega _I(z)$ are double and the form admits a
unique (up to an overall sign) square root $h_\nu (z)$ which is a
holomorphic 1/2 form. The prime form is a $-1/2$ form in both variables
$z$ and $w$, defined by
\bea
E(z,w) \equiv {\tet [\nu ] (z-w, \Omega) \over h_\nu (z) h_\nu (w)}
\eea
where the argument $z-w$ of the $\tet$-functions stands for the Abel map
of (\ref{abelmap}) with $z_1=z$, $z_2=w$ and $q_1=-q_2=1$. The form
$E(z,w)$ defined this way is actually independent of
$\nu$. It is holomorphic in $z$ and $w$ and has a unique simple zero at
$z=w$. It is single valued when $z$ is moved around $A_I$ cycles, but has
non-trivial monodromy when $z\to z'$ is moved around $B_I$ cycles,
\bea
E(z',w) = - \exp \biggl ( -i \pi \Omega _{II} + 2 \pi i \int ^z _w \!
\omega _I \biggr ) E(z,w) \, .
\eea
The combination $\p_z \p_w \ln E(z,w)$ is a single valued meromorphic
differential (Abelian of the second kind) with a single double pole at
$z=w$. Its integrals around homology cycles are given by
\bea
\label{prf}
\oint _{A_I} \! dz \p_z \p_w \ln E(z,w) & = & 0
\nonumber \\
\oint _{B_I} \! dz \p_z \p_w \ln E(z,w) & = & 2 \pi i \omega _I(w)
\eea
and will be of use throughout.

\subsection{Holomorphic differentials}

The covariant derivatives on forms of rank $n$ will be denoted by $\nabla
^{(n)} _z$ and $\nabla ^{(n)} _{\bar z}$. In complex coordinates adapted
to the metric, we may simply use $\p_z$ and $\p _{\bar z}$ instead,
whenever no confusion is possible. For $n\in {\bf Z} +1/2$, proper
definition of these forms requires specification of a spin structure. The
elliptic operators $\nabla ^{(n)} _{\bar z}$ have kernels with the
following dimensions $\Upsilon (n) \equiv {\rm dim} \ {\rm Ker} \nabla
^{(n)} _{\bar z} $
\be
\Upsilon (n) 
=
\left \{ 
\matrix{
0 & n < 0\, , \ {\rm and} \ n=1/2 \ {\rm even \ spin \ structure} \cr  
1 & n=0\, ,   \ {\rm and} \ n=1/2 \ {\rm odd \ spin \ structure}  \cr 
h & n=1 \cr  (2n-1) (h-1) & n \geq 3/2 } 
\right . 
\ee
while the cokernels have dim coKer $\nabla ^{(n)} _{\bar z}=
\Upsilon (1-n)$. (The dimensions listed for $n=1/2$ are for generic
moduli and are valid for exceptional moduli mod 2.) A set of basis 
holomorphic differentials are denoted by $\phi _a ^{(n)}$, $a=1, \cdots ,
\Upsilon (n)$, and are section of the line bundles $T^n$, ($n$-th power of
the canonical bundle $T$), for which the number of zeros and poles are
related by 
\be
(\# \ {\rm zeros} - \# \ {\rm poles}) \ \phi ^{(n)} _a(z) = c_1(T^n) =
2n(h-1) \, .
\ee 
For $n=0$, they are just constants, for $n=1/2$ and $\nu $ odd they are
denoted by $h_\nu (z)$, while for $n=1$ they are the Abelian
differentials usually denotes by $\omega _I$, $I=1,\cdots ,h$.

\medskip

Given any set of $\Upsilon (n)$ points $z_1, \cdots , z_{\Upsilon (n)}$
on the surface, we may choose a basis $\phi ^{(n)*} _a$ for the
holomorphic $n$-differentials normalized at the points $z_b$ by
\be
\label{diffnorm}
\phi _a ^{(n)*} (z_b) = \delta _a ^b\, .
\ee
The holomorphic differentials with this normalization may be exhibited
explicitly in terms of the prime form $E(z,w)$, the $h/2$ differential
$\sigma (z)$ and $\vartheta$-functions. For $n \geq 3/2$, we have 
\be
\label{diffexp}
\phi _a ^{(n)*} (z)
= 
{\vartheta [\delta ](z-z_a + \sum z_b -(2n-1)\Delta) \over 
 \vartheta [\delta ] ( \sum z_b -(2n-1)\Delta)}
{\prod _{b\not= a} E(z,z_b) \over \prod _{b\not= a} E(z_a,z_b)}
\biggl ( {\sigma (z) \over \sigma (z_a)} \biggr ) ^{2n-1}\, .
\ee
To simplify notation, we shall not exhibit the spin structure dependence
of these differentials. 
Here, $\sigma (z)$ is a tensor of rank $h/2$ without zeros or poles, and
which may be defined up to a constant by the ratio
\bea 
{\sigma (z) \over \sigma (w)} & = & 
{\vartheta (z-\sum p_i +\Delta) \over \vartheta (w-\sum p_i +\Delta)}
\prod _{i=1} ^h {E(w,p_i) \over E(z,p_i)}
\eea
where $p_i$, $i=1,\cdots , h$ are arbitary points on the surface. Note
that $\sigma(z)$ is single valued around $A_I$ cycles but multivalued
around $B_I$ cycles in the following way
\bea
\sigma (z') &=& \sigma (z) \exp \bigl \{ -i\pi (h-1) \Omega _{II} + 2 \pi i
\Delta _{Iz} \bigr \}
\eea
Besides the $\Upsilon(n)-1$ zeros $z_b$, $b\not=a$, the differential $\phi
_a ^{(n)*}(z)$ has $h$ additional zeros. The tensor  $\phi _a ^{(n)*}(z)$
is of rank $n$ in $z$, rank $-n$ in $z_a$ and rank 0 in $z_b$ with
$b\not=a$. For $n=1$, we have 
\be
\phi ^{(1)*} _a (z)
= 
{\tet  (z-z_a  + \sum z_b -w_0 - \Delta ) \over 
 \tet (  \sum z_b -w_0 - \Delta ) \ E(z,w)} \
\prod _{b\not= a} {E(z,z_b) \over E(z_a,z_b)} 
{ E(z_a,w_0) \over E(z, w_0)}
\ {\sigma (z) \over \sigma (z_a)} \, .
\ee

\subsection{Meromorphic differentials : Green's functions}

Meromorphic Green's functions $G_{n}(z,w)= G_n (z,w;z_1,\cdots,
z_{\Upsilon(n)})$ for the operators $\nabla _{\bar z} ^{(n)}$ with  
$n\geq 3/2$ for general spin structure and $n=1/2$ for even spin
structure may be defined by the following relations
\bea
\nabla _{\bar z} ^{(n)} G _n (z,w)  & = & + 2 \pi \delta (z,w)
\\
\nabla _{\bar w} ^{(1-n)} G_n(z,w) & = & - 2 \pi \delta (z,w)
+ 2\pi \sum _{a=1} ^{\Upsilon (n)} \phi _a ^{(n)*} (z) \delta (w,z_a)\, . 
\eea
The properly normalized holomorphic $n$-differentials $\phi _a ^{(n)*}(z)$
are defined in (\ref{diffnorm}) and (\ref{diffexp}). Setting $z=z_a$, we
have $\nabla _{\bar w} ^{(1-n)} G_n(z_a,w)=0$, so that $G_n(z_a,w)=0$.
Explicit expressions for the Green's function are 
\be
G_n [\delta](z,w) 
= 
{\vartheta [\delta ]\big (z-w + \sum z_b -(2n-1)\Delta \big ) \over 
 \vartheta [\delta ]\big ( \sum z_b -(2n-1)\Delta \big ) \ E(z,w)} \
{\prod _{a} E(z,z_a) \over \prod _{a} E(w,z_a)} \
\biggl ( {\sigma (z) \over \sigma (w)} \biggr ) ^{2n-1}
\ee
In particular, for $n=1/2$, this reduces to the standard form of the
Szeg\" o kernel, usually denoted by
\be
S_\delta (z,w) = 
{\vartheta [\delta ]\big (z-w  \big ) \over 
 \vartheta [\delta ]\big ( 0 \big ) \ E(z,w)} \, .
\ee
For $n=1$, the Green's function $G_1(z,w)=G_1(z,w;z_1,\cdots ,z_h, w_0)$
is the Abelian differential of the third kind, satisfying
\bea
\nabla _{\bar z} ^{(1)} G _1 (z,w)  & = & + 2 \pi \delta (z,w) - 2\pi
\delta (z,w_0)
\\
\nabla _{\bar w} ^{(0)} G_1 (z,w) & = & - 2 \pi \delta (z,w)
+ 2\pi \sum _{a=1} ^h \phi _a ^{(1)*} (z) \ \delta (w,z_a)\, , 
\eea
and explicitly given by the following expression
\be
G_1(z,w) 
= 
{\vartheta \big (z-w-w_0 + \sum z_b - \Delta \big ) \over 
 \vartheta \big ( -w_0 + \sum z_b - \Delta \big ) \ E(z,w)} \
{\prod _{a} E(z,z_a) E(w,w_0) \over \prod _{a} E(w,z_a) E(z, w_0)} \
 {\sigma (z) \over \sigma (w)} \, .
\ee

\subsection{Variational Formulas}

The variation $\delta _{ww} \phi$ of any object $\phi$ under a variation
of the metric, parametrized by a Beltrami differential $\mu$ is defined as
follows
\bea
\delta \phi 
\equiv {1 \over 2\pi} \int _\Sigma d^2w \mu _{\bar w}{}^w \delta _{ww}
\phi 
\qquad \qquad
 \mu _{\bar w}{}^w \equiv \half g_{w \bar w} \delta g^{ww} \, .
\eea
The variational formulas for the covariant derivatives on rank $n$ forms
are
\begin{eqnarray}
\delta \nabla ^z 
&=& 
+\half \delta g^{zz} \nabla _z +{n \over 2} (\nabla _z \delta g ^{zz})
\nonumber \\
\delta \nabla _z 
&=& 
-\half \delta g_{zz} \nabla ^z +{n \over 2} (\nabla ^z \delta g _{zz})
\end{eqnarray}
ignoring the Weyl variation part. From these formulas, one derives the
following variational formulas for differentials and periods,
\begin{eqnarray}
\delta _{ww} \omega _I (z) &=& \omega _I(w) \pz \pw \ln E(z,w)
\nonumber \\
\delta _{ww} \Omega _{IJ} &=& 2 \pi i \omega _I(w) \omega _J (w)
\nonumber \\
\delta _{ww} \Delta _{Iz} &=& + \half \pw \omega _I (w) - \omega _I (w)
\pw \ln \psi(w,z)
\nonumber \\
\delta _{ww} \ln \sigma (z) &=& {1 \over 2 h-2} \biggl ((\pw \ln
\psi(w,z))^2 - \pw ^2 \ln \psi (w,z) \biggr )
\nonumber \\
\delta _{ww} \ln E(x,y) &=& - \half \biggl ( \partial _w \ln {E(x,w)
\over E(y,w) } \biggr ) ^2
\end{eqnarray}
Here, we have $\psi(w,z) = \sigma (w) E(w,z)^{h-1}$. 
Under an analytic coordinate change $u \to w(u)$ the variation $\delta
_{uu} \ln \sigma (z)$ transforms with a Schwarzian derivative $\{u,w\}$, 
\be
\delta _{ww} \ln \sigma (z) = \biggl ( {du\over dw} \biggr )^2 \delta
_{uu}
\ln \sigma (z) + {1 \over 2h-2} \{u,w\}\, .
\ee
since $\sigma (z)$ is the carrier of the gravitational anomaly.

\medskip

For the holomorphic differentials $\phi _a ^{(n)*}$ introduced above  with
normalizations $\phi ^{(n)*} _a (z_b) =\delta _a ^b$ and the Green's
functions $G_n $ with normalization $G_n(z_a,w)=0$, we have the following
variational formulas
\bea
\label{vardiff}
\delta _{ww} \phi ^{(n)*} (z)
= n \nabla _w G_n (z,w) \phi ^{(n)*} (w)
+(n-1) G_n (z,w) \nabla _w \phi ^{(n)*} (w)
\nonumber \\
\delta _{ww} G_n (z,y)
= n \nabla _w G_n (z,w) G_n (w,y)
+(n-1) G_n (z,w) \nabla _w G_n (w,y)
\eea
Clearly, the variations preserve the normalization conditions.

\section{Appendix: $\N=1$ Supergeometry Formulas}
\setcounter{equation}{0}

It is useful to define a $\N=1$ supergeometry in 2 dimensions in the
following way. A detailed account, including supercomplex structures and
complete Wess-Zumino expressions may be found in \cite{dp88}. We begin
with a space of real dimension $2|2$, and local coordinates $v^M = (v,\bar
v | \theta , \bar \theta)$, where the index $M$ is a super-Einstein index.
We also have a U(1) gauge group of (Euclidean) frame rotations, and we
classify superfields according to their U(1) charge, which for our
purposes is $n\in {\bf Z} /2$. The supergeometry data are a superframe
$E_M {}^A$ and a U(1) connection $\Omega _M$. The inverse of $E_M {}^A$
will be denoted by $E_A {}^M$. The frame index $A=(a|\alpha)$ runs over
$2|2$ values, customarily denoted by $(z,\bar z|+,-)$.

\subsection{Supergeometry and super derivatives}

On a superfield $V$ of U(1) weight $n$, the superderivatives are defined
by
\bea
\D _A ^{(n)} V \equiv E_A {}^M (\p_M V + i n \Omega _M V)
\eea
and the torsion $T_{AB}{}^C$ and curvature $R_{AB}$ tensor are defined by
\bea
[\D _A , \D _B ] V = T_{AB} {}^C \D _C V + i n R_{AB} V
\eea
and $[ \ , \ ]$ is a commutator unless both $A$ and $B$ are spinor
indices, in which case it is an anti-commutator. The torsion constraints
are
\bea
T_{\alpha \beta} {}^\gamma = T_{ab}{}^c =0,
\qquad 
T_{\alpha \beta }{}^c = 2 \gamma _{\alpha \beta }{}^c
\eea
We assume $\{ \gamma ^a , \gamma ^b\} = - \delta ^{ab}$, so
that $\gamma _{++} ^z = \gamma _{--}^{\bar z} =1$ and $\gamma ^a _{+-} =
\gamma ^a _{-+}=0$.
The torison constraints imply that the odd superderivatives have very
simple anti-commutation relations on a superfield of U(1) weight $n$,  
\bea
\D_+ ^2 V = \D _z V \qquad 
\D_- ^2 V = \D _{\bar z} V \qquad 
\{ \D_+ , \D _- \} V = i n R_{+-} V
\eea

\medskip

In Wess-Zumino gauge, the algebraic components of sDiff($\Sigma$),
sWeyl($\Sigma$) and U(1) are eliminated and the remaining independent
fields are the ordinary frame $e_m {}^a$, the gravitino field $\chi _m
{}^\alpha$ and an auxiliary scalar $A$, 
\bea
\label{supergeometry}
E_M{}^A & = & e_m{}^a + \theta \gamma ^a \chi _m -{i \over 2} \theta \bar
\theta A e_m {}^a
\nonumber \\
\sdet E_M{}^A & = & e \bigl (1 + \half \theta \gamma ^m \chi _m
- {i \over 2} A + {1 \over 8} \theta \bar \theta \epsilon ^{mn} \chi _m
\gamma _5 \chi _n \bigr )
\eea
The superderivatives acting on a superfield $V$ of U(1) weight $n$,
\bea
 V= V_0 + \theta V_+ + \bar \theta V_- + i \theta \bar \theta V_1
\eea
are given by
\bea
\label{superderivatives}
{\cal D} ^{(n)} _+ V 
&=&
V_+ + \theta \biggl ( \pz V_0 + \half \chi _z {}^- V_- \biggr )
+ \bar \theta \biggl ( iV_1 + {i \over 2} n A V_0 \biggr )
 \\
&& 
+ \theta \bar \theta \biggl ( {i \over 4} (1-2n) AV_+ - {1 \over 4}
\chiz \chi _z {}^- V_+ + \half \chi _z {}^- \p _{\bar z} V_0 + \p _z V_- 
- n \p _{\bar z} \chi _z {}^- V_0 \biggr )  
\nonumber\\
{\cal D} ^{(n)} _- V 
&=&
V_- + \bar \theta \biggl ( \p_{\bar z} V_0 + \half \chiz V_+ \biggr ) +
\theta \biggl ( -iV_1 + {i \over 2} n A V_0 \biggr )
\nonumber \\
&& 
+ \theta \bar \theta \biggl ( {i \over 4} (1+2n) AV_- - {1 \over 4}
\chiz \chi _z {}^- V_- - \half \chiz \pz V_0 - \p _{\bar z} V_+ 
- n \pz \chiz V_0 \biggr )  
\nonumber
\eea

\subsection{Holomorphic superdifferentials}

We seek to solve the equation $\D _- ^{(n)} \Phi =0$ with $n\geq 3/2$ both
for the cases of even and odd valued $\phi$, in Wess-Zumino gauge. The
starting point  is the generic expression for $\Phi$ as a superfunction
\be
\Phi (z,\theta,\bar \theta)
=
\Phi _0 (z)+ \theta \Phi _+(z) + \bar \theta \Phi _-(z) + i \theta \bar
\theta \Phi _1(z)
\ee
and the expression for the operator $\D_- ^{(n)}$ of
(\ref{superderivatives}).
As a result, $\Phi _-=0$ and $\Phi _1 = {n \over 2} A \Phi _0$, so that
the remaining non-trivial equations are
\bea
\label{superdiffeq}
0 &=& \nabla _{\bar z} ^{(n)} \Phi _0 + \half \chiz \Phi _+ 
\nonumber \\
0 &=& \nabla _{\bar z} ^{(n+1/2)} \Phi _+ + \half \chiz \nabla _z \Phi _0 
+n  (\nabla _z \chiz) \Phi _0\, .
\eea 
The general solution is recovered from iterating the equivalent set of
integral equations
\bea
\label{superdiffsol}
\Phi _0 (z) 
&=& \phi _0 (z) -  {1 \over 2\pi} \int d^2w\ G_n (z,w) \
\half \chiw \Phi _+ (w) 
 \\
\Phi _+ (z) 
&=&  \phi _+ (z) - {1 \over 2 \pi} \int d^2w\ G_{n+1/2} (z,w) \ \biggl (
\half
\chiw \nabla _w
\Phi _0 + n (\nabla _w \chiw) \Phi _0 \biggr )(w)
\nonumber
\eea
where $\phi _0(z)$ and $\phi _+(z)$ satisfy the bosonic equations
$\nabla _{\bar z} ^{(n)} \phi _0=0$ and $\nabla _{\bar z} ^{(n+1/2)} \phi 
_+=0$, i.e. they are holomorphic $n$ and $n+1/2$ differentials respectively.

\medskip

The Green's functions $G_n$ and $G_{n+1/2}$ are not unique in general,
as explained in the previous subsection. However, the arbitrariness may
be aborbed by a shift in $\phi _0$ and $\phi _+$. Since these spaces are
generated by $\psi _\alpha \equiv \phi _\alpha ^{(n)}$ with
$\alpha =1,\cdots ,\Upsilon (n)$ and $\phi _a\equiv \phi _a ^{(n+1/2)}$
with
$a=1,\cdots, \Upsilon (n+1/2)$ respectively, we may define the following
basis of all differentials
\bea
\label{evenodd}
{\sl EVEN} \quad \Phi _{\alpha 0} (z) &=& 
\psi _\alpha  (z) -  {1 \over 2 \pi } \int d^2w\ G_n (z,w) \ \half \chiw
\Phi _{\alpha +} (w) 
\\
\Phi _{\alpha +} (z) &=& 
- {1 \over 2 \pi } \int d^2w\ G_{n+1/2} (z,w) \ \biggl ( \half \chiw \nabla _w
\Phi _{\alpha 0} + n (\nabla _w \chiw) \Phi _{\alpha 0} \biggr )(w)
\nonumber \\
& & \nonumber \\
{\sl ODD} \ \quad \Phi _{a 0} (z) &=& 
 -  {1 \over 2 \pi } \int d^2w\ G_n (z,w) \ \half \chiw \Phi _{a +} (w) 
\\
\Phi _{ a +} (z) &=&  \phi _a  (z)
- {1 \over 2 \pi } \int d^2w\ G_{n+1/2} (z,w) \ \biggl ( \half \chiw \nabla _w
\Phi _{a 0} + n (\nabla _w \chiw) \Phi _{a 0} \biggr )(w)
\nonumber
\eea

It will be convenient to normalize the differentials and to fix the
Green's functions in a consistent way as we did in the preceding
subsection. To this end, we introduce arbitrary points $q_\alpha $,
$\alpha =1, 
\cdots , \Upsilon (n)$ and arbitrary points $p_a$, $a=1,\cdots, \Upsilon
(n+1/2)$, and denote the normalized holomorphic differentials with $*$ 
superscripts~:
\bea
\psi _\alpha ^*  (q_\beta) &=& \delta _{\alpha \beta} \qquad \qquad
G_n(q_\alpha ,w)=0 
\nonumber \\
\phi _a ^* (p_b) &=& \delta _{ab} \qquad \qquad G_{n+1/2} (p_a,w)=0\, .
\eea
>From these normalizations, it follows that 
\bea
\label{pointdif}
\Phi _{\alpha 0} ^* (q_\beta) &=& \delta _{\alpha \beta} 
\qquad \qquad 
\Phi _{\alpha +} ^* (p_b) \ = \ 0
\nonumber \\
\Phi _{a +} ^* (p_b) &=& \delta _{ab}
\qquad \qquad \
\Phi _{a 0} ^* (q_\beta)\ =\ 0 \, .
\eea
As a result, the differential $\psi _\alpha ^*(z)$ is a differential of 
degree 
$n$ in $z$ and of degree $-n$ in $q_\alpha$, while it is a scalar in 
$q_\beta$  when $\beta \not= \alpha$. Similarly for $\phi _a (z)$ which
is a differential  of degree $n+ 1/2$ in $z$, $-n-1/2$ in $p_a$ and 0 in
$p_b$ when $b\not=a$.

The cases of interest to string theory have $n=3/2$, so that $\alpha
=1,\cdots, 2h-2$ for even 3/2 superdifferentials and $a=1,\cdots ,3h-3$
for odd 3/2 superdifferentials, as well as Abelian differentials which we 
now  spell out in more detail.

\subsection{Abelian Super-Differentials -- Even Spin Structures}

The case of Abelian super-differentials is special, mainly because they 
have a  natural normalization on homology cycles, since they may be
integrated. For  even spin structure, there are $h$ odd superholomorphic
1/2 differentials, denoted  by $\hat \omega _I$, $I=1,\cdots ,h$, and
there are no even ones. The odd  differentials are normalized by 
\be
\oint _{A_I} \hat \omega _J = \delta _{IJ}
\qquad \qquad  
\oint _{B_I} \hat \omega _J = \hat \Omega  _{IJ}\, . 
\ee
Since the $\chi$-dependent corrections to the differential must integrate 
to 0  along $A_I$-cycles, the relevant Green's function on 1-forms to be
used is the  one built out of the prime form. Thus, we have (including
the auxiliary field $A$ of the supergeometry (\ref{supergeometry}),
\be
\hat \omega _I (z, \theta, \bar \theta)
= \hat \omega _{I0} (z) + \theta \hat \omega _{I+} (z) +{i \over 4} 
\theta \bar 
\theta A \hat \omega _{I0}(z)
\ee
and the solution is given by the implicit equations,
\bea
\label{abelian}
\hat \omega_{I 0} (z) &=& 
 -  {1 \over 4 \pi } \int d^2w\ S_\delta (z,w) \chiw \hat \omega  _{I +} 
(w) 
\\
\hat \omega_{I +} (z) &=&  \omega _I  (z)
- {1 \over 4 \pi } \int d^2w\ \pz \pw \ln E (z,w)  \chiw \hat \omega _{I0} (w)
\, .
\nonumber
\eea
which may be solved by iterating precisely $h-1$ times.
Here, $S_\delta (z,w)$ is the Szego kernel for (even) spin structure 
$\delta$. The superperiod matrix is given by
\be
\label{defsuperperiod}
\hat \Omega _{IJ} = \Omega _{IJ} 
- {i \over 8 \pi } \int d^2u \int d^2v \omega _I (u) \chi _{\bar u} {}^+
S_\delta (u,v) \chi _{\bar v}{}^+ \hat \omega _{J+}(v) \, . 
\ee
which may also be solved by iterating $h-1$ times. For genus 2, the full
solution is simply obtained by setting $\hat \omega _{J+}= \omega _J$ on
the rhs.

\subsection{Super-Beltrami differentials}

We spell out the signs that arise when dealing with the
superdifferential. First, using (\ref{susy}), we have the following
result for the super-Beltrami differentials in Wess-Zumino gauge
\cite{dp88},
\bea
H = H_- {}^z = \bar \theta \bigl ( e_{\bar z} {}^m \delta e_m {}^z 
- \theta \delta \chiz \bigr )
\eea
When evaluating the derivatives $H_A$ in Wess-Zumino gauge, we use the
definitions $H_A \equiv \bar \theta (\mu _A - \theta \nu_A)$. To compute
this object, the infinitesimal  differentials $\delta m^A$ must be
extracted, giving rise to new sign changes, which we list for convenience,
\bea
\mu _a & = & + e_{\bar z} {}^m {\p e_m {}^z \over \p m^a}
\hskip 1in 
\chi _a =  + {\p \chiz \over \p m^a}
\nonumber \\
\mu _\alpha & = & - e_{\bar z} {}^m {\p e_m {}^z \over \p \zeta ^\alpha}
\hskip 1in 
\chi _\alpha  =  + {\p \chiz \over \p \zeta ^\alpha} 
\eea
In terms of the differential $\hat \mu$, we have $\mu _\alpha = \p \hat
\mu / \p \zeta ^\alpha$.

The inner product with the superghost field $B= \beta + \theta b$ is
given by
\bea
\< H_a |B \> &=& + \< \mu _a | b\> - \< \nu _a | \beta \>
\nonumber \\
\< H_\alpha |B \> &=& - \< \mu _\alpha | b\> - \< \nu _\alpha | \beta \>
\eea

\subsection{Variational Formulas}

The variation $\delta _{w+}\Phi$ of any object $\Phi$ under a variation
of the supergeometry is defined by
\be
\delta \Phi = {1 \over 2 \pi} 
\int _{s\Sigma} d^{2|2}w E \ H_- ^w \ \delta _{w+} \Phi\, .
\ee
The variation under $H$ of the covraiant derivatives are given by $\delta
\D _+ ^{(n)} =0$ and
\bea
\delta \D _- ^{(n)} &=& - H \D _z ^{(n)} + \half (\D_+ H) \D_+ ^{(n)} -
n \D _z H \, .
\eea
The variation $\delta \hat \omega _I$ of
any holomorphic Abelian differential $\hat \omega _I$ satisfies
\be
\D _- \delta \hat \omega _I + \D _+ \bigl ( H \D _+ \hat \omega _I -
\half (\D_+ H) \hat \omega _I \bigr ) =0\, .
\ee
Using the fact that the super-prime form \cite{dp89,dp90} satisfies
\be
\label{primekernel}
\D _- ^\z \biggl ( \dzplus \dwplus \ln \E_\delta (\z,\w) \biggr ) = -2\pi
\dzplus \delta ^{(2|2)} (\z,\w)
\ee
as well as the canonical normalizations of $\hat \omega _I$ and the
following integrals 
\bea
\oint _{A_I} d\z \ \dzplus \dwplus \ln \E(\z,\w) & = & 0
\nonumber \\
\oint _{B_I} d\z \ \dzplus \dwplus \ln \E(\z,\w) & = & 2 \pi i \hat
\omega _I(\w)
\eea
we obtain a unique expression for the variation of the holomorphic
differentials
\be
\delta \hat \omega _I (\z) 
=
{1 \over 2 \pi} \int d^{2|2}w \dzplus \dwplus \ln \E(\z,\w) \biggl (
H \D_+ \hat \omega _I - \half (\D_+ H) \hat \omega _I \biggr ) (\w)
\, ,
\ee
which may alternatively be expressed as
\be
\delta _{w+} \hat \omega _I (\z) 
=
\half \dzplus \dwplus \ln \E(\z,\w) \D_+ \hat \omega _I(\w) +
\half \dzplus \D_w \ln \E(\z,\w) \ \omega _I(\w) \, .
\ee
The variational formula for the super-period matrix automatically follows
\bea
\label{varsuperperiod}
\delta \hat \Omega _{IJ} & = & \< H | \Phi _{IJ} \> = 
 \int d^{2|2}w \ H \Phi _{IJ}
\nonumber \\
\delta _{w+} \hat \Omega _{IJ} & = & 2 \pi \Phi _{IJ} \hskip .15in  =
- \pi i \biggl ( \hat \omega _J \D _+ \hat \omega
_I + \hat \omega _I \D _+ \hat \omega _J \biggr ) (\w)
\eea

\section{Appendix: The superdeterminant}
\setcounter{equation}{0}

We have the following two equivalent forms of the superdeterminant
\bea
\sdet \bigg (\matrix{L & M \cr N & P}\bigg ) &=&
\det (L-MP^{-1} N) (\det P)^{-1} 
=
\det L \ \det (P-NL^{-1} M)^{-1} \qquad
\eea
We prove both formulas by showing that the
variations of both sides are equal, using 
\be
\delta \ln \sdet \bigg (\matrix{L & M \cr N & P}\bigg )
= \str \bigg (\matrix{L & M \cr N & P}\bigg ) ^{-1}
\bigg (\matrix{\delta L & \delta M \cr \delta N & \delta P}\bigg )\, .
\ee
The existence of these two seemingly different formulas is due to the
existence of two seemingly different inverses. 
We use the abbreviations $\tilde L = L-MP^{-1}N$ and
$\tilde P=P-NL^{-1}M$. 
The first form of the inverse is
\be
\bigg (\matrix{L & M \cr N & P}\bigg ) ^{-1} =
\bigg (\matrix{
\tilde L^{-1}  & - \tilde L^{-1} MP^{-1} \cr 
 - P ^{-1} N \tilde L^{-1} & 
P ^{-1} + P^{-1} N \tilde L ^{-1} MP^{-1} }\bigg )\, .
\ee
and the resulting variational formula is
\bea
\delta \ln \sdet \bigg (\matrix{L & M \cr N & P}\bigg )
&= &
\tr \tilde L^{-1} \delta L
- \tr \tilde L^{-1} M P ^{-1} \delta N  \\
&& \quad 
+\tr P ^{-1} N \tilde L^{-1} \delta M 
- \tr (P ^{-1} + P^{-1} N \tilde L^{-1} MP^{-1} ) \delta P
\nonumber 
\eea
The second form of the inverse is
\be 
\bigg (\matrix{L & M \cr N & P}\bigg ) ^{-1} =
\bigg (\matrix{
L^{-1} + L^{-1} M\tilde P^{-1}  NL^{-1} & -L^{-1} M
\tilde P ^{-1}
\cr  -\tilde P ^{-1} N L^{-1} & \tilde P ^{-1} }\bigg )\, .
\ee
and the resulting variational formula is 
\bea
\delta \ln \sdet \bigg (\matrix{L & M \cr N & P}\bigg )
&= &
\tr (L^{-1} + L^{-1} M\tilde P^{-1}  NL^{-1})\delta L
- \tr L^{-1} M \tilde P ^{-1} \delta N  \\
&& \quad 
+\tr \tilde P ^{-1} N L^{-1} \delta M - \tr \tilde P ^{-1} \delta P
\nonumber 
\eea
which proves both formulas.
Thus, addition of multiples of columns or rows is
immaterial
\bea
\label{sdet}
\sdet \bigg (\matrix{L & M +LD \cr N & P+ND}\bigg ) 
=
\sdet \bigg (\matrix{L & M \cr N & P}\bigg ) 
\eea

\vfill\eject

\section{Appendix: Slices, Forms and Vector Fields}
\setcounter{equation}{0}

If $x^1,\cdots,x^n$ are local coordinates on an $n$-dimensional manifold
$M$, then the corresponding tangent vectors $H_j$, $1\leq j\leq n$,
are vector fields, normalized on 1-forms $dx^j$,
\bea
H_j={\partial\over\partial x^j}
\hskip 1in 
\< H_k | dx^j \> =\delta_k^j.
\eea
Consider now the case of a slice for supermoduli at genus 2, parametrized
by the coordinates $(\hO_{IJ},\zeta^{\alpha})$, and let the  corresponding
tangent vectors be $H_{IJ},H_{\alpha}$ and the corresponding covectors be
$d\hO_{IJ}$, $d\zeta^{\alpha}$, $I\leq J=1,2$, $\alpha =1,2$. The
preceding duality relations become
\bea
\< H_{KL} | d\hO_{IJ} \> & = & \delta_{IK}\delta_{JL}
\hskip 1in 
\<  H_{KL} | d\zeta^{\alpha} \> =0
\nonumber \\
\< H_{\alpha} | d\hO_{IJ} \> &=& 0
\hskip 1.4in
\< H_{\alpha} | d\zeta^{\beta} \>=\delta_\alpha{}^\beta.
\eea
These duality relations have the following practical use. If the $H_A$ are
known, then they determine completely the forms $dm^A$ within the class of
superholomorphic $3/2$-differentials. If the $dm^A$ are known, then they
determine completely the {\sl equivalence classes} $[H_A]$ of the
superBeltrami differentials $H_A$. Note however that the $H_A$ themselves
are not determined, a distinction of importance in the sequel.

\medskip

For the covectors $d\hO_{IJ}$, we have a concrete description.
Recall the deformation formula (\ref{varsuperperiod}) for the superperiod
matrix resulting from a variation $H_-{}^z$ of a supergeometry
\bea
\delta\hO_{IJ}
= - {i \over 2}
\int d^{2|2}{\bf z} H \biggl (\hat\omega_I{\cal D}_+\hat\omega_J+
\hat\omega_J{\cal D}_+ \hat\omega_I \biggr )
\eea
The left hand side should be interpreted as a pairing
$\<H| d\hO_{IJ} \>$ between the covector $d\hO_{IJ}$ and the vector
$H$ on supermoduli space. The right hand side is just the
usual pairing between super Beltrami differentials and
super quadratic differentials. Thus we may write
\bea
d\hO_{IJ} =  \Phi _{IJ} = - {i \over 2}(\hat\omega_I{\cal D}_+
\hat\omega_J + \hat\omega_J{\cal D}_+ \hat\omega_I) 
\eea
giving a realization of the covector $d\hat\Omega_{IJ}$
in terms of superholomorphic 3/2 differentials.

\subsection{Matrix Elements of $H_A$}

Let $H_A = \bar\theta(\mu_A - \theta \chi _A) $ be the expression of the
super Beltrami differentials $H_A$ in terms of a Beltrami differential
$\mu_A$ and a gravitino variation $\chi _A$. As preparation for the
component formalism, we seek now the matrix elements of $\mu_A$ and $\chi
_A$ themselves. 

\subsubsection{The case of $H_{\alpha}$}

The case of $H_{\alpha}$ is easier, since $\chi _{\alpha}$ is the
gravitino slice, given as an input. The equivalence class $[\mu_{\alpha}]$
can then be determined from the condition  $ \< H_{\alpha}| \Phi _{IJ}
\>=0$, which may be recast in components of $\hat \omega _I = \hat \omega
_{I0} + \theta \hat \omega _{I+}$ as
\bea
\< \mu_{\alpha} | -\hat \o_{I0} \pz \hat \o_{J0} - \hat \o_{J0} \pz
\hat \o_{I0} +2\hat \o_{I+} \hat \o_{J+} \>
+\< \chi_{\alpha} | \hat \o_{I0} \hat \o_{J+} + \hat \o_{J0}
\hat \o_{I+} \> =0
\eea
Since $\hat \o_{I+}=\o_I+\O (\zeta^1 \zeta^2)$, and $\mu_{\alpha}$  and
$\hat \omega_{I0}$ are both of order $\zeta$, we have, in terms of the
holomorphic forms $\o_I$ defined by $\Omega_{IJ}$,
\be
\label{dnine}
\< \mu_{\alpha} |\o_I\o_J \>
=
{1\over 8\pi}
\int d^2zd^2w \chi_{\alpha\bar z}{}^+
S_{\delta}(z,w)\chiw \biggl ( \o_J(z) \o_I(w)
+ \o_I(z) \o_J(w) \biggr )\, .
\ee
which uniquely characterizes the class $[\mu _\alpha]$ of $\mu _\alpha$.

\medskip

We observe that $[\mu_{\alpha}]$ can also be obtained by the following
different argument. For fixed $\hO_{IJ}$ and varying $\zeta_{\alpha}$,
$\mu_{\alpha}$ can be recognized as the Beltrami differential which
shifts  the period matrix $\Omega(\hO,\zeta)$ to $\Omega(\hO,\zeta
+\delta\zeta)$. From the formula (\ref{defsuperperiod}) giving
$\Omega(\hO,\zeta)$, we find readily
\be
\delta \Omega_{IJ}
=
{i\over 8\pi} \int d^2vd^2w
\chi_{\alpha}(v)S_{\delta}(v,w)\chiw \biggl (\o_I(w) \o_J(v) + \o_J(w)
\o_I(v) \biggr ) 
\ee
Comparing with the standard variational formula
$\delta\Omega_{IJ}=i\int d^2z \hat \mu_{\bar z}{}^z\o_I\o_J$,
we obtain again the previous relation (\ref{dnine}).

\subsubsection{The case of $H_{IJ}$}

In components, the equations for $H_{IJ}$ can be rewritten as
\bea
\< \mu_{KL}|\Phi _{IJ+} \> - \< \chi_{KL}|\Phi _{IJ0}\>
&=& \delta_{KI}\delta_{JL}
\nonumber \\
\<\mu_{KL}|\Phi _{\beta +} \>- \< \chi _{KL}|\Phi _{\beta 0}\ >&=&0.
\eea
Although these equations can be solved, they lead to complicated
expressions involving Green's functions, and it is not
easy to see how they can be put to practical use.

\vfill\eject

\end{document}